\documentclass[%
 reprint,
superscriptaddress,
 amsmath,amssymb,
 aps,
]{revtex4-1}

\usepackage{graphicx}
\usepackage{dcolumn}
\usepackage{bm}
\usepackage{soul}
\usepackage{float}
\newcommand{\vectornorm}[1]{\left|\left|#1\right|\right|}

\usepackage{caption}
\usepackage{subcaption}
\newcommand{\beginsupplement}{%
        \setcounter{table}{0}
        \renewcommand{\thetable}{S\arabic{table}}%
        \setcounter{figure}{0}
        \renewcommand{\thefigure}{S\arabic{figure}}%
     }

\usepackage{color} 

\begin{document}

\preprint{APS/123-QED}

\title{The Robustness of Cluster Expansion: Assessing the Roles of Relaxation and Numerical Error}

\author{Andrew H. Nguyen}
\affiliation{%
 Department of Physics and Astronomy, Brigham Young University, 
 Provo, Utah, 84602, USA
}%
 
\author{Conrad W. Rosenbrock}%
\affiliation{%
 Department of Physics and Astronomy, Brigham Young University, 
 Provo, Utah, 84602, USA
}%

\author{C. Shane Reese}%
\affiliation{%
 Department of Statistics, Brigham Young University, 
 Provo, Utah, 84602, USA
}%

\author{Gus L. W. Hart}%
 \email{gus.hart@gmail.com}
\affiliation{%
 Department of Physics and Astronomy, Brigham Young University, 
 Provo, Utah, 84602, USA
}%

\date{\today}

\begin{abstract}

Cluster expansion (CE) is effective in modeling the stability of metallic 
alloys, but sometimes cluster expansions fail. Failures are often attributed to
atomic relaxation in the DFT-calculated data, but there is no metric for
quantifying the degree of relaxation. Additionally, numerical errors can also
be responsible for slow CE convergence. We studied over one hundred different
Hamiltonians and identified a heuristic, based on a normalized mean-squared
displacement of atomic positions in a crystal, to determine if the effects of
relaxation in CE data are too severe to build a reliable CE model. Using this
heuristic, CE practitioners can determine a priori whether or not an alloy
system can be reliably expanded in the cluster basis. We also examined the
error distributions of the fitting data. We find no clear relationship between
the type of error distribution and CE prediction ability, but there are clear
correlations between CE formalism reliability, model complexity, and the number 
of significant terms in the model. Our results show that the \emph{size} of the
errors is much more important than their distribution.

\end{abstract}

\maketitle


\section{\label{sec:level1}Introduction}

Increasing computational power and algorithmic advancements are making many
computational materials problems more tractable. For example, density
functional theory (DFT) is used to assess the stability of potential metal
alloys with high accuracy. However, DFT computational burdens prevent feasible
exploration of all possible configurations of a system. In certain cases, one
can map first-principles results on to a faster Hamiltonian, the cluster 
expansion (CE) \cite{Sanchez1984,Sanchez1993,Sanchez2010}. Over the past 30 
years, CE has been used in combination with first-principles calculations to 
predict the stability of metal alloys \cite{Connolly1983, 
Ferreira1987,Ferreira1988,Ferreira1989,Lu1991,Wolverton1997,
VandeWalle2002,Barabash2006,Asato2007,Nelson2013a,Gargano2015,
Aldegunde2016,Chinnappan2016}, to study the stability of oxides
\cite{Ceder2000,Seko2006,Seko2008,Tanaka2010a,Tanaka2010}, and to model
interaction and ordering phenomena at metal surfaces
\cite{Muller2003,Drautz2003,Muller2006,Herder2015,Tanaka2015}. 
Numerical error and \emph{relaxation} effects decrease the predictive power of
CE models. The aim of this paper is to demonstrate the effects of both and to
provide a heuristic for knowing when a reliable CE model can be expected for a
particular material system.
 
 CE treats alloys as a purely configurational problem, i.e., a problem
 of decorating a fixed lattice with the alloying elements
 \cite{Sanchez1984,Sanchez1993}. However, CE models are usually
 constructed with data taken from ``relaxed'' first-principles
 calculations where the individual atoms assume positions that
 minimize the total energy, displaced from ideal lattice
 positions. Unfortunately, cluster expansions of systems with larger
 lattice relaxation converge more slowly than cluster expansions for
 unrelaxed systems \cite{Laks1992}. In fact, CE with increased
 relaxation may fail to converge altogether. No rigorous description
 of conditions for when the CE breakdown occurs exists in the literature. 

A persistent question in the CE community regards the impact of
relaxation on the accuracy of the cluster expansion. Proponents of CE
argue that the CE formalism holds even when the training structures
are relaxed because there is a one-to-one correspondence in
configurational space between relaxed and unrelaxed structures. In this paper,
we demonstrate a relationship between relaxation and loss of sparsity in the CE
model. As sparsity decreases, the accuracy of CE prediction decreases.

In addition to the effects of relaxation, we also examine the impact of
numerical error on the reliability of the CE fits. There are several
sources of numerical error: approximations to the physics of the
model, the number of $k$-points, the smearing method, basis set sizes and
types, etc. Most previous studies \cite{Arnold2010,Diaz-Ortiz2007a,
Diaz-Ortiz2007} only examine the effect of Gaussian errors on the CE model,
but Arnold et al. \cite{Arnold2010} also investigated systematic
error (round-off and saturation error). They showed that,
above a certain threshold, the CE model fails to recover the correct
answer, that is, the CE model started to incorporate spurious terms
(i.e., sparsity was reduced).  A primary question that we seek to answer is
how the shape of the error distribution impacts predictive performance of a 
CE model.
 
In this study, we quantify the effects of: 1) relaxation, by comparing CE fits
for relaxed and unrelaxed data sets, and 2) numerical error, by adding
different error distributions (i.e., Gaussian, skewed, etc.) to ideal CE
models. We study more than one hundred Hamiltonians ranging from very simple
pair potentials to first-principles DFT Hamiltonians. We present a heuristic
for judging the quality of the CE fits. We find that a small mean-squared
displacement is indicative of a good CE model. In agreement with past studies,
we show that the predictive power of CE is lowered when the level of error is
increased. We find that there is no clear correlation between the shape of the
error profile and the CE predictive power. It is possible to decide whether the
computational cost of generating CE fitting data is worthwhile by examining the
degree of relaxation in a smaller set of 50--150 structures. 

\section{\label{sec:level2}Relaxation}

\begin{figure}[H]
\centerline{\includegraphics[width=0.55\textwidth]{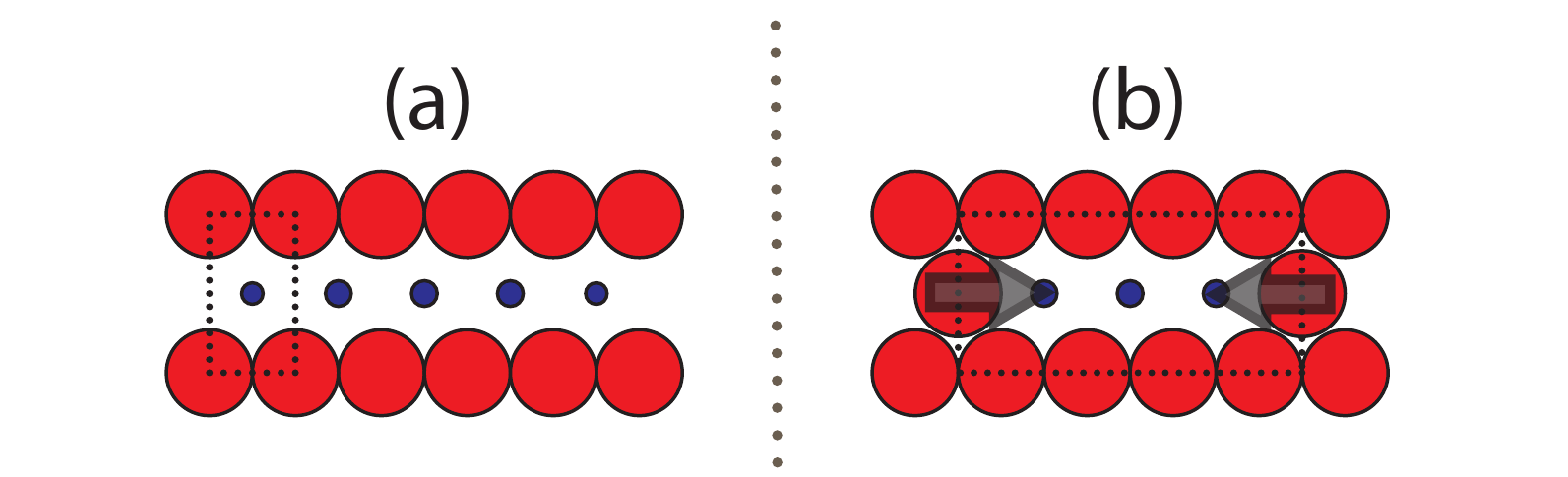}}
\caption{\label{fig:symmetry_allowed_relaxations} (color online) Symmetry-
allowed distortions for two different unit cells. The atomic positions of the 
cell on the left do not have any symmmetry-allowed degrees of freedom, but the 
aspect ratio of the unit cell is allowed to change. For the unit cell on the 
right, the horizontal positions of the atoms in the middle layer may change 
without destroying the symmetry. (The unit cell aspect ratio may also change.)}
\end{figure}

Relaxation is distinct from numerical error---it is not an error---but it has a
similar negative effect. When relaxations are significant, it is less likely
that a reliable CE model exists. Relaxation is a systematic form of distortion,
the local adjustment of atomic positions to accommodate atoms of different 
sizes. Atoms ``relax'' away from ideal lattice sites to reduce the energy, with
larger atoms taking up more room, smaller atoms giving up volume. The type of
 relaxations (i.e., the distortions that are possible) for a particular unit
 cell are limited by the symmetry of the initially undistorted case, as shown
in  Fig. \ref{fig:symmetry_allowed_relaxations}. In the rectangular case 
(left), the unit cell aspect ratio may change without changing the initial
rectangular symmetry. At the same time, the position of the blue atom is 
\emph{not allowed} to change because doing so would destroy rectangular
symmetry. In contrast, the two blue atoms in the similar structure shown in the
right panel of the figure can move horizontally without reducing the symmetry.  

\begin{figure}[H]
\centerline{\includegraphics[width=0.55\textwidth]{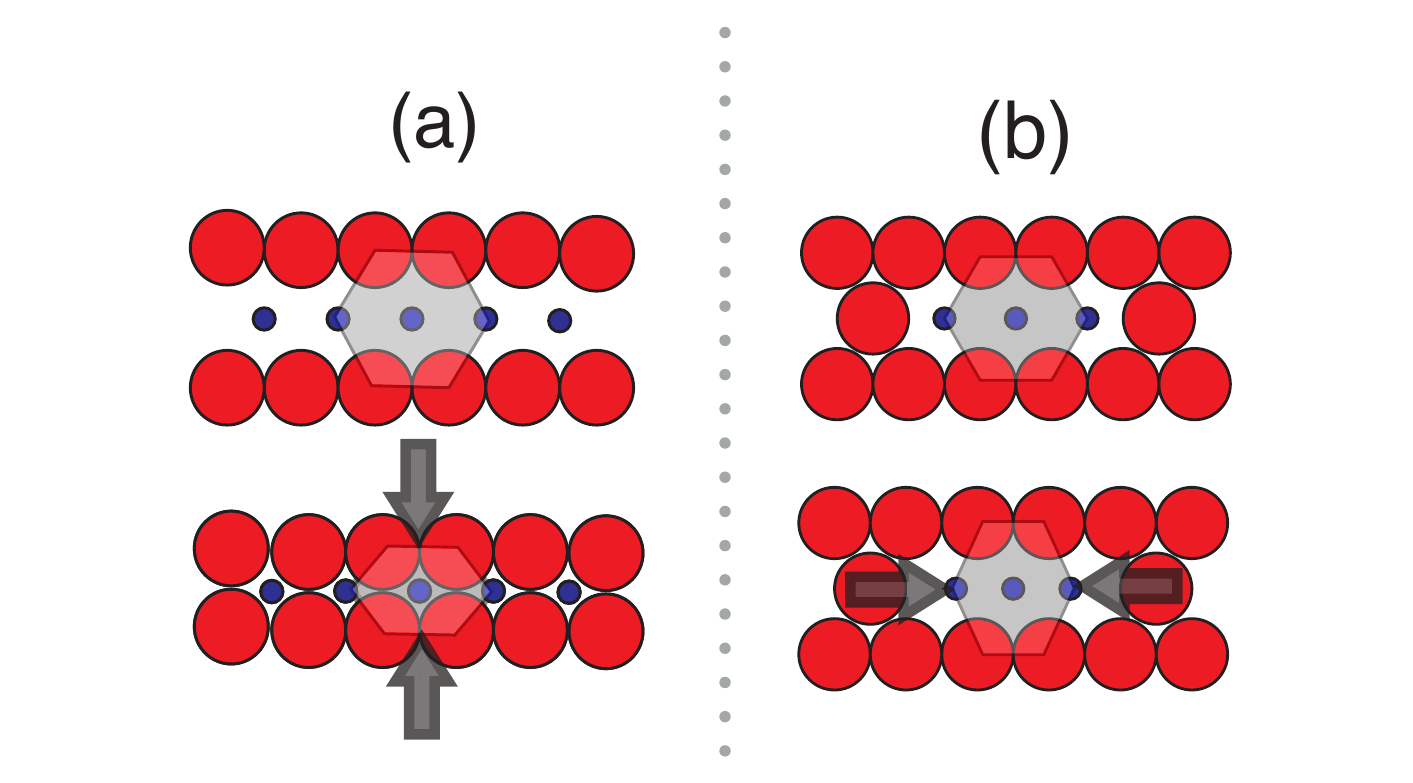}}
\caption{\label{fig:relaxEX} (color online)  Relaxation scheme. The top images
show the original unrelaxed configurations, while the bottom figures show the 
relaxed configuration.  The left images (a) shows the relaxation where the 
hexagon is contracted as shown by the black arrows in bottom left figure. The 
relaxation in the right images (b) is restricted to displacement of the blue 
atoms as shown by the black arrows in bottom right figure. }
\end{figure}

Conceptually, the cluster expansion is a technique that describes the local
environment around an atom and then sums up all the ``atomic energies'' 
(environments in a unit cell) to determine a total energy for the unit cell. 
For the cluster expansion model to be sparse---to be a predictive model with
few parameters---it relies on the premise that any specific local
neighborhood contributes the same atomic energy to the total energy regardless 
of the crystal in which it is embedded. For example, the top row of Fig.
\ref{fig:relaxEX} shows the same local environment (denoted by the hexagon 
around the central blue atom) embedded in two distinct crystals. If the
contribution of this local environment to the total energy is the same in both
 cases, then the cluster expansion of the energy will be sparse.   

The effect of relaxation on the sparsity becomes clear in the bottom row of
 Fig.~\ref{fig:relaxEX}. In the left-hand case [panel (a)], the crystal relaxes 
dramatically and the central blue atom is now \emph{four-fold coordinated}
entirely by red atoms. 
By contrast, in the right-hand case [panel (b)], a collapse of the layers is
not possible and the blue atoms are allowed by symmetry to move closer to each
other. From the point of view of the cluster expansion, the local environments
of the central blue atom are the same for both cases.
This fact, that two different relaxed local environments have identical
descriptions in the cluster expansion basis, leads to a slow convergence of
cluster expansion models. The problem is severe when the atomic mismatch is 
large and relaxations are significant (i.e., when atoms move far from the 
ideal lattice positions.)

\subsection{Methodology}
We investigated the predictive power of cluster expansions using data
from more than one hundred Hamiltonians generated from Density Functional
Theory (DFT), the embedded atom method, Lennard-Jones potential and
Stillinger-Weber potential. To investigate the
effects of relaxation, we examined different metrics to measure the
degree of atomic relaxation in a crystal configuration. 

\subsubsection{Hamiltonians}

First-principles DFT calculations have been used to simulate metal alloys and
for building cluster expansion models
\cite{Ferreira1989,Wolverton1997,VandeWalle2002,Barabash2006,Asato2007,
Nelson2013a,Gargano2015}.  However, DFT calculations are too expensive to
extensively examine the relaxation in many different systems (lattice
mismatch). Thus, we examine other methods such as the embedded atom method
(EAM) which is a multibody potential. The EAM potential is a semi-empirical 
potential derived from first-principles calculations. EAM potentials of metal 
alloys such as Ni-Cu, Ni-Al, and Cu-Al have been parameterized from DFT 
calculations and validated to reproduce their experimental properties such as 
bulk modulus, elastic constants, lattice constants, etc. \cite{Foiles1986}. EAM
potentials are computationally cheaper, allowing us to explore the effects of 
relaxation for large training sets; however, we are limited by the number of 
EAM potentials available.

Therefore, we also selected two classical potentials, Lennard-Jones (LJ) and 
Stillinger-Weber (SW), to adequately examine various degrees of relaxation, 
which can be varied using free parameters in each model. The Lennard-Jones 
potential is a pairwise potential. Using the LJ potential, we can model a 
binary ($\mathrm{A_x}\mathrm{B_{1-x}}$) alloy with different lattice mismatch 
and interaction strength between the A and B atoms by adjusting the $\sigma$ 
parameter in the model.  Additionally, we also examined the Stillinger-Weber 
potential which has a pair term and an angular (three-body) term.  In 
attempting to determine the conditions under which the CE formalism breaks 
down, we implemented a set of parameters in the SW potential where the angular 
dependent term could be turned on/off using the $\lambda$ coefficient 
\cite{Stillinger1985}. For example, depending on the strength of $\lambda$, the
local atomic environment in 2D could switch between 3-, 4- and 6-fold 
coordination by changing a single parameter. Thus, when the system relaxes to
a different coordination, the CE fits would no longer be valid or at least not
sparse.  
       
All first-principles calculations were performed using the Vienna Ab initio 
Simulation package (VASP) \cite{Kresse1993,Kresse1994,Kresse1996a,Kresse1996}. 
We used the projector-augmented-wave (PAW) \cite{Blochl1994} potential and the 
exchange-correlation functional proposed by Perdew, Burke, and Ernzerhof (PBE) 
\cite{Perdew1996}. In all calculations, we used the default settings implied by
the high-precision option of the code. Equivalent $k$-point meshes were used 
for Brillioun zone integration to reduce numerical errors \cite{Froyen1989}. We
used 1728 ($12^3$) $k$-points for the pure element structures and an equivalent
mesh for the binary alloy configurations. Each structure was allowed to fully 
relax (atomic, cell shape and cell volume). 

Relaxation was carried out using molecular dynamics simulations for EAM, LJ and
SW potentials. Two molecular dynamics packages were used to study the 
relaxation: GULP \cite{Gale1997,Gale2003} and LAMMPS \cite{Plimpton1995}. 
Details for the LJ, SW and EAM potentials and the DFT calculations can be found
in the supplementary materials\footnote{See supplemental materials for details about the different potentials and their parameters.}. 

\subsubsection{Cluster Expansion Setup}
The universal cluster expansion (UNCLE) software 
\cite{Hart2008,Hart2009,Lerch2009} was used to generate 1000 derivative 
superstructures each of  face-centered cubic (FCC), body-centered cubic (BCC) 
and hexagonal closed-packed (HCP) lattice. For the DFT calculations, we used 
only 500 structures instead of 1000 due to the computational cost. We generated
a set of 1100 clusters, ranging from 2-body up to 6-body interactions. 100 
independent CE fits were performed for each system (Hamiltonian and lattice). 

We performed cluster expansions using the UNCLE software \cite{Lerch2009}.  We 
briefly discuss some important details about cluster expansion here, but for a 
more complete description, see the supplementary materials \footnote{See 
supplemental materials for more details on cluster expansion.} and past works 
\cite{Connolly1983,Sanchez1984,Ferreira1991,Zunger2002,VandeWalle2002,
VandeWalle2013,Nelson2013,Nelson2013a}. Cluster expansion is a generalized 
Ising model with many-body interactions. The cluster expansion formalism allows
one to map a physical property, such as E, to configuration ($
\overrightarrow{\sigma}$):
\begin{equation}
 E_i^{\mathrm{CE}} =\Sigma_i J_i \Pi_i(\overrightarrow{\sigma})
\end{equation} 
 where E is energy, $\Pi$ is the correlation matrix (basis), and $J$ is 
 coefficient or effective cluster interaction (ECI). 
 
When constructing a CE model, we are solving for the effective cluster 
interactions, or $J$s. We used the compressive sensing (CS) framework to solve 
for these coefficients \cite{Nelson2013,Nelson2013a}. The key assumption in 
compressive sensing is that the solution vector has few nonzero components, 
i.e., the solution is sparse \cite{Candes2006,Candes2008}. The CS framework 
guarantees that the sparse solution can be recovered from a limited number of 
DFT energies. Using the $J$s, we can build a CE model to interpolate the
configuration space.

Each CE fit used a random selection of 25\% of the data for training and 75\% 
for validation. Results were averaged over the 100 CE fits with error bars 
computed from the standard deviation. We defined the percent error as a ratio 
of the  prediction root mean squared error (RMS) over the standard deviation of
the input energies, $\mathrm{percent\ error} = \mathrm{RMS}/ 
\mathrm{STD(E_{input})} \times 100\%.$ This definition of percent error allowed
us to consistently compare different systems.

\subsubsection{Relaxation Metrics}
Currently, there is no standard measure to indicate the degree of relaxation. 
We evaluated different metrics as a measure of the relaxation: normalized mean-
squared displacement, Ackland's order parameter \cite{Ackland2006}, difference 
in Steinhardt order parameter ($D_6$) \cite{Steinhardt1983}, SOAP 
\cite{Bartok2013}, and the centro-symmetry parameter \cite{Kelchner1998}. We 
compared the metrics across various Hamiltonians to find a criterion that is 
independent of the potentials and systems \footnote{See supplemental materials
for details about the different relaxation metrics.}. We found that none of
these metrics are descriptive/general enough except for the normalized mean-
squared displacement. 

\subsubsection{Normalized Mean-Squared Displacement (NMSD)}
To measure the relaxation of each structure/configuration, we used the mean-
squared displacement (MSD) to  measure the displacement of an atom from its 
reference position, i.e., the unrelaxed atomic position. The MSD metric is 
implemented in the LAMMPS software \cite{Plimpton1995}, which also incorporates
the periodic boundary conditions to properly account for displacement across a 
boundary. The MSD is the total squared displacement averaged over all atoms in 
the crystal:

\begin{equation}
\mathrm{MSD} = \frac{1}{N_{\mathrm{atom}}} \sum_{\mathrm{atom}} \sum_{X=x,y,z} 
(X[t] - X[0])^2 
\end{equation}

\noindent where $t$ is the final relaxed configuration and 0 is the initial 
unrelaxed configuration. Additionally, we defined a normalized mean-square 
displacement (NMSD) percent:
 \begin{equation}
 \mathrm{NMSD} = \frac{\mathrm{MSD}}{V^{2/3}} \times 100\%
\end{equation}
\noindent which is the ratio of MSD to volume of the system. This allows for a 
relaxation comparison parameter that is independent of the overall scale.
  
\subsection{Results and Discussions}

To explore the effects of relaxation on CE predictability, we examine 
relaxation in various systems from very high accuracy (DFT) to very simple, 
tunable systems (LJ and SW potentials). We examine more than one hundred 
different Hamiltonians and we find several common trends among the different 
systems. 
 
\begin{figure}[H]
\centerline{\includegraphics[width=0.5\textwidth]{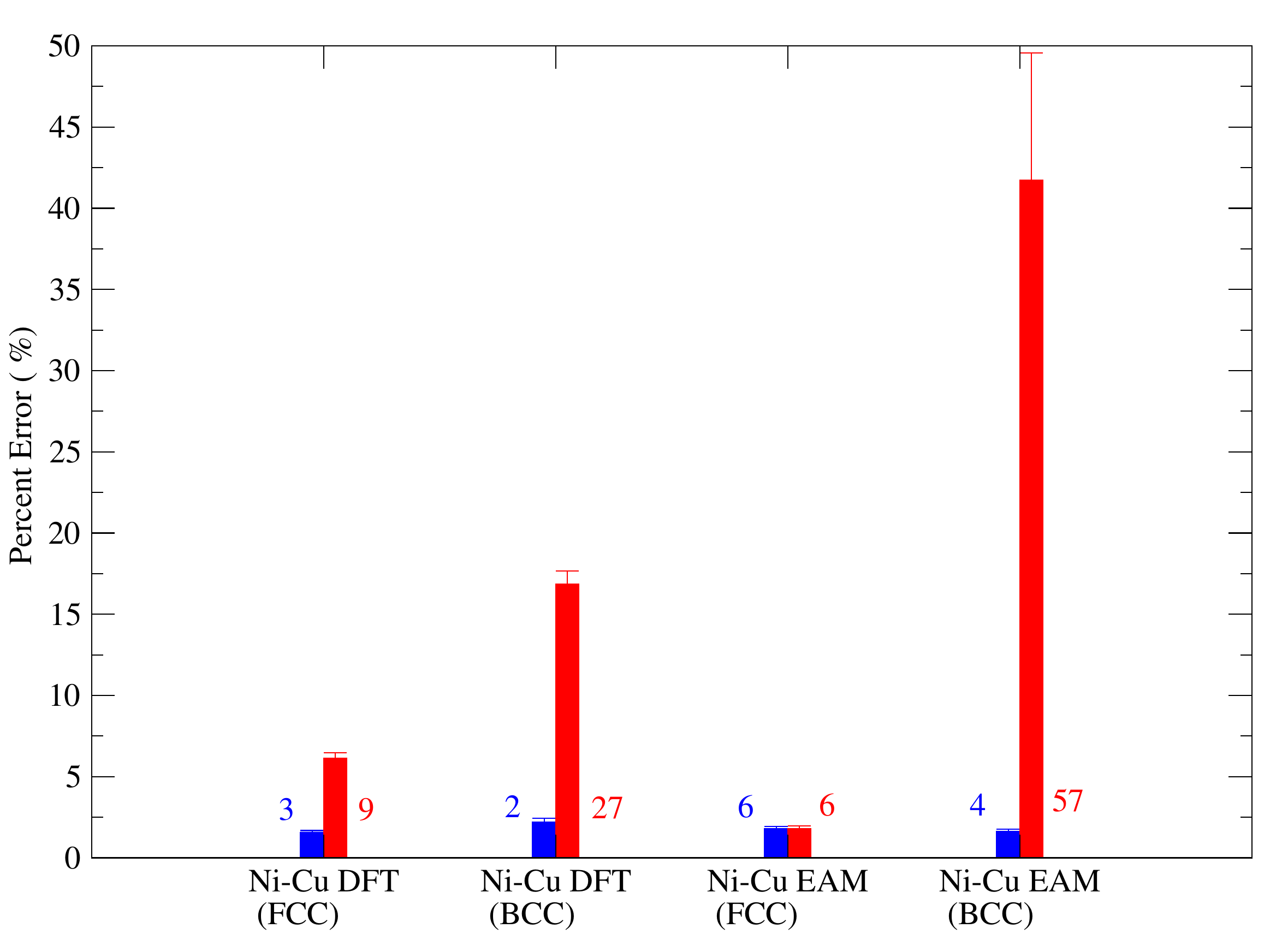}}
\caption{\label{fig:Unrel-rel-VASP-EAM} (color online)  Cluster expansion fits 
for Ni-Cu alloy using DFT or EAM potential. Each bar represents the average 
percent error and error bar (standard deviations) for 100 independent CE fits. 
The blue bars represent the unrelaxed CE fits, while the red bars represent the
relaxed CE fits. The colored number represents the average number of 
coefficients used in the CE models. When the configurations are relaxed, we 
find that the CE fits are often worse (higher prediction error and higher 
number of $J$s) than unrelaxed system. However, we show that in one case (Ni-Cu
 EAM) the unrelaxed and relaxed CE fits are identical (same error and same 
number of coefficients) and this is due to a small relaxation.}
\end{figure}

In most cases, we find that the relaxed CE fits are worse (higher prediction 
error and higher number of coefficients) than the unrelaxed. For example, 
Fig. \ref{fig:Unrel-rel-VASP-EAM} shows the cluster expansion fitting for 
unrelaxed and relaxed data sets of Ni-Cu alloy system using DFT and EAM with 
two different primitive lattices.  Though it seems strange for us to model Ni-
Cu using a BCC primitive lattice when Ni-Cu is closed-packed, this is a method 
for us to evaluate the relaxation of Ni-Cu for a highly relaxed system. As 
Fig. \ref{fig:Unrel-rel-VASP-EAM} shows, Ni-Cu alloy fitting for a FCC 
lattice is below 10\% error, while BCC fitting result in more $J$s and higher 
percent error (above 10\%) \footnote{In our experience, a percent error above 
10\% often gives unreliable CE model.}.  We find similar results in the 
relaxation of Ni-Cu alloy using first-principles DFT and EAM potential. The 
difference between relaxed and unrelaxed CE fits are negligible when 
relaxations are small. This is shown in fig. \ref{fig:Unrel-rel-VASP-EAM} for 
the relaxation of FCC superstructures using a Ni-Cu EAM potential.

\begin{figure*}[]
     \raggedright 
    \begin{subfigure}[b]{0.55\textwidth}
        \caption{Histogram of clusters in \ref{fig:VASP-FCC-CE} }
        \includegraphics[width=\textwidth]{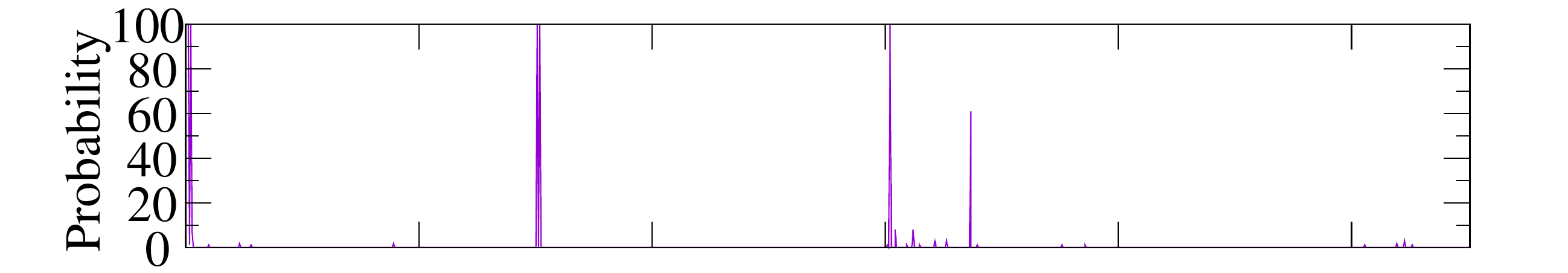}
        \label{fig:VASP-FCC-prob}
    \end{subfigure}
    \begin{subfigure}[b]{0.55\textwidth}
        \includegraphics[width=\textwidth]{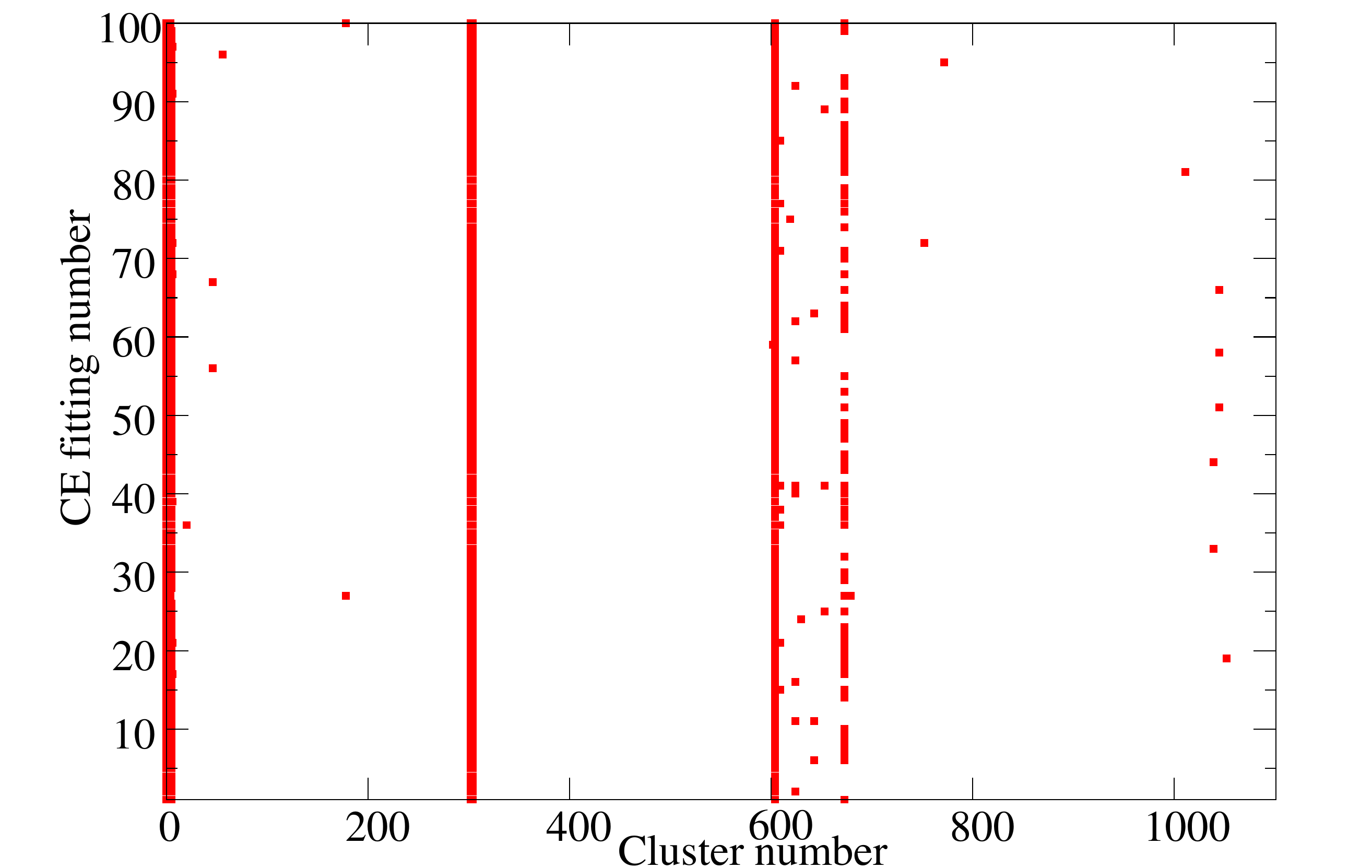}
        \caption{CE fits vs clusters for FCC parent lattice} 
        \label{fig:VASP-FCC-CE}
    \end{subfigure}
    ~ 
    \begin{subfigure}[b]{0.180\textwidth}
        \includegraphics[width=\textwidth]{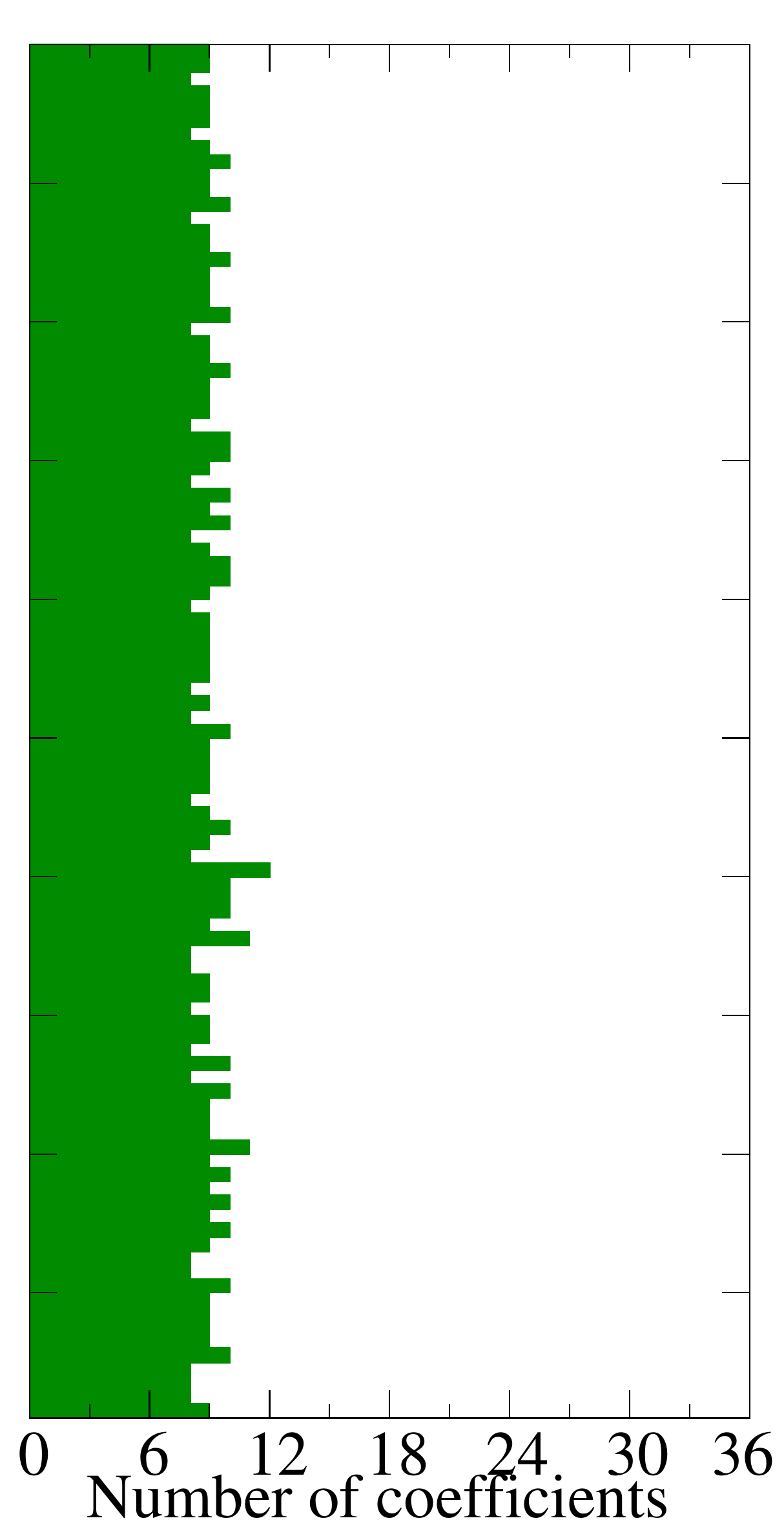}
        \caption{CE fits vs $J$s}
        \label{fig:VASP-FCC-Js}
    \end{subfigure}
    ~ 
    \begin{subfigure}[b]{0.178\textwidth}
        \includegraphics[width=\textwidth]{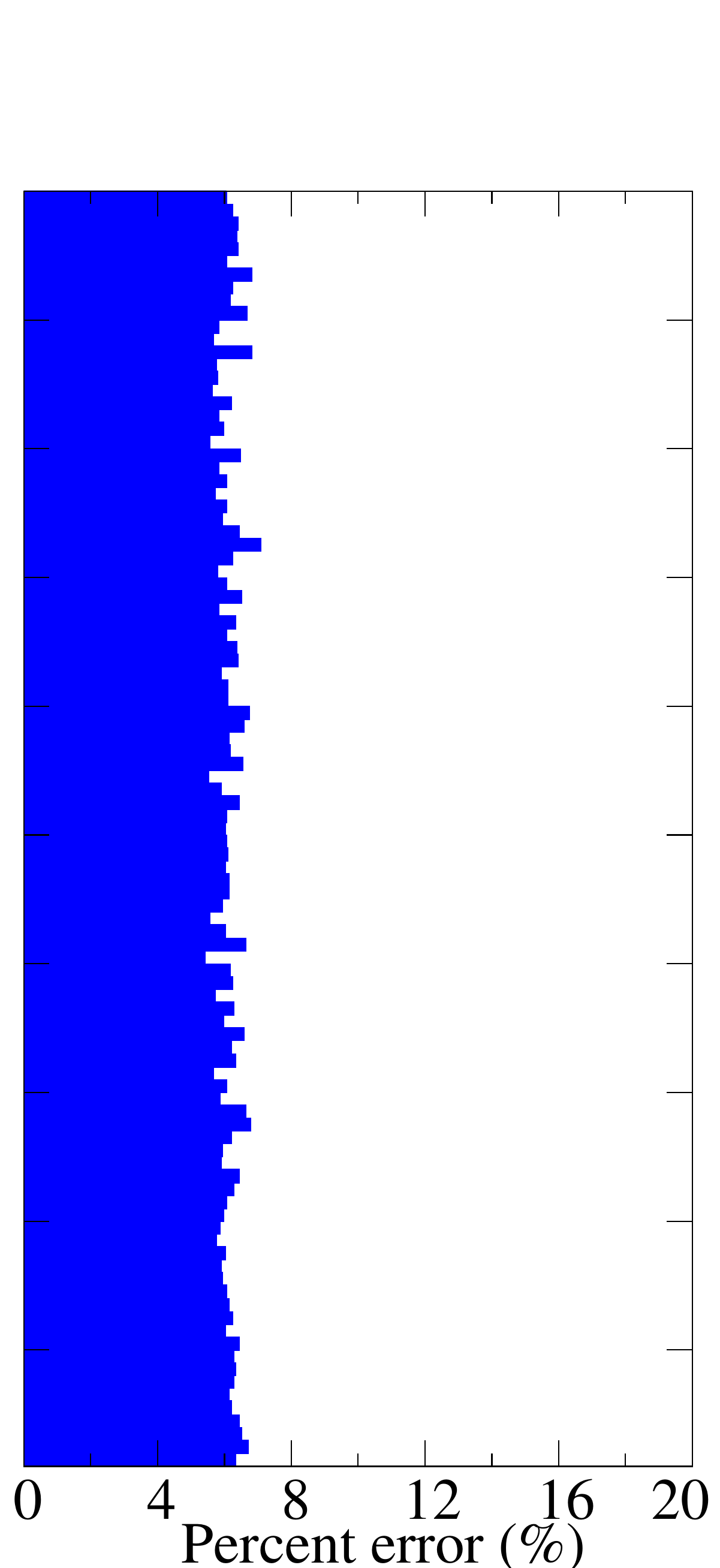}
        \caption{CE fits vs errors }
        \label{fig:VASP-FCC-error}
    \end{subfigure}
    \begin{subfigure}[b]{0.55\textwidth}
        \caption{Histogram of clusters in \ref{fig:VASP-BCC-CE}}
        \includegraphics[width=\textwidth]{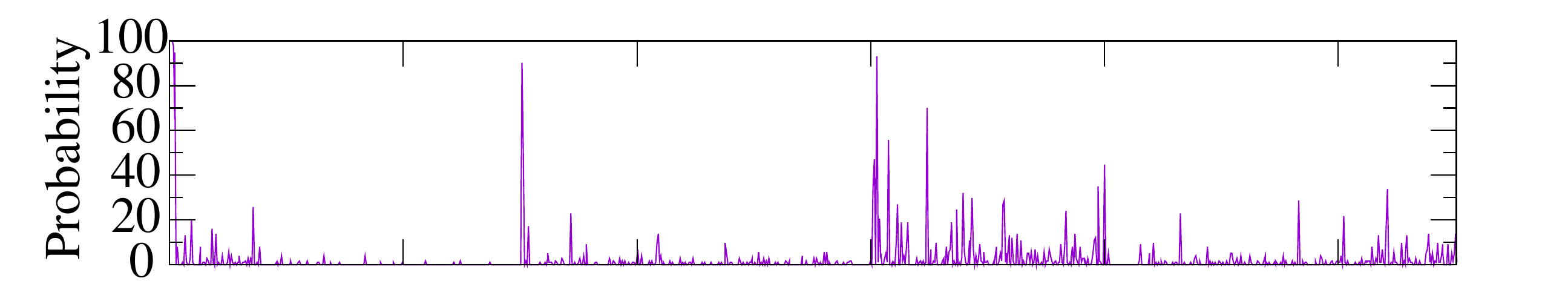}
        \label{fig:VASP-BCC-prob}
    \end{subfigure}
\begin{subfigure}[b]{0.55\textwidth}
        \includegraphics[width=\textwidth]{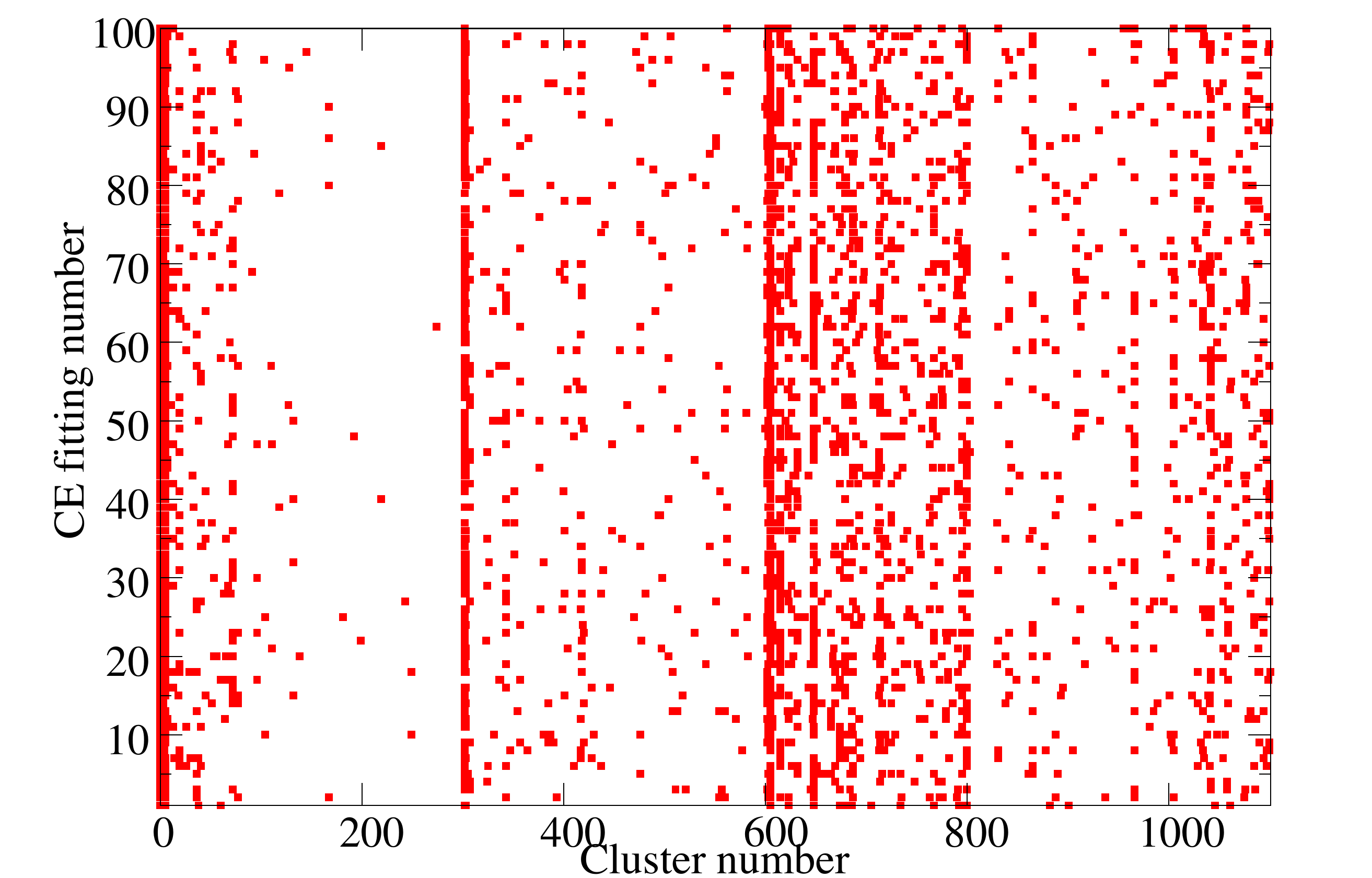}
        \caption{CE fits vs clusters for BCC derivative structures}
        \label{fig:VASP-BCC-CE}
    \end{subfigure}
    ~ 
    \begin{subfigure}[b]{0.178\textwidth}
        \includegraphics[width=\textwidth]{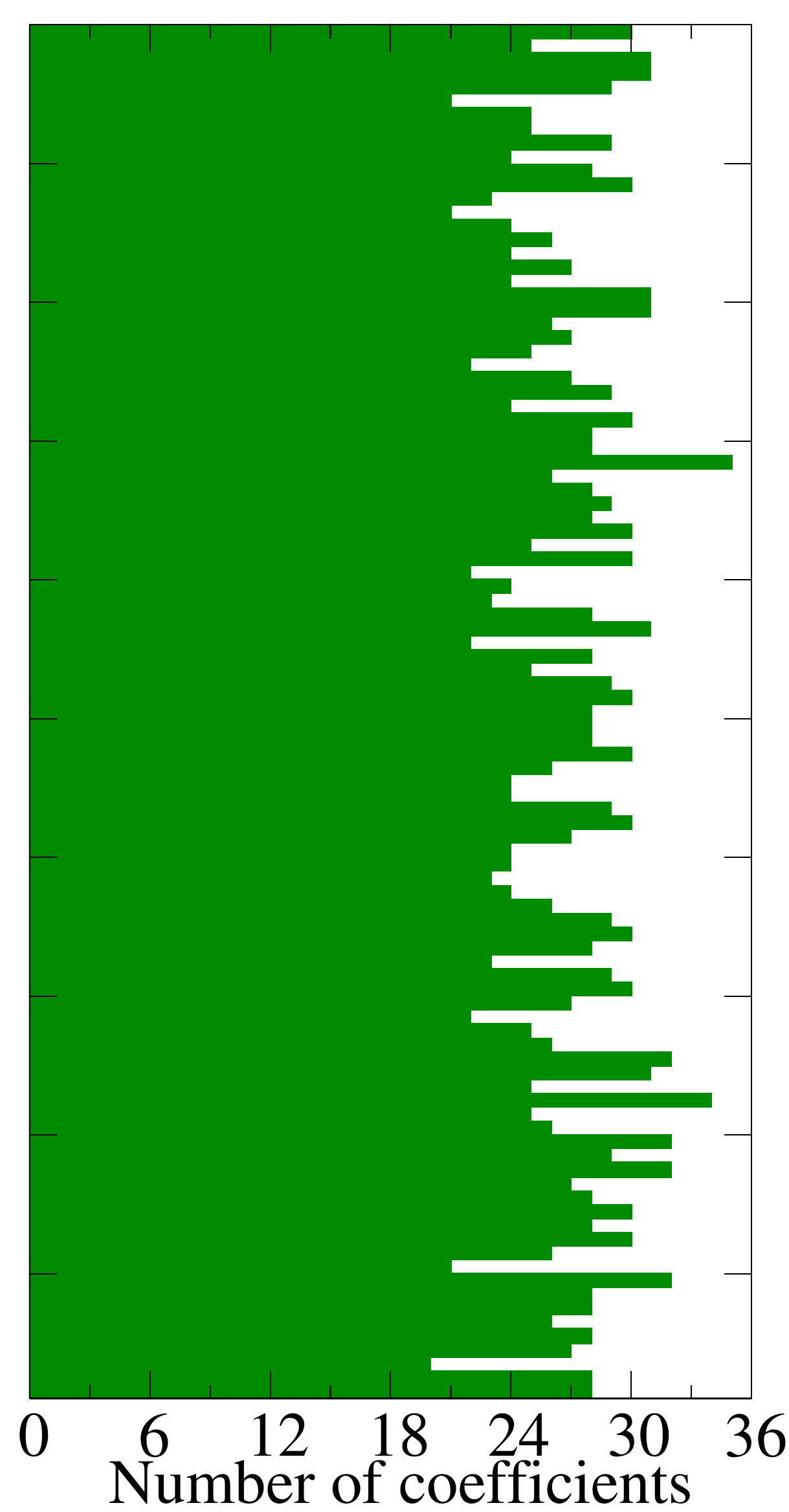}
        \caption{CE fits vs $J$s}
        \label{fig:VASP-BCC-Js}
    \end{subfigure}
    ~ 
    \begin{subfigure}[b]{0.192\textwidth}
        \includegraphics[width=\textwidth]{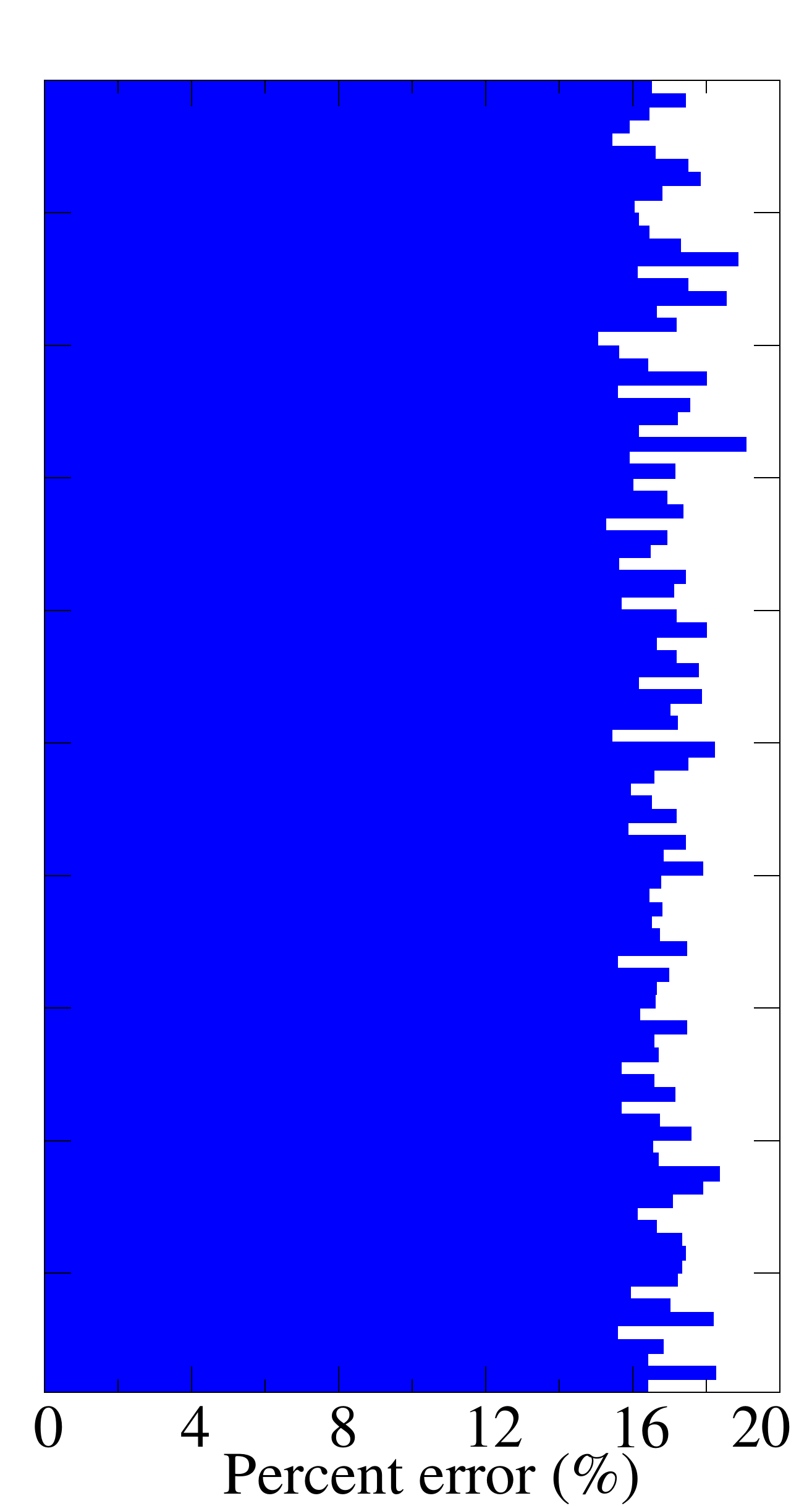}
        \caption{CE fits vs errors}
        \label{fig:VASP-BCC-error}
    \end{subfigure}  
    \captionsetup{justification=RaggedRight}  
    \caption{(color online) CE fitting and relaxation of Ni-Cu alloys
      using FCC derivative superstructures and BCC derviative
      superstructures. Shown in Fig. \ref{fig:VASP-FCC-CE} and
      \ref{fig:VASP-FCC-prob} are the 100 CE fits and the histogram of
      the clusters used for the FCC lattices, while plot
      \ref{fig:VASP-BCC-CE} and \ref{fig:VASP-BCC-prob} are for the
      BCC lattice. The errors and coefficients are shown in
      \ref{fig:VASP-FCC-Js} and \ref{fig:VASP-FCC-error} for the FCC
      structures, and in \ref{fig:VASP-BCC-Js} and
      \ref{fig:VASP-BCC-error} for the BCC lattice. The plot shows
      that the number of clusters used in fitting is small when
      cluster expansion fitting is good (error is on average 6.03\%
      for FCC derivative structures). However, the CE fitting of BCC
      parent lattice is worse at 16.70\% compared to FCC
      at 6.03\%. More coefficients are used when CE fails. The
      increased number of $J$s and error indicate a bad CE fitting
      model as shown by plots \ref{fig:VASP-BCC-Js} and
      \ref{fig:VASP-BCC-error}. Fig. \ref{fig:VASP-BCC-prob} shows
      only a few significant terms with many other
      clusters used sparingly in the fits.  }
      \label{fig:All-VASP-FCC-BCC-relaxed}
\end{figure*}

\begin{figure}[H]
    \centering
     \begin{subfigure}[b]{0.5\textwidth}
        \includegraphics[width=\textwidth]{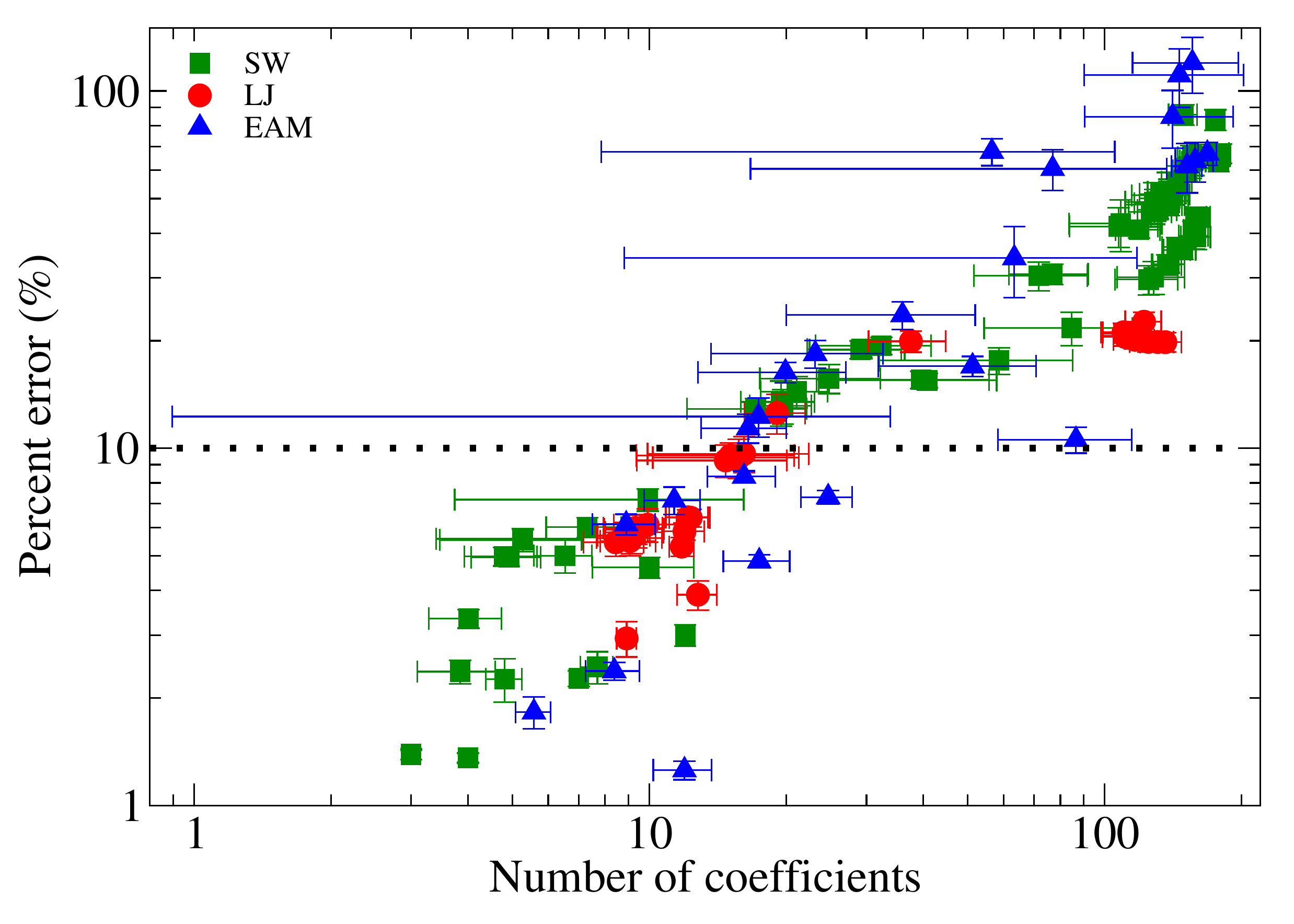}
        \caption{Error vs $J$s}
        \label{fig:rms-js}
    \end{subfigure}
    ~ 
    \begin{subfigure}[b]{0.5\textwidth}
        \includegraphics[width=\textwidth]{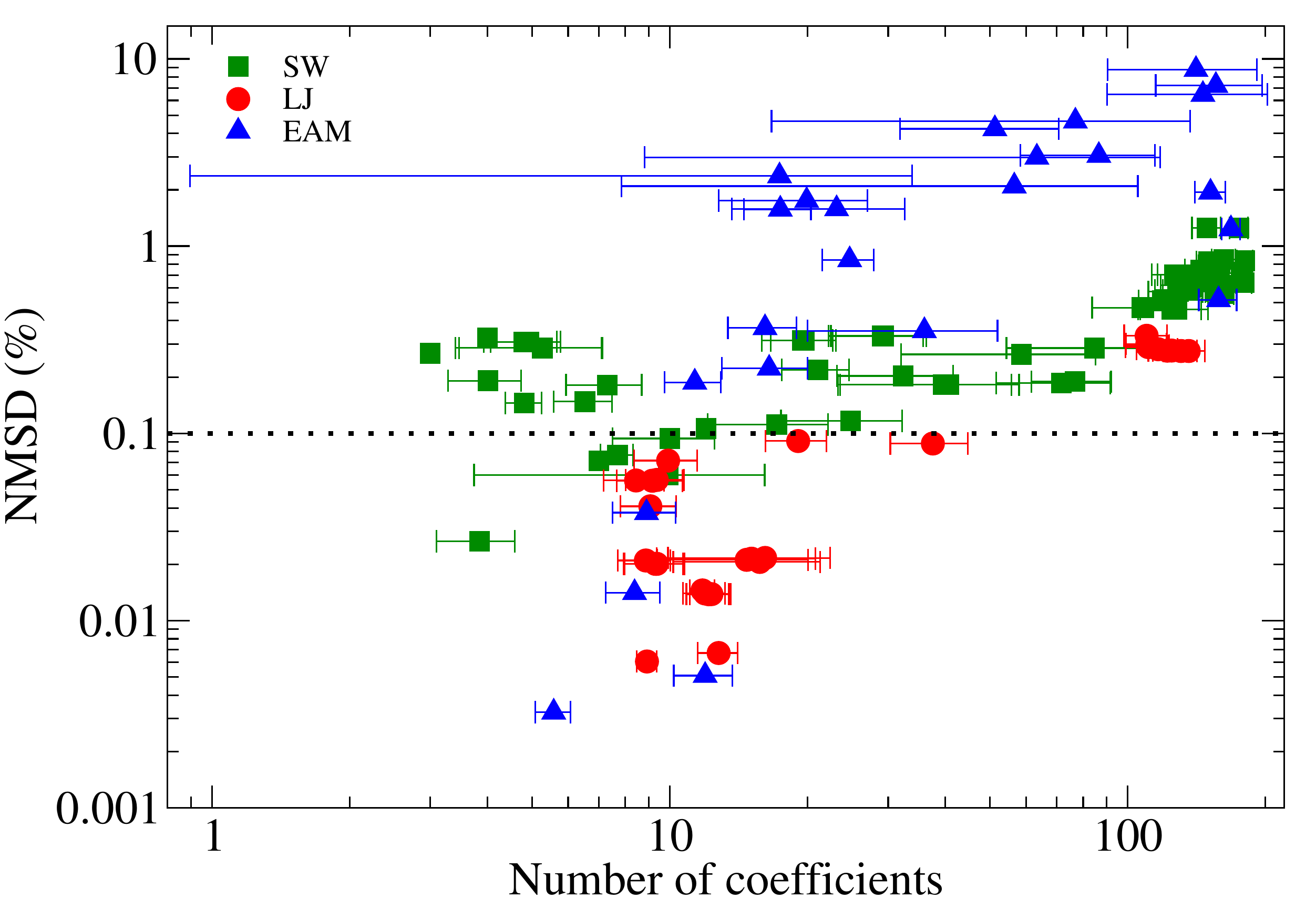}
        \caption{Relaxation vs $J$s }
        \label{fig:msd-js}
    \end{subfigure}
    \caption{(color online) Plot \ref{fig:rms-js} displays the CE fitting error 
vs the number of coefficients, while plot \ref{fig:msd-js} highlights the 
relationship between number of coefficients and relaxation. The dashed line 
approximates what we consider as the maximum acceptable error for a CE model 
(10\%). The dashed line in Fig. \ref{fig:msd-js} marks the estimated 
threshold for acceptable relaxation level. Each symbol represents 100 
independent CE fittings for each Hamiltonian. Higher error correlates with a 
higher number of coefficients.} \label{fig:Relaxed-jsall-error}
\end{figure}

Fig. \ref{fig:Unrel-rel-VASP-EAM} shows that increased relaxation is
associated with reduced sparsity (increased cardinality of $J$s). One
possible implication is that number of coefficients ($J$) could be
used to evaluate the predictive performance of the CE fits. The number
of coefficients used in the fits (such as in fig.
\ref{fig:Unrel-rel-VASP-EAM}) is a simple way to determine whether or
not a CE fit can be trusted. Fig. \ref{fig:VASP-FCC-CE} and
\ref{fig:VASP-BCC-CE} show similar clusters across the 100 independent
CE fittings; thus, vertical lines indicate the presence of the same cluster 
across all CE fits.  When the fit is good, only a small subset of clusters 
is needed (Fig. \ref{fig:VASP-FCC-CE}).  On the other hand, Fig. 
\ref{fig:VASP-BCC-CE} shows some common clusters in all of the CE fits with 
several additional clusters.  Fig. \ref{fig:rms-js} shows the correlation of 
the percent error with the number of terms in the expansion. We find that as 
the number of coefficients increases the percent error increases. However, this
is not a sufficient metric as shown in Fig. \ref{fig:rms-js} where the number 
of coefficient varies a lot. Nonetheless, the number of coefficients may be 
used as a general, quick test.     

The degree of relaxation is crucial to define whether or not the CE
model is accurate or not. However, there is no standard for
\emph{when} cluster expansion fails due to relaxation. Thus far, we
have made some remarks about relaxation and CE fits. But the question
of how much relaxation is allowed has not been addressed. By examining
a few metrics: NMSD, SOAP \cite{Bartok2013}, D6
\cite{Steinhardt1983},Ackland \cite{Ackland2006}  and
centro-symmetry \cite{Kelchner1998}, we find that there is a
relationship between degree of relaxation and the quality of CE
fits. As shown in the supplementary information, we have used these
metrics to investigate over 100+ systems (different potentials,
lattice mismatches, and interaction strengths). Here, we present a
heuristic to measure the degree of relaxation based on the NMSD.

\begin{figure}[H]
\centering
\includegraphics[width=0.5\textwidth]{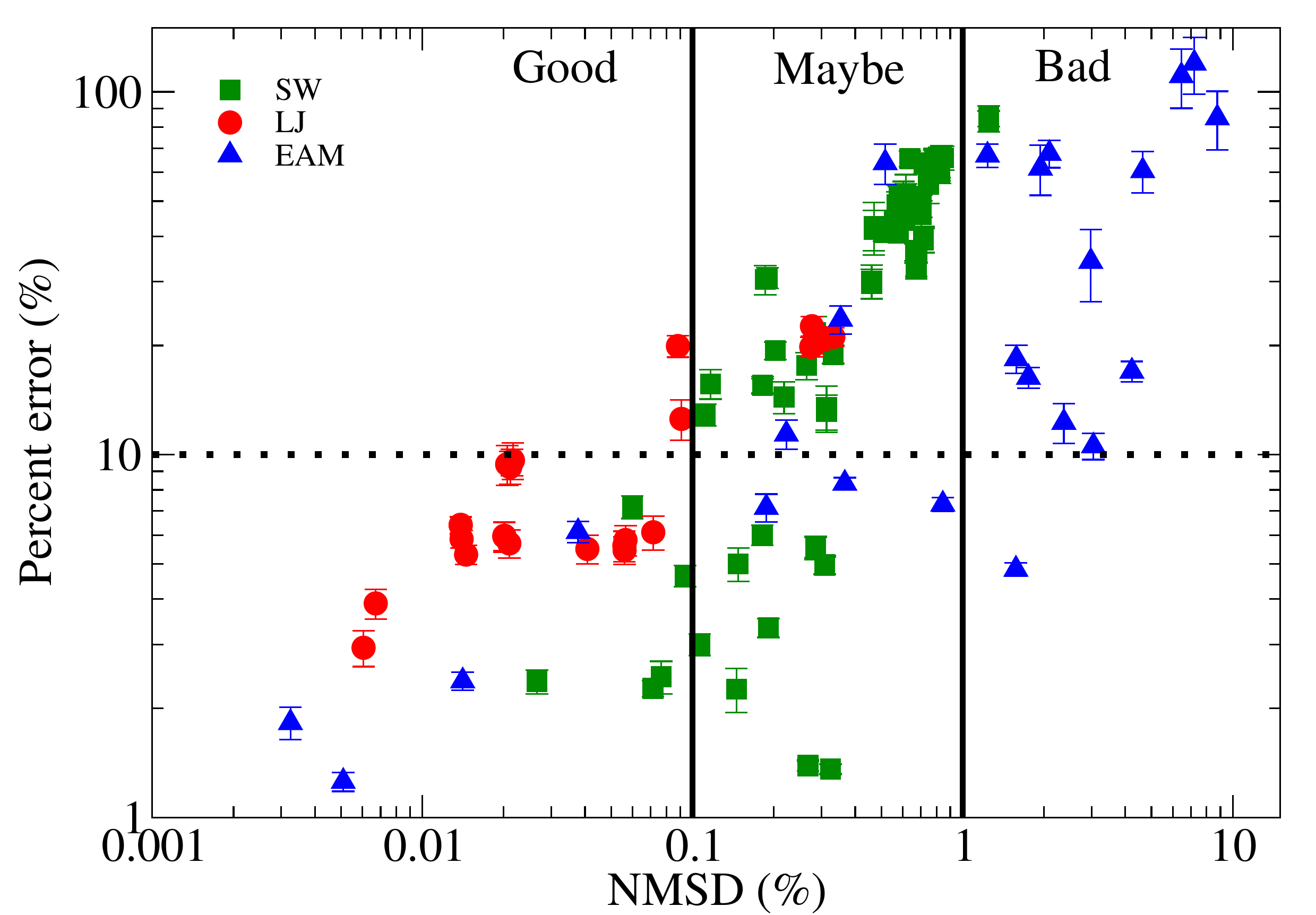}
\caption{(color online) Relationships between relaxation and CE fitting reveal 
a heuristic for determining the quality of a CE model. This graph shows the CE 
fitting error vs normalized mean square displacement (NMSD). Each mark 
represents 100 individual CE fittings for each system (potentials and 
parameters). As the NMSD (relaxation) increases, the CE fitting error increases
 for various systems and potentials. Using the relaxation metric, the quality/
reliability of the CE fits can be divided into three regions: good, maybe and 
bad CE model. The solid black lines indicates these three areas.  
}\label{fig:rms-msd} 
 
\end{figure}

In general, cluster expansion will fail when the relaxation is
large. Figure \ref{fig:msd-js} shows that a small NMSD weakly
correlates with a small number of coefficients. However, Figure
\ref{fig:rms-msd} highlights the correlation between degree of
relaxation and prediction error. There is a roughly linear
relationship between the degree of relaxation and the CE
prediction. We partition the quality of the CE models into three
regions: good (NMSD $<$ 0.1\%), maybe (0.1\% $\leq$ NMSD $\leq$ 1\%)
and bad (NMSD $>$ 1\%). The ``maybe'' region is the gray area where the CE
fit can be good or bad. This metric provide a heuristic to evaluate
the reliability of the CE models, i.e., any systems that exhibit high
relaxation will fail to provide an accurate CE model.   
  
\section{\label{sec:level3}Numerical Error}

As we have shown in the previous section, greater relaxation results in
worse CE fitting. In addition to the effects of relaxation, we now
investigate the effects of numerical error on reliability of CE
models. Numerical error arises from various sources such as
 the number of $k$-points, the smearing method, minimum force tolerance, basis
  set sizes and types, etc. These errors are not stochastic error or 
 measurement errors; they arise from tuning the numerical methods. We
assume that the relaxation-induced change in energy for each structure is an
  \emph{error term} that the CE fitting algorithm must handle. The collection
  of these  ``errors'' from all structures in the alloy system then form an
   error profile  (or distribution). Using the simulated relaxation error
  profiles from the previous section together with common analytic
  distributions, we built ``toy'' CE models with known
  coefficients. We then examined whether or not the shape of the
error distribution affects the CE predictive ability.

\subsection{Methodology}

\begin{figure}[]
\centerline{\includegraphics[width=0.5\textwidth]{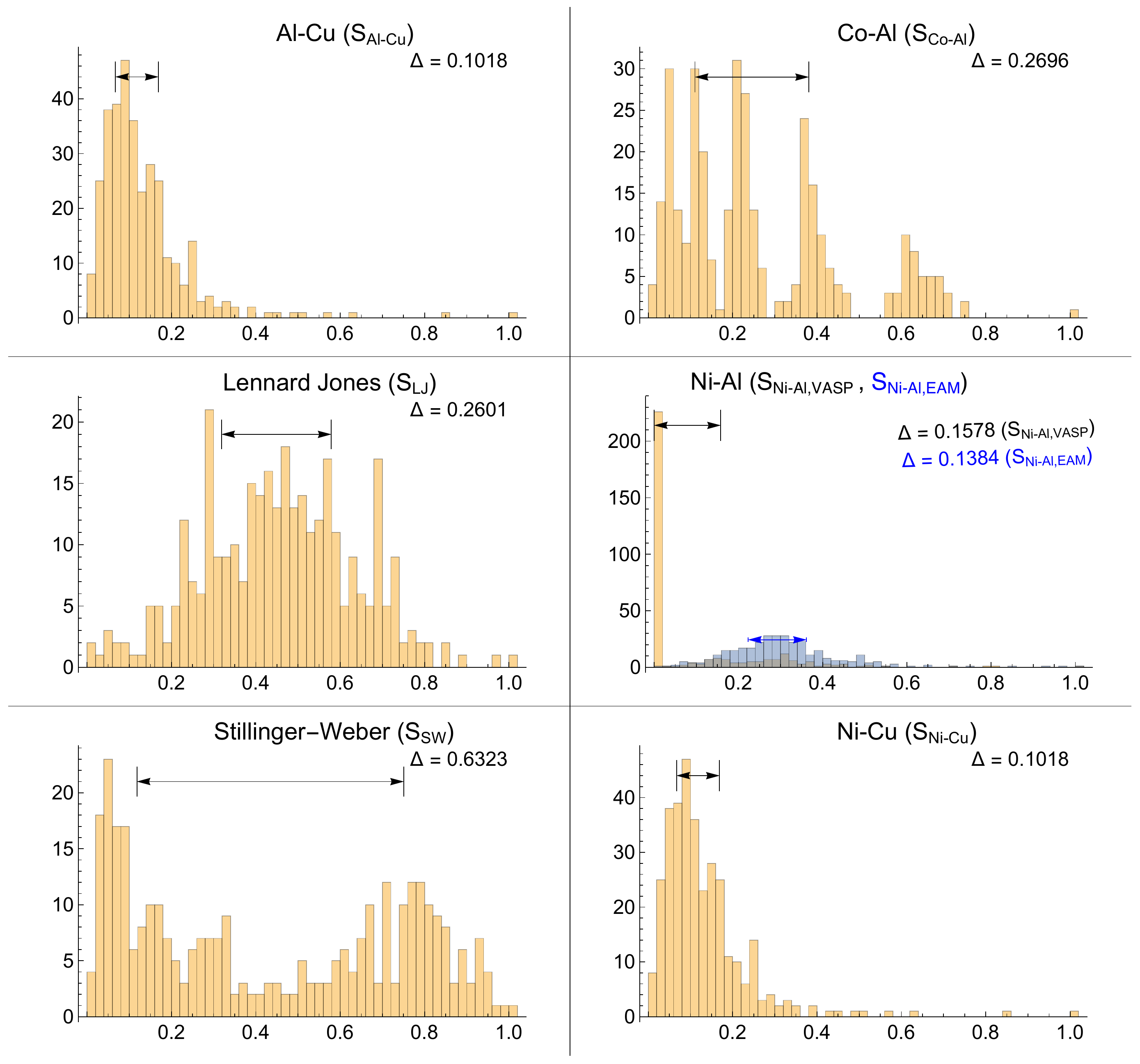}}
\captionsetup{justification=RaggedRight}
\caption[]{\label{fig:ExperimentalDistrbutions}(color online) Distributions 
from real relaxations using classical and semi-classical potentials, as well as
DFT calculations. The distributions are all normalized to fall within 0 and 1. 
The widths, $\Delta$, were calculated by taking the difference between the 
25$^{\textrm{th}}$ and 75$^{\textrm{th}}$ percentiles.}
\end{figure}

\begin{figure}[]
\centerline{\includegraphics[width=0.5\textwidth]{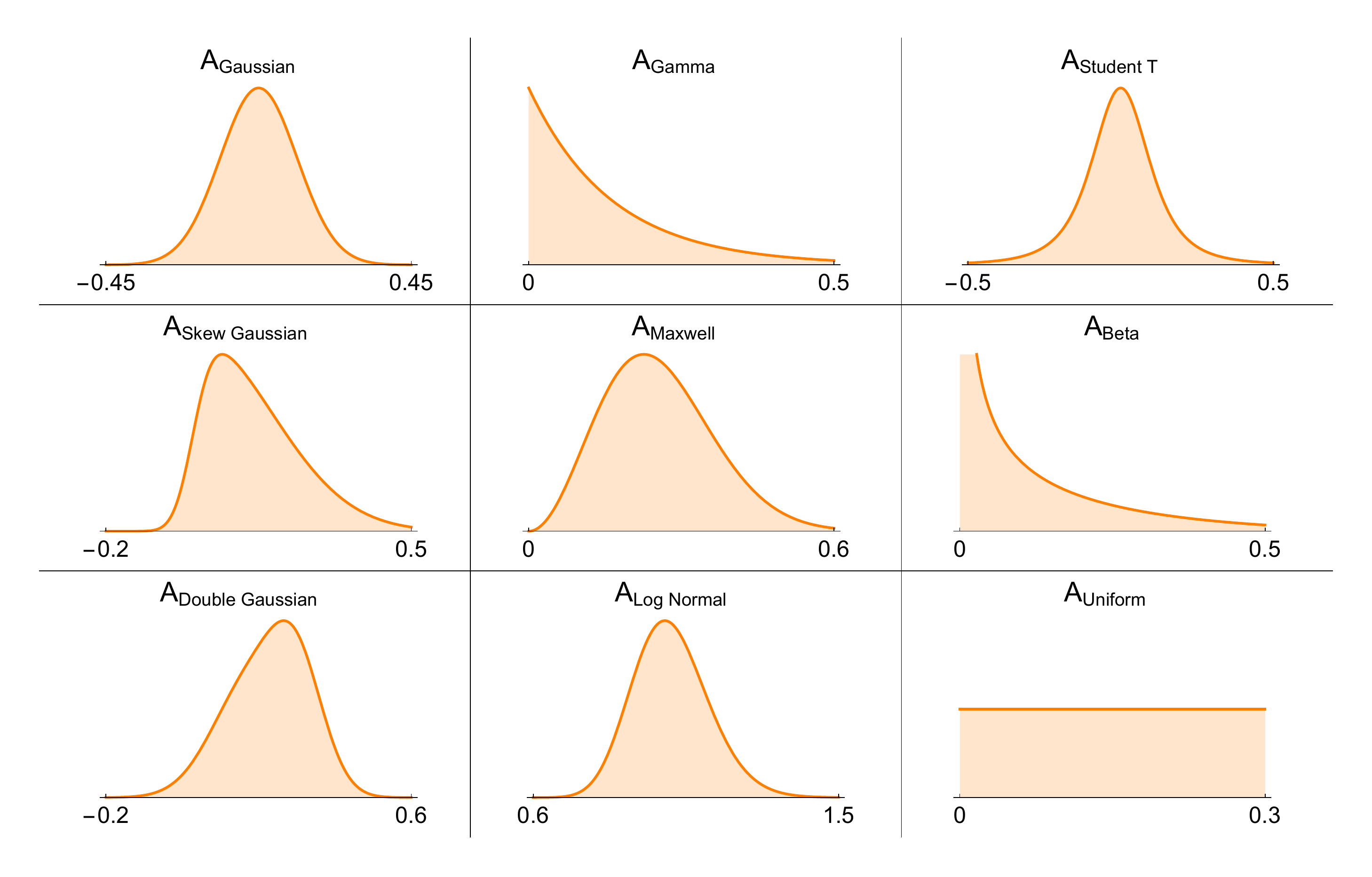}}
\captionsetup{justification=RaggedRight}
\caption[]{\label{fig:AnalyticEqualWidthDistributions} (color online) The 
analytic, equal width distributions used for adding error to the toy model CE 
fit.}
\end{figure}

The numerical errors in DFT calculations are largely understood, but it is 
difficult to disentangle the effects of different, individual error sources.
Instead of studying the effects of errors separately, we added different
distributions of error to a  ``toy'' model in order to imitate the aggregate 
effects of the numerical error on CE models.  
Hence, we opt to simplify the problem by creating a ``toy'' problem for which
the exact answer is known.  To restrict the number of independent variables, 
we formulated a ``toy'' cluster expansion model by selecting five non-zero
values for a subset of the total clusters. Using this toy CE, we predicted 
a set of energies $y$ for 2000 known derivative superstructures of an FCC
lattice, These $y$ values are used as the true energies for all
subsequent analysis. We added error to $y$, chosen from either: 
1) ``simulated'' distributions obtained by computing the difference between
 relaxed and unrelaxed energies predicted by either DFT, EAM, LJ or SW models
(Fig. \ref{fig:ExperimentalDistrbutions}); or 2) common analytic
distributions (Fig. \ref{fig:AnalyticEqualWidthDistributions}).  

To generate the simulated distributions, we chose a set of identical
structures and fitted them using a variety of classical and
semi-classical potentials, and quantum mechanical calculations using
VASP. For each of the potentials we selected, we calculated an
unrelaxed total energy $y$ for each structure and then performed
relaxation to determine the lowest energy state, $\tilde{y}$. The
difference between these two energies ($\Delta y = \tilde{y}-y$) was
considered to be the ``relaxation'' error.

Certain assumptions are usually made about the error in the signal, namely that 
it is Gaussian. The original CS paradigm proves that the $\ell_2$ error for 
signal recovery obeys \cite{Candes2008}:
\begin{equation}
\label{eq:BoundsOnL2Error}
\vectornorm{x^{*}-x}_{\ell_2} \le C_0 \cdot
\vectornorm{x-x_S}/\sqrt{S}+C_1 \cdot \epsilon,
\end{equation}

\noindent where $\epsilon$ bounds the amount of error in the data,
$x^{*}$ is the CS solution, $x$ is the true solution, and $x_S$ is the
vector $x$ with all but the largest $S$ components set to zero. This
shows that, \emph{at worst, the error in the recovery is bounded by a
  term proportional to the error}. For our plots of this error, we
first normalized $\Delta{y}$ so that $\epsilon \equiv
\mathrm{normalized}(\Delta y) \in [0,1]$ using 
\begin{equation}
\label{eq:NoiseNormalizationEquation}
\epsilon = \frac{y-\mathrm{min}(y)}{\mathrm{max(y)} - \mathrm{min}(y)}.
\end{equation}

\noindent Not surprisingly, the various potentials produced different error
 profiles. 

The expectation value of the distributions was set to be a percentage
of the average, unrelaxed energy across all structures. Thus, ``15\%
error'' means that each unrelaxed energy was changed by adding a
randomly drawn value from a distribution with an expectation value of
15\% of the mean energy. We performed CE fits as a function of the
\%-error added (2,5,10 and 15\%) for each distribution. Although we
only present the 15\% error results in the next section, all results
at different error levels can be found in the supporting
information. For each data point, we performed 100 independent CE fits
and used the mean and standard deviation to produce the values and
error bars for the plots.

\subsection{Results and Discussions}

As shown in Fig. \ref{fig:EqualWidthSummary}, the error is weakly uniform 
across all (analytic and simulated) distributions, implying that there is no 
correlation between specific distribution and error. None of the normal 
quantifying descriptions of distribution shape (e.g. width, skewness, kurtosis,
standard deviation, etc.)  show a correlation with the CE prediction error. The
error increased proportionally with the level of error in each system (2, 5, 10
and 15\% error). We therefore turn to the compressive sensing (CS) formalism 
for insight.

\begin{figure}[H]
\centerline{\includegraphics[width=0.5\textwidth]{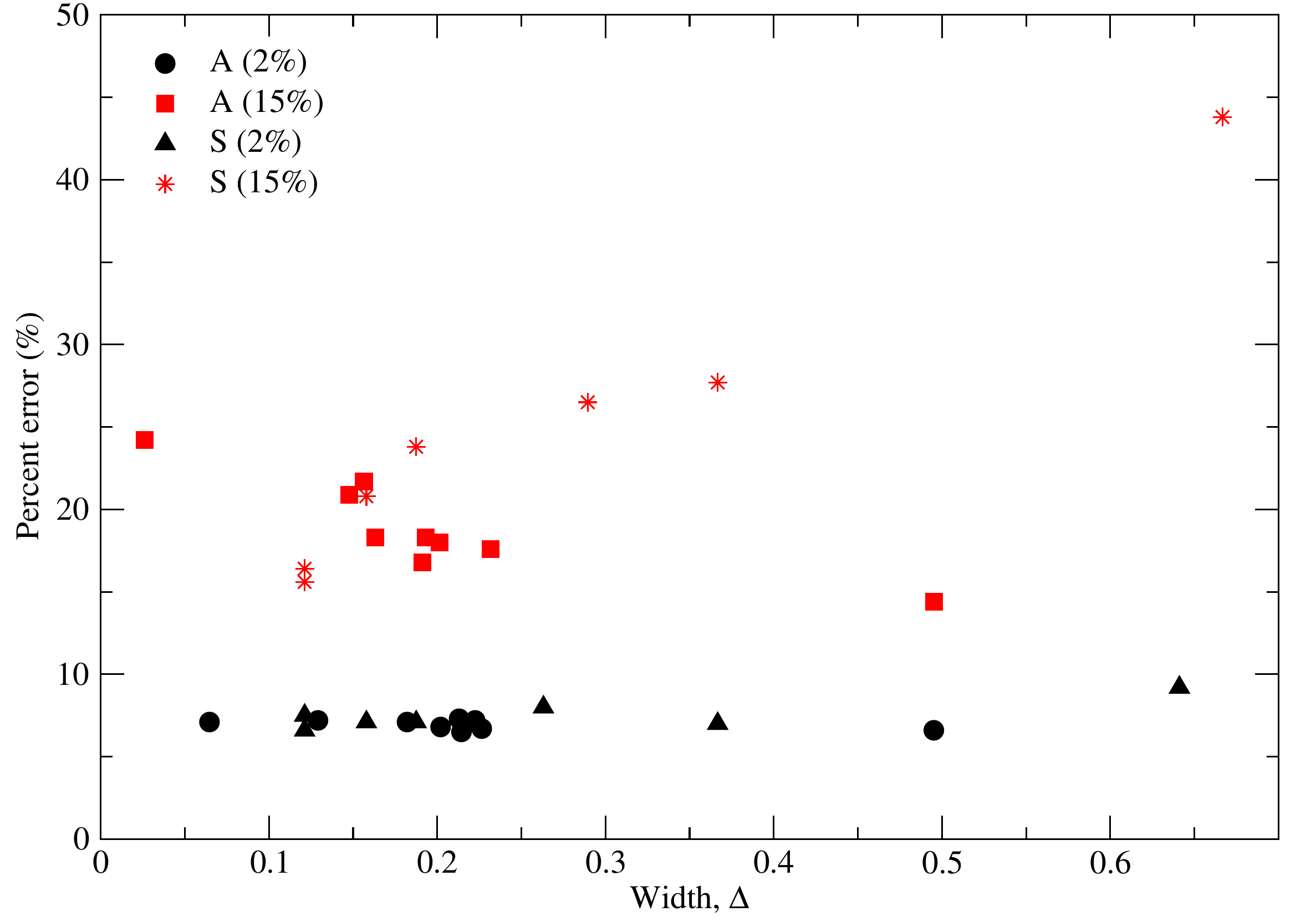}}
\caption[]{\label{fig:EqualWidthSummary} (color online) Comparison of
  the predictive error in CE fits as the shape of relaxation error
  changes. (A) refers to the analytic distribution while (S) refers to
  simulated distribution. The fits are ordered from lowest to highest
  distribution width. Fits were averaged over 100 randomly selected
  subsets with 500/2000 data points used for training; the remaining
  1500 were used to verify the model's predictions. The black and red
  colored symbols represent 2\% and 15\% error levels, respectively.
  The circles and triangles represent the analytical and simulated
  distributions, respectively.  Higher error produces higher prediction
  errors. }
\end{figure}

The theorems of Tao and Cand\'es \cite{Candes2006} guarantee that the solution 
for an underdetermined CS problem can be recovered \emph{exactly} with 
overwhelming probability provided:
\begin{itemize}
\item The solution is sparse within the chosen representation basis.
\item Sufficient data points, sampled independent and identically distributed 
(i.i.d).
\item The sensing and representation bases are maximally incoherent.
\end{itemize}

If all of these conditions are met, we know that CS will provide a solution 
that is very close to the true answer. Conversely, if CS cannot converge to a 
good solution, it means that one of these conditions has been violated. We have
control over the number of training points, and the incoherence of the sensing-
representation bases. However, we \emph{cannot} control whether the true 
physical solution is sparsely represented for relaxed systems. This suggests a 
useful connection between the CS framework and the robustness of CE: if CS 
cannot reproduce a good CE fit (quantified below), then sparsity has been lost.

In the CS framework, the foundational assumption is that of sparsity, meaning 
that the compressed signal (or cluster expansion) requires only a few terms to 
accurately represent the true signal (physics). Thus, the number of terms 
recovered by CS to produce the CE is a good measure of the quality of the CS 
fit. This begs the question: can we use the number of terms within the CS 
framework to heuristically predict \emph{in advance} whether the CE fit will 
converge well?

In answering the question of predictability for a good CE fit, we define three 
new quantities:

\begin{enumerate}
\item\label{item:XiDefinition} $\Xi$: total number of unique clusters used over 
100 CE fits of the same dataset. We also call  this the model complexity.
\item\label{item:NotInDefinition} $\not\in$: number of ``exceptional'' 
clusters. These are clusters that show up fewer than 25 times across 100 fits, 
implying that they are not responsible for representing any real physics in the
signal, but are rather included because the CE basis is no longer a sparse 
representation for the relaxed alloy system. They are sensitive to the 
training/fitting structures.
\item\label{item:LambdaDefinition} $\Lambda$: number of \emph{significant} 
clusters in the fit; essentially just the total number of unique clusters minus 
the number of ``exceptional'' clusters, $\Lambda = \Xi - \not\in$.
\end{enumerate}

In the relaxation section, we showed that the average number of coefficient is 
not sufficient to determine the quality of the CE model. Here, we decompose the 
number of $J$s into three new quantities to provide additional insights into 
the reliability of the CE fits. In Fig. \ref{fig:ErrorVsModelComplexity}, we 
plot the CE error, ordered by model complexity and show that it reproduces the 
trend identified by the number of coefficients (indeed they are intimately 
related, $\Xi$ being the statistically averaged number of coefficients across 
many fits). An ordering by the number of exceptional clusters $\not\in$ 
produces an identical trend, showing that it may also serve to quantify a good 
fit \footnote{See supplemental materials for additional plots at different levels of error.}.

\begin{figure}[H]
\centerline{\includegraphics[width=0.5\textwidth]{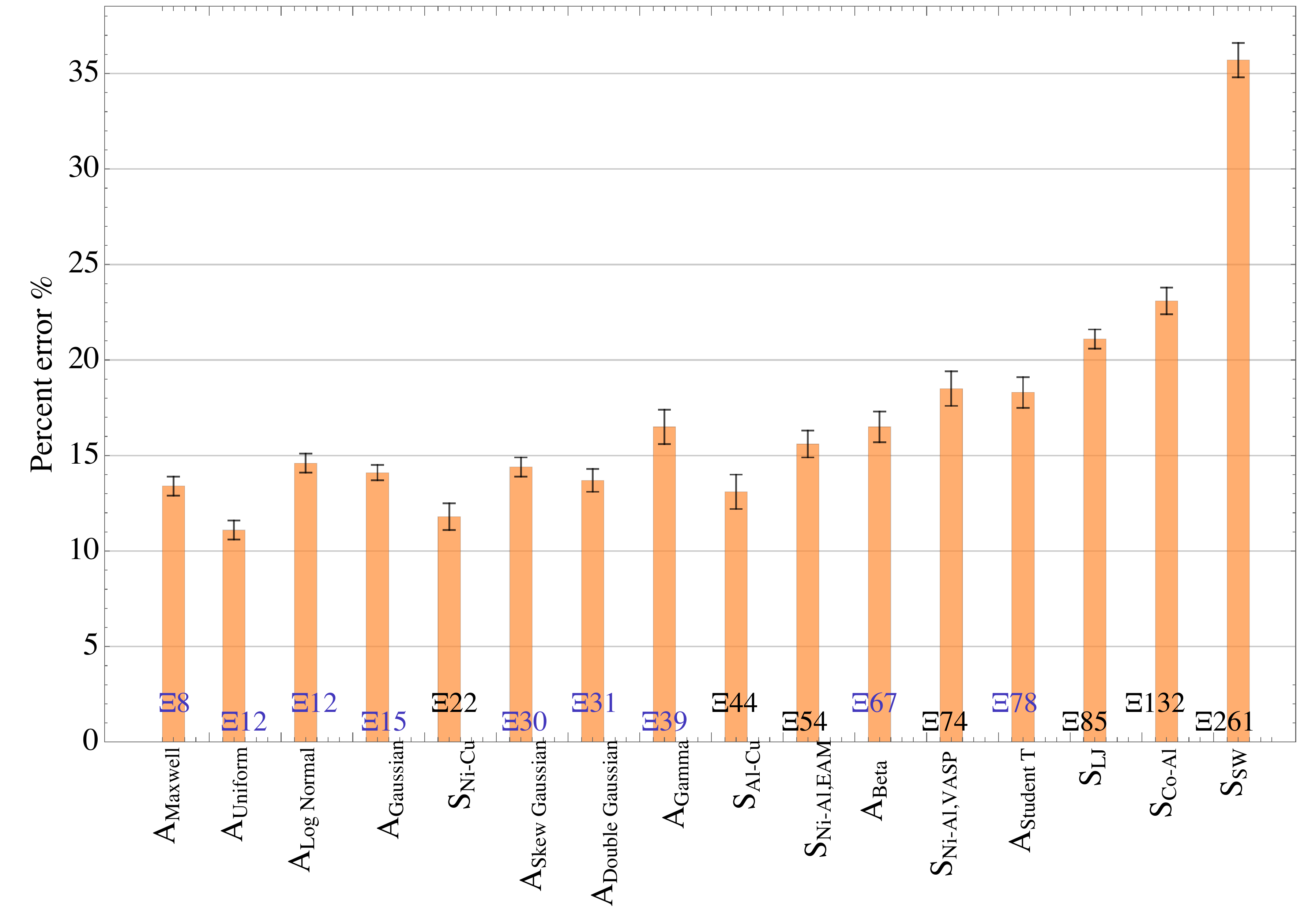}}
\captionsetup{justification=RaggedRight}
\caption[]{\label{fig:ErrorVsModelComplexity} (color online) Prediction error 
over 65\% of the structures for the ``toy'' cluster expansion (at 15\% error 
added). The systems are ordered by $\Xi$, which is the total number of unique 
clusters used by any of the 100 CE fits for the system. This ordering shows a 
definite trend with increasing $\Xi$.}
\end{figure}

As indicated earlier, all these experiments were performed for a \emph{known} 
CE model that had 5 non-zero terms. Additional insight is gained by plotting 
the errors, ordered by $\Lambda$, the number of significant clusters (Fig. 
\ref{fig:ErrorVsSignificantTerms}). Fig. \ref{fig:ErrorVsSignificantTerms} 
shows that in almost all cases, once we remove the exceptional clusters $\not
\in$, the remaining model is almost \emph{exactly} the known CE model that we 
started with. The CS framework provides a rigorous mathematical framework for 
this statement because it guarantees to \emph{exactly} recover the original 
function with high probability as long as we have enough measurements and our 
representation basis is truly sparse. Once the cluster expansion stops 
converging, we lose sparsity and CS fails. This gives us confidence to use the 
CS framework as a predictive tool for CE robustness.

\begin{figure}[H]
\centerline{\includegraphics[width=0.5\textwidth]{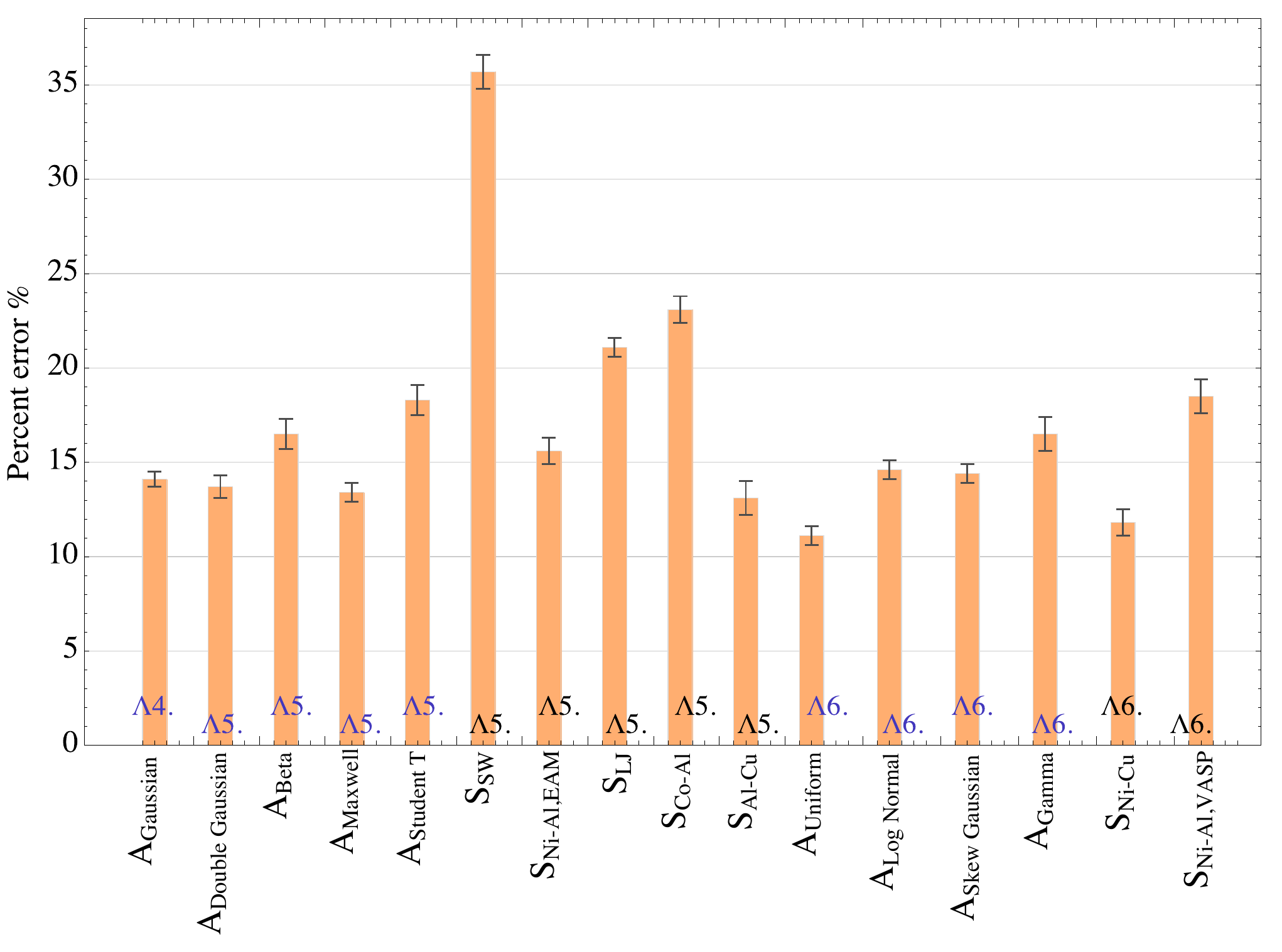}}
\captionsetup{justification=RaggedRight}
\caption[]{\label{fig:ErrorVsSignificantTerms} (color online)
  Prediction error over 65\% of the structures for the ``toy'' CE
  model (at 15\% error added). The errors are ordered by $\Lambda$,
  the number of significant terms in the expansion. As expected, the
  values are close to the known model complexity (5 terms) and the
  ordering once more appears random.}
\end{figure}

Provided the training structures are independent and identically distributed, 
we do not necessarily need hundreds of costly DFT calculations to tell us that 
the CE will not converge. Using our toy CE model, we discovered that for all 
error distributions, a training set size of 50 data points was sufficient to 
recover the actual model complexity (5 terms) \footnote{The supplementary data 
has many plots from the hundreds of CE fits that we performed for this 
analysis.}. For actual DFT calculations, where relaxation was known to disrupt 
CE convergence, we saw a similar trend with about 100 data points needed to 
identify whether the CE would converge with more data or not.

\begin{figure}[]
\centerline{\includegraphics[width=0.5\textwidth]{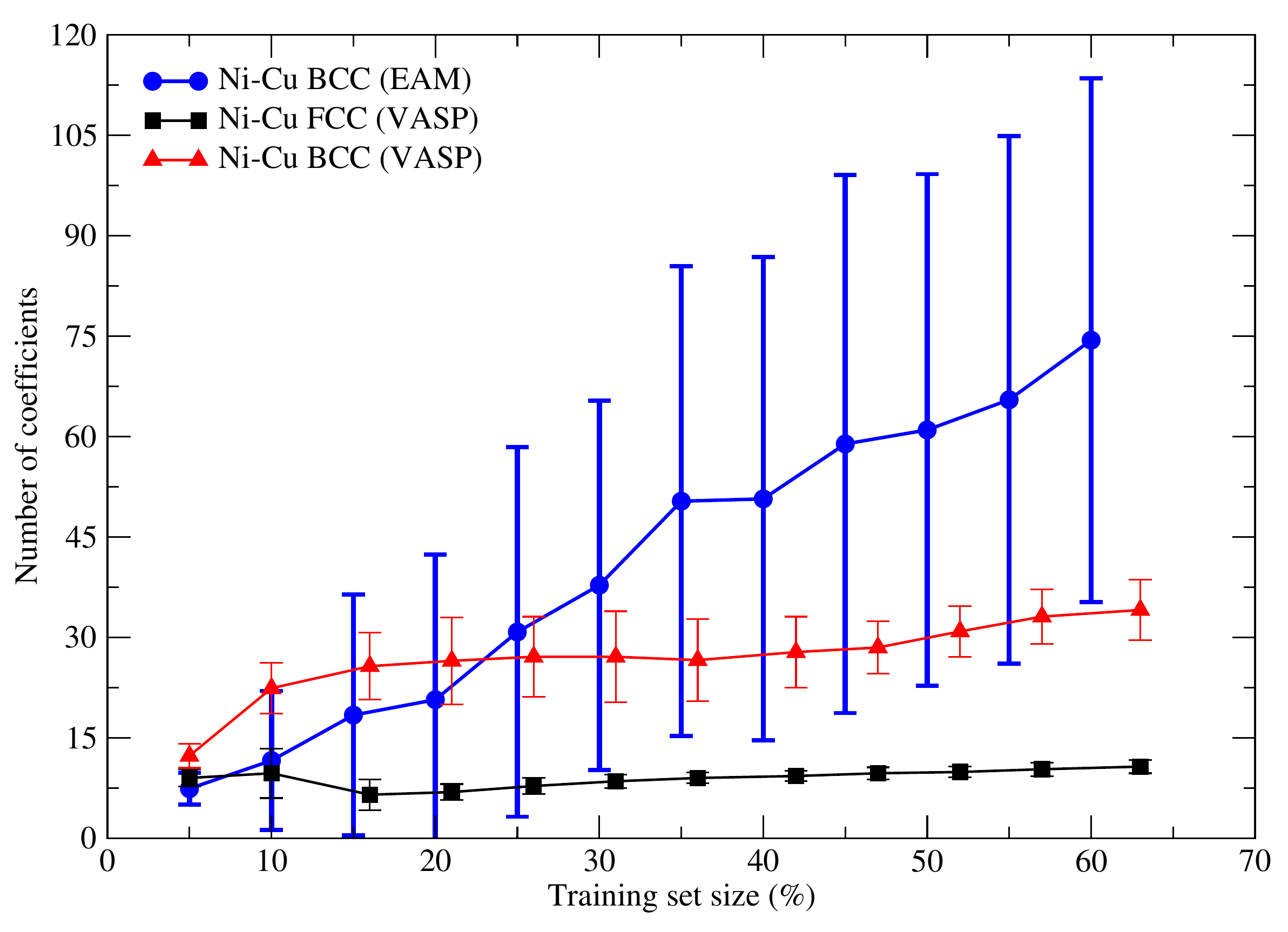}}
\captionsetup{justification=RaggedRight}
\caption[]{\label{fig:norm-training} (color online) For a reliable CE
  model, the number of coefficients converges as a function of the
  training set size.  A total of 500 structure were available for
  training. The number of coefficients in a fit and its error bars
  give us an indication of the predictive power of CE with only a
  small training set. The black curve represents a good CE fit; only
  25 to 50 (or 5 to 10\%) of training structures were needed.  On the
  other hand, the red and blue curves show that CE fails to fit the
  data due to a slowly converging expansion.  The error bars on the
  blue points indicate extremely bad
  fitting.}
\end{figure}

We conclude that CE robustness for relaxed systems can be predicted with a 
\emph{much} smaller number of data points than is typically needed for a good 
CE fit (on the order of 5-10\% from our experience) \footnote{For typical 
binary CEs we typically need about 300-500 structures to get a good fit, which 
is then verified using an additional 200+ DFT calculations.}. The proposed 
heuristic to verify convergence of the relaxed CE, when trained with a limited 
dataset, is to examine the values of $\Lambda$ and $\Xi$ over a large number of
independent fits. If the number of the exceptional clusters $\not\in$ is 
significant compared to $\Lambda$, then it is likely that the CE will 
\emph{not} converge on a larger dataset as shown in Fig. 
\ref{fig:ErrorVsSingleShow}. Figure \ref{fig:norm-training} highlights the CE 
fitting as function of training set size. We observe small relaxation (black 
curve) correlates with a small number of coefficients; thus the CE can fit 
using a small number of $J$s even with 5\% (25) to 10\% (50) of the structures.
On the other hand, red and blue curves which have high relaxation, do not 
converge. By using a small relaxed dataset (50 to 100 structures), we can 
predict whether or not the computational cost of relaxing \emph{many} 
structures is fruitful. 

\begin{figure}[]
\centerline{\includegraphics[width=0.5\textwidth]{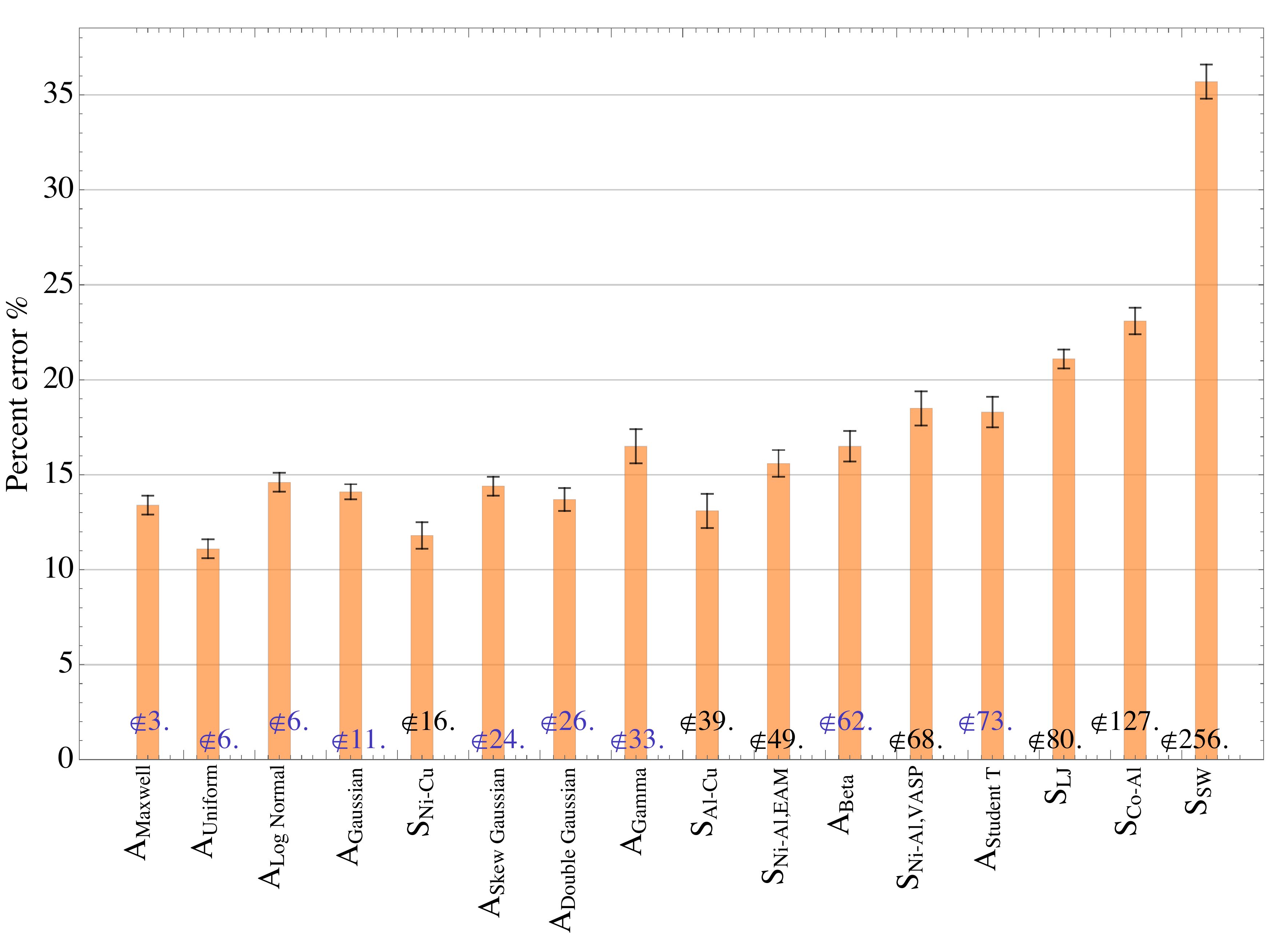}}
\captionsetup{justification=RaggedRight}
\caption[]{\label{fig:ErrorVsSingleShow} (color online) Plot of predictive 
error over 65\% of the structures for the ``toy'' problem (at 15\% error 
added). The systems are ordered by $\not\in$ the number of clusters that were 
used less than 25 times across all 100 CE fits. These are considered exceptions
to the overall fit for the system. As for Figure 
\ref{fig:ErrorVsModelComplexity}, there is a definite trend toward higher error
for systems with more exceptional clusters.}
\end{figure}

\section{\label{sec:level4}Conclusions}

Relaxation and error decrease the reliability of the cluster expansion
fit because the CE model is no longer sparse. Nevertheless,
until now, there has been no measure of relaxation
that provides a heuristic as to when the CE fitting data is reliable.
Using four different Hamiltonians (first-principles, Lennard-Jones,
Stillinger-Weber and embedded atom method), we show that the
normalized mean-squared displacement of alloy configuration is a good
measure of relaxation and CE predictability.  A small displacement
percent, e.g., less than 0.1\%, will usually generate
a reliable CE model. The number of cluster terms in the CE models is
also a good indicator of how well cluster expansions can perform. As
the number of clusters increases, the predictability of CE model
decreases. CE tends to fail when the number of $J$s exceeds 80.
          
In our error analysis, we investigated the ability of the compressive sensing 
framework to obtain fits to a toy, cluster expansion model as the energy of 
relaxation changes in a predictable way. We used 16 relaxation error 
distributions (both analytic and simulated) and compared the prediction errors 
of the resulting CE fits for the relaxed vs. unrelaxed case. No clear 
correlation appears between the statistical measures of distribution shape and 
the predictive errors. However, there are clear correlations between the 
predictive error, the complexity of the resulting CE model, and the number of 
significant terms in that model.

We cannot use the relaxation distributions alone to determine the viability of 
a CE fit in advance. However, the analysis does reveal that the majority of the 
clusters used by the unrelaxed CE fit will also be present in the relaxed case 
(albeit with adjusted $J$ values) if the CE fit is viable. This suggests that 
it may be possible to  decide whether the computational cost of full CE is 
worthwhile by making predictions for a few relaxed systems (50-100) and 
determining whether the error remains small enough.

\section{\label{sec:level4.5}Acknowledgments}
The authors thank Volker Blum, Lance J. Nelson and Mark K. Transtrum for useful 
discussions. This work was supported by funding from ONR (MURI 
N00014-13-1-0635). We thank the Fulton Supercomputing Lab at Brigham Young 
University for allocation of computing time. 

\bibliographystyle{unsrt}

%

\newpage

\beginsupplement

\section{\label{sec:levelSup}Supplementary materials}
\subsection{\label{sec:levelS0}Cluster Expansion}

Cluster expansion is a generalized Ising model with many-body interactions. The
CE model provides a fast, accurate way to compute physical properties which are
function of the configuration. Consider a binary alloy $\mathrm{A}_x\mathrm{B}
_{1-x}$, the alloy is treated as a lattice problem. Each site of a given 
lattice is assigned a occupation variable, $\sigma_i$ (i=1,2,...,N) with 
$\sigma_i$ = --1 or +1 depending on the site $i$ being occupied by an A or a B
atom. Any atomic configuration on a given lattice can then be specified using a
 vector of the occupation variable, $\bm{\sigma} = [\sigma_1,\sigma_2,...,
 \sigma_N]$. A physical quantity such as energy can be expressed as a linear 
 combination of basis function:
\begin{equation}
E(\sigma) = \sum_i J_i \Pi_i(\bm{\sigma})
\end{equation} 
\noindent where the argument to the function is a vector of occupation 
variable, $\bm{\sigma}$.  The $\Pi_i$ are the basis function or often referred
 to as the cluster functions.  Each cluster function corresponds to a cluster 
 of lattice sites. The coefficients $J_i$ are the effective cluster 
 interactions or ECI's.  The main task of building a CE model is to find the 
 $J$s and their values. We can solved for the $J$s using the structure 
 inversion method \cite{Connolly1983}. However, we use a new approach based on 
 compressive sensing to solve for these coefficients 
 \cite{Nelson2013a,Nelson2013}.  

\subsection{\label{sec:levelS1}Relaxation}

Here, we present additional information and metrics for the more than one 
hundred Hamiltonians to sample the effect of relaxation.

\subsubsection{Molecular Dynamics}
This is a more extensive version of the method present in the paper including 
the various parameters, forms of the potential and relaxation metrics. Two molecular dynamics packages were used to study the relaxation: GULP and 
LAMMPS. GULP (general utility lattice program) is written to perform a variety 
of tasks based on force field methods such molecular dynamics, Monte Carlo and 
etc \cite{Gale1997}. GULP is a general purpose code for the modeling of solids, 
clusters, embedded defects, surfaces, interfaces and polymers 
\cite{Gale1997,Gale2003}. LAMMPS (large-scale atomic molecular massively 
parallel simulator) is a widely used molecular dynamics program 
\cite{Plimpton1995}.  We used these two programs to minimize/relaxed each 
structure. We computed the energy of each structure (unrelaxed and relaxed). 
Relaxation of each structure was obtained by minimization of total energy using
GULP or LAMMPS via a conjugate gradient scheme. Molecular dynamics simulations 
were carried for the embedded atom method (EAM), Lennard-Jones (LJ) and 
Stillinger-Weber (SW) potentials. 

\subsubsection{Embedded Atom Method (EAM)}
 The EAM potential is a semi-empirical potential derived from first-principles 
 calculations. The embedded atom method (EAM) potential has following form:  
\begin{equation}
E_i = F_\alpha \left(\sum_{j \neq i}\ \rho_\beta (r_{ij})\right) +
   \frac{1}{2} \sum_{j \neq i} \phi_{\alpha\beta} (r_{ij})
\end{equation}
EAM potentials of metal alloys such Ni-Cu, Ni-Al, Cu-Al have been 
parameterized from first-principle calculations and validated to reproduce 
experimental properties, bulk modulus, elastic constants, lattice constants, 
etc \citep{Foiles1986}. Compared to first-principles calculations, EAM 
potentials are computationally cheaper. Thus, this allows us to explore the 
effect of relaxation for large training sets. Nonetheless, we are limited by 
the number of EAM potentials available. We used various EAM potentials to 
study the relaxation; these binary EAM potentials are shown in table 
\ref{eam-parameter}. 

\begin{table}[H]
\centering
\caption{EAM potentials used to study the relaxation. Lattice mismatch is 
shown in percentage. Lattice mismatch = ($a_{\mathrm{A}} - a_{\mathrm{B}}$)/
(($a_{\mathrm{A}} + a_{\mathrm{B}}$)/2) $\times 100 \%$, where $a$ is the 
lattice constant of the pure element.}
\label{eam-parameter}
\begin{tabular}{|c|c|}
\hline
EAM potential & lattice mismatch (\%)  \\
\hline
Al-Cu      & 11.5\%  \\
\hline
Al-Fe      & 34.1\%   \\
\hline
Al-Mg 	   & 23.1\%  \\
\hline
Al-Pb 	   & 20.0\%  \\
\hline
Co-Al      & 46.9\%  \\
\hline  
Cu-Ag 	   & 12.5\%  \\
\hline
Cu-Zr      & 11.1\%  \\
\hline
Fe-Cr      & 0.35\%  \\
\hline
Fe-Ni 	   & 20.3\%  \\
\hline
Ni-Al      & 14.0\%  \\
\hline
Ni-Co 	   & 33.5\%  \\
\hline
Ni-Cu      & 2.50\%  \\
\hline  
Ni-Zr 	   & 8.6\%  \\
\hline
Pb-Cu 	   & 31.3\%  \\
\hline
Ti-Al	   & 31.4\%  \\
\hline
V-Fe      & 5.10\%  \\
\hline   
\end{tabular}
\end{table}

These 16 EAM potentials represented different lattice mismatch ranging from 
0.35\% (Fe-Cr) to 46.9\% (Co-Al). Ni-Cu EAM potential was used to compare/
validate the relaxation using molecular dynamics to first-principles DFT 
calculations as shown in the main text.

\subsubsection{Lennard-Jones (LJ)}
We selected the Lennard-Jones potential to adequately examine various degrees 
of relaxation, which can be tuned using free parameters in the model.
The Lennard-Jones potential is a pairwise potential with a repulsive and 
attractive part. The functional form of Lennard-Jones potential is given by
\begin{equation}
  u_{LJ}(r) = 4\epsilon \left[\left(\frac{\sigma}{r} \right)^{12} - 
  \left(\frac{\sigma}{r}
  \right)^{6} \right],
\end{equation}
where $\epsilon$ is the well depth of the pair interaction and  $\sigma$ is 
the onset of the repulsive wall where $u_{LJ}(r)$ = 0. A cutoff distance of 12 
\AA was used in the interaction potential and long range correction to the 
energy was included. We varied the parameters to mimic the lattice mismatch of 
5, 10, 15, 20, 30 and 40\%.  Also we varied the interaction strength to 
simulated systems with strong and weak attraction between atomic species.  The 
LJ parameters are shown in \ref{lj-parameter}.

\begin{table}[H]
\centering
\caption{LJ parameters used to study the relaxation. $\epsilon $ is in eV and 
$\sigma$ is in unit of \AA. The LJ1 system was used in next section to 
evaluate the relaxation metrics.}
\label{lj-parameter}
\begin{tabular}{|l|l|l|l|l|l|}
\hline
$\epsilon_{AA}$ & $\sigma_{AA}$ & $\epsilon_{BB}$ & $\sigma_{BB}$ & 
$\epsilon_{AB}$ & $\sigma_{AB}$ \\
\hline
0.25      & 0.975     & 0.25  & 1.025 & 0.50 & 1.0    \\
\hline
0.25      & 0.95     & 0.25  & 1.05 & 0.50 & 1.0      \\
\hline
0.25      & 0.925     & 0.25  & 1.075 & 0.50 & 1.0    \\
\hline
0.25      & 0.900    & 0.25  & 1.10 & 0.50 & 1.0    \\
\hline
0.25      & 0.85     & 0.25  & 1.15 & 0.50 & 1.0    \\
\hline
0.25      & 0.80     & 0.25  & 1.20 & 0.50 & 1.0    \\
\hline
0.2625      & 0.95     & 0.2375  & 1.05 & 0.375 & 1.0    \\
\hline 
0.2625      & 0.95     & 0.2375  & 1.05 & 0.400 & 1.0    \\
\hline 
0.2625      & 0.95     & 0.2375  & 1.05 & 0.450 & 1.0    \\
\hline 
0.255      & 0.95     & 0.245  & 1.05 & 0.50 & 1.0      \\
\hline   
0.2625      & 0.95     & 0.2375  & 1.05 & 0.50 & 1.0      \\
\hline
0.27      & 0.95     & 0.23  & 1.05 & 0.50 & 1.0      \\
\hline
2.625      & 0.95     & 2.375  & 1.05 & 3.750 & 1.0    \\
\hline
0.185 (LJ1) & 0.215 &  0.336 & 0.290 & 0.5 & 0.5 \\
\hline
0.235 & 0.243 & 0.265 & 0.258 & 0.5 & 0.5 \\
\hline
0.0981 & 0.157 & 0.336 & 0.290 & 0.5 & 0.5 \\
\hline
\end{tabular}
\end{table}

The uses of classical potentials allow us to use molecular dynamics to relax 
each structure computationally cheaper and faster. The most appealing factor 
of using classical potential is ability to modify the potential in a way that 
we can simulated a highly relaxed structure, i.e., going off the lattice. Such 
as in the Lennard-Jones potentials where we can modify the interaction between 
the particles as well as adjust the size to vary the lattice mismatch in the 
binary alloy.

\begin{figure}[H]
    \centering
    \begin{subfigure}{0.5\textwidth}
        \includegraphics[width=\textwidth]{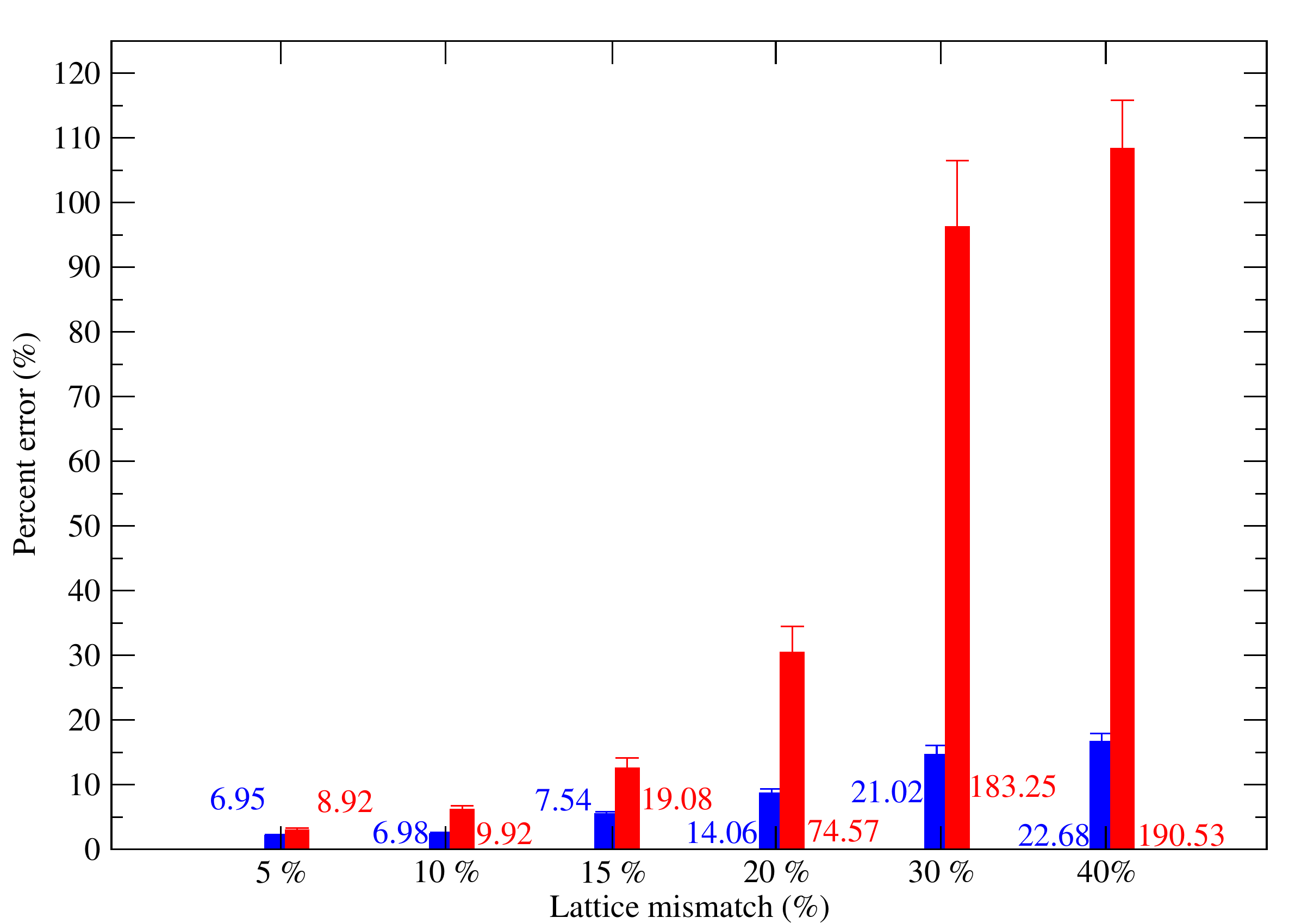}
        \caption{Relaxation of FCC derivative structure }
        \label{fig:LJ-FCC}
    \end{subfigure}
    ~ 
    \begin{subfigure}{0.5\textwidth}
        \includegraphics[width=\textwidth]{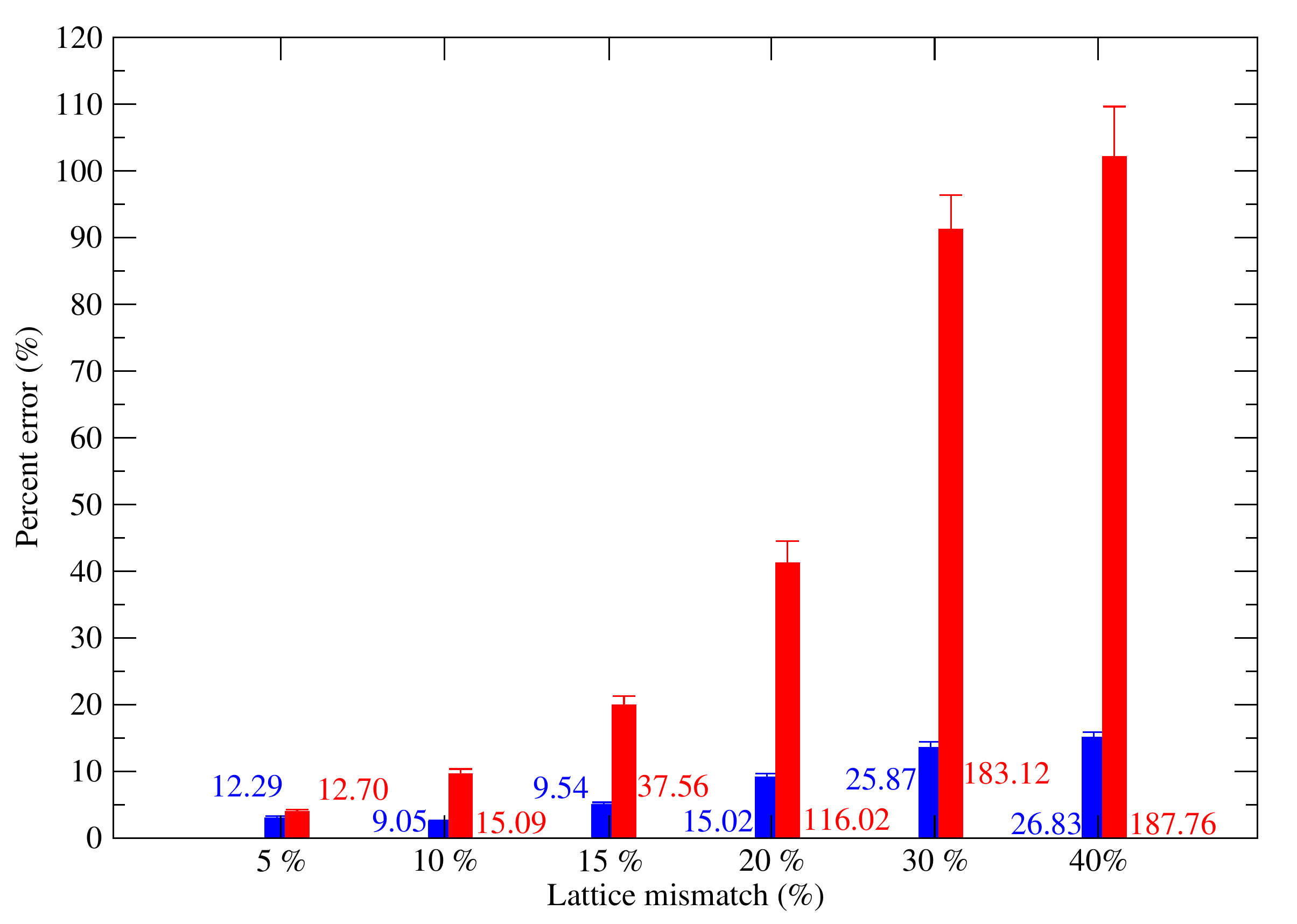}
        \caption{Relaxation of HCP derivative structure}
        \label{fig:LJ-HCP}
    \end{subfigure}
    ~ 
    \begin{subfigure}{0.5\textwidth}
        \includegraphics[width=\textwidth]{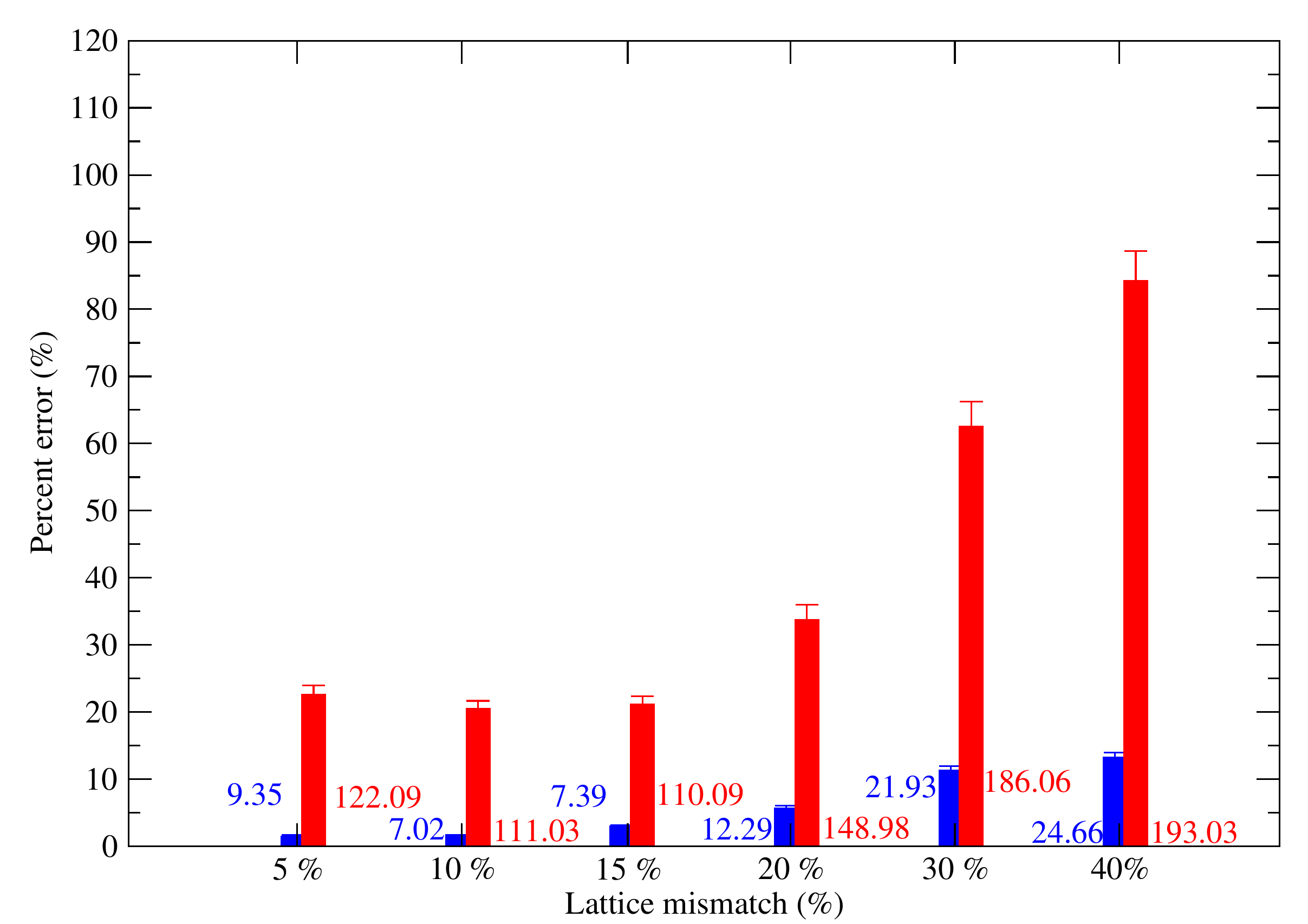}
        \caption{Relaxation of BCC derivative structure}
        \label{fig:LJ-BCC}
    \end{subfigure}
    \caption{(color online) Cluster expansion fittings for a binary alloy 
    using a Lennard-Jones potential. Blue bars represent the unrelaxed CE 
    fits, while red bars show the relaxed CE fits. The colored numbers show 
    the averaged number of coefficients for unrelaxed (in blue) and relaxed 
    (in red) systems. As the lattice mismatch increases, the reliability of CE
     fits decreases.}\label{fig:Unrel-rel-LJ}
\end{figure}

We computed the unrelaxed and relaxed using GULP and LAMMPS for FCC, BCC and 
HCP structures.  Fig. \ref{fig:Unrel-rel-LJ} shows the fitting result of 
cluster expansion with Bayesian compression sensing. Cluster expansion can fit 
any unrelaxed energy computed from Lennard-Jones potential. All three 
unrelaxed crystals are within 1 to 2\% prediction error. As the lattice 
mismatch increases, the predictive power of CE decreases (higher prediction 
errors and higher number of coefficients). 

\subsubsection{Stillinger-Weber}
In addition to the LJ potential, we study the effect of relaxation using the 
Stillinger-Weber potential. Similar to the LJ potential, the Stillinger-Weber 
potential has a pair interaction but there is an additional angular 
(three-body) term. The parametric form of the SW potential is written as a 
sum of two-body and an anisotropic three-body interaction term as: 
\begin{equation}
U=\sum_{\mathrm{pair}}\phi_2 +\lambda\sum_{\mathrm{triplet}}\phi_3
\end{equation}
\noindent Here $\phi_2$ depends only on the pair separation of atoms and 
$\phi_3$ depends on pair distances as well as angle formed by any three atoms 
and $\lambda$ controls the strength of the angular terms. Complete expressions 
for $\phi_2$ and $\phi_3$ can be given
as \cite{Stillinger1985}
\begin{equation}
  \phi_2(r_{ij})=A\epsilon\left[ B\left(\frac{\sigma}{r_{ij}}\right)^p-\left(\frac{\sigma}{r_{ij}}\right)^q \right]
                  \exp\left(\frac{\sigma}{r_{ij}-a\sigma}\right),
\end{equation}
\begin{eqnarray}
  \phi_3(r_{ij},r_{ik},\theta_{ijk})=\epsilon(\cos\theta_{ijk}-\cos\theta_0)^2 
  \nonumber \\
 \exp \left(\frac{\gamma\sigma}{r_{ij}-a\sigma}\right) 
 \exp\left(\frac{\gamma\sigma}{r_{ik}-a\sigma}\right),
\end{eqnarray}

where the parameters are $A$ = 7.049556277, $B$ = 0.6022245584, $p$ = 4, 
$q$ = 0, $\gamma$ = 1.20 and $a$ = 1.80. We varied the $\lambda$ values 
from 0 to 1. Table \ref{SW-parameter} shows the parameters for a set of 
Stillinger-Weber potential. Similar to the LJ potential, the three systems 
have 5\%, 10\% and 15\% lattice-mismatch. Additional modifications of 
Stillinger-Weber parameters are found in table \ref{SW-parameter}. We 
performed extensive studies of the CE fit using SW potential. Some of the 
parameters are very similar or only vary by one or two parameters to examine 
the interaction strength, lattice mismatch, and angular dependence in the 
conditions that will lead to a myriad of relaxation. Additionally, we have a 
series of SW potentials where we varied the three-body contribution from 0 to 
20 (see table \ref{SW-parameter0t20}). As we increased the $\lambda$ 
parameter, the relaxation is higher allowing us to map the CE fits in the 
highly relaxed configurations.     

\begin{table}[H]
\centering
\caption{SW parameters used to study the relaxation. $\epsilon$ is in eV and 
$\sigma$ is in unit of \AA. $\lambda$ is equal to 0 (three-body contribution 
is off) or 1 (three-body contribution is on). System A, B and C are selected 
to show the effect of relaxation and to evaluate the relaxation metrics.}
\label{SW-parameter}
\begin{tabular}{|l|l|l|l|l|l|l|}
\hline
$\epsilon_{AA}$ & $\sigma_{AA}$ & $\epsilon_{BB}$ & $\sigma_{BB}$ & $
\epsilon_{AB}$ & $\sigma_{AB}$ & $\theta$ \\
\hline
0.21683      & 1.990     & 0.21683  & 2.1998 & 0.4336 & 2.0951 & 60      \\
\hline  
0.21683      & 1.938     & 0.21683  & 2.252 & 0.4336 & 2.0951 & 60         \\
\hline  
0.21683     & 2.0427     & 0.21683  & 2.1740 & 0.4336 & 2.0951 & 60        \\
\hline 
2.1683      & 2.0427     & 2.1683  & 2.1683 & 2.3813 & 2.0951 & 60         \\
\hline 
2.1683      & 2.0427     & 2.1683  & 2.1683 & 2.3813 & 2.0951  & 60       \\
\hline 
0.21683      & 2.0427     & 0.21683  & 2.1683 & 0.23813 & 2.0951 & 60     \\
\hline
0.21683      & 2.0427     & 0.21683  & 2.1683 & 0.26020 & 2.0951  & 60     \\
\hline
0.21683      & 2.0427     & 0.21683  & 2.1683 & 0.27103 & 2.0951  & 60    \\
\hline
0.21683      & 2.0427     & 0.21683  & 2.1683 & 0.28188 & 2.0951  & 60     \\
\hline
0.21683      & 2.0427     & 0.21683  & 2.1683 & 0.30356 & 2.0951  & 60     \\
\hline
0.21683      & 2.0427     & 0.21683  & 2.1683 & 0.32525 & 2.0951  & 60     \\
\hline  
0.21683 (A)    & 2.0427     & 0.21683  & 2.1740 & 0.4336 & 2.0951 & 109.5 \\
\hline
0.21683 (B)   & 1.990     & 0.21683  & 2.1998 & 0.4336 & 2.0951 & 109.5     \\
\hline  
0.21683 (C)   & 1.938     & 0.21683  & 2.252 & 0.4336 & 2.0951 & 109.5    \\
\hline 
2.1683      & 2.0427     & 2.1683  & 2.1683 & 2.3813 & 2.0951 & 109.5     \\
\hline 
2.1683      & 2.0427     & 2.1683  & 2.1683 & 2.3813 & 2.0951  & 109.5    \\
\hline 
0.21683      & 2.0427     & 0.21683  & 2.1683 & 0.23813 & 2.0951 & 109.5  \\
\hline
0.21683      & 2.0427     & 0.21683  & 2.1683 & 0.26020 & 2.0951  & 109.5 \\
\hline
0.21683      & 2.0427     & 0.21683  & 2.1683 & 0.27103 & 2.0951  & 109.5 \\
\hline
0.21683      & 2.0427     & 0.21683  & 2.1683 & 0.28188 & 2.0951  & 109.5 \\
\hline
0.21683      & 2.0427     & 0.21683  & 2.1683 & 0.30356 & 2.0951  & 109.5 \\
\hline
0.21683      & 2.0427     & 0.21683  & 2.1683 & 0.32525 & 2.0951  & 109.5  \\
\hline

\end{tabular}
\end{table}

\begin{table}[H]
\centering
\caption{SW parameters used to study the relaxation with $\lambda$ varying 
from 0 (no interaction) to 20. $\epsilon $ is in eV and $\sigma$ is in unit 
of \AA. $\lambda$ is three-body contribution where $\lambda$ equals zero they 
are no interaction and as $\lambda$ increases the three-body contribution is 
higher and higher relaxation. }
\label{SW-parameter0t20}
\begin{tabular}{|l|l|l|l|l|l|l|}
\hline
$\epsilon_{AA}$ & $\sigma_{AA}$ & $\epsilon_{BB}$ & $\sigma_{BB}$ & 
$\epsilon_{AB}$ & $\sigma_{AB}$ & $\theta$ \\
\hline
1.1683      & 1.990     & 2.80  & 2.1998 & 0.4336 & 2.0951 & 109.5      \\
\hline  
2.125 & 1.990      & 1.5285   & 2.200 & 3.0570 & 2.095 & 109.5      \\
\hline 
\end{tabular}
\end{table}

The Stillinger-Weber potential has several tunable parameters to simulate 
very high level of relaxations. CE fitting for BCC structures yields lower 
error than FCC or HCP as shown in table \ref{result-SW-SW-like}.

\begin{table}[H]
\centering
\caption{Cluster expansion fittings using a Stillinger-Weber potential (system
 B in table \ref{SW-parameter}) at 10\% lattice mismatch.}
\label{result-SW-SW-like}
\begin{tabular}{|l|l|l|l|}
\hline
Lattice & simulation & percent error (RMS/std(y) \%) & $J$s \\
\hline
FCC     & unrelaxed     &   0.07 \%          &  3    \\
\hline
FCC     & relaxed    &    41.2 \%          &  110   \\
\hline
BCC      & unrelaxed    &      0.03 \%         &  3    \\
\hline
BCC     & relaxed     &        18.91 \%          &   42     \\
\hline 
HCP      & unrelaxed     &       0.23 \%      &  4     \\
\hline      
HCP     & relaxed     &       72.64 \%      &  167     \\
\hline            
\end{tabular}
\end{table} 

\subsubsection{Order Parameters (OPs)}
In order to distinguish and measure the relaxation of the atoms from their 
ideal positions, we examined several metrics (order parameters) to quantify 
the relaxation: normalized mean-squared displacement or NMSD (see the method 
in the  main article), Ackland's order parameter \cite{Ackland2006}, $D_6$, 
SOAP, and centro-symmetry. We found that some of these order parameters are 
not descriptive/general enough for all cases (potentials and crystal lattices). 
 
We used the Ackland's order parameter to identify the crystal structure after 
relaxation.  Ackland's OP identify each atomic local environment and assign it 
as FCC, BCC, HCP and Unknown \cite{Ackland2006}. We used this OP to determine 
which structures remain the same or on lattice and which structures undergo a 
structural change.  We can use this order parameter to separate/sort those 
structure that remain the same to examine the robustness of CE due the 
relaxation of crystal structure. Similar to the Ackland's order parameter, the 
centro-symmetry identifies the crystal structure of each atom based on the 
local arrangement (neighbors). 

\begin{figure}[H]
\centering
    \begin{subfigure}{0.5\textwidth}
  \includegraphics[width=\textwidth]{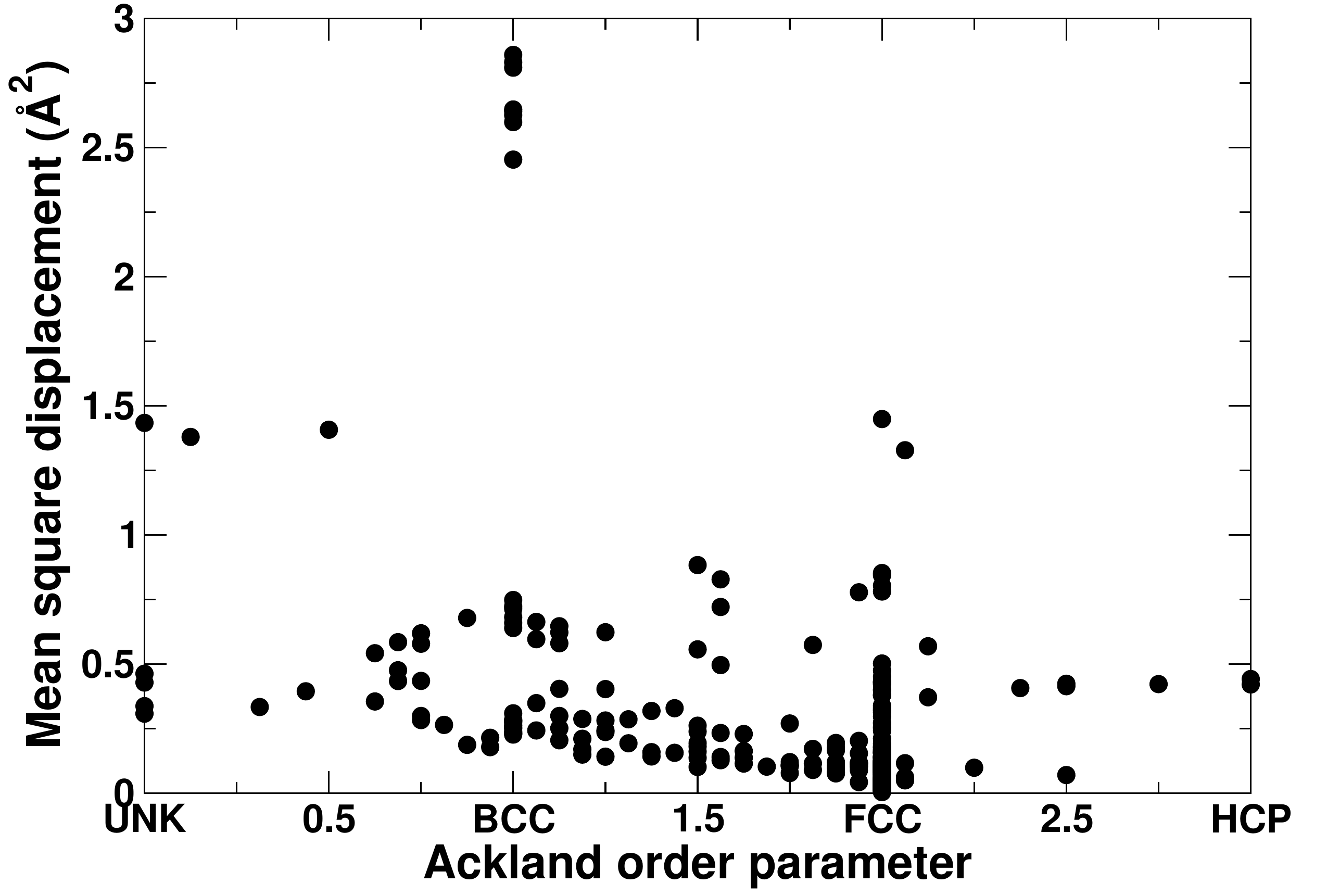} 
   \caption{5\% lattice mismatch}
    \label{fig:5percent-lm-SW}
    \end{subfigure} 
    \centering
    \begin{subfigure}{0.5\textwidth}
   \includegraphics[width=\textwidth]{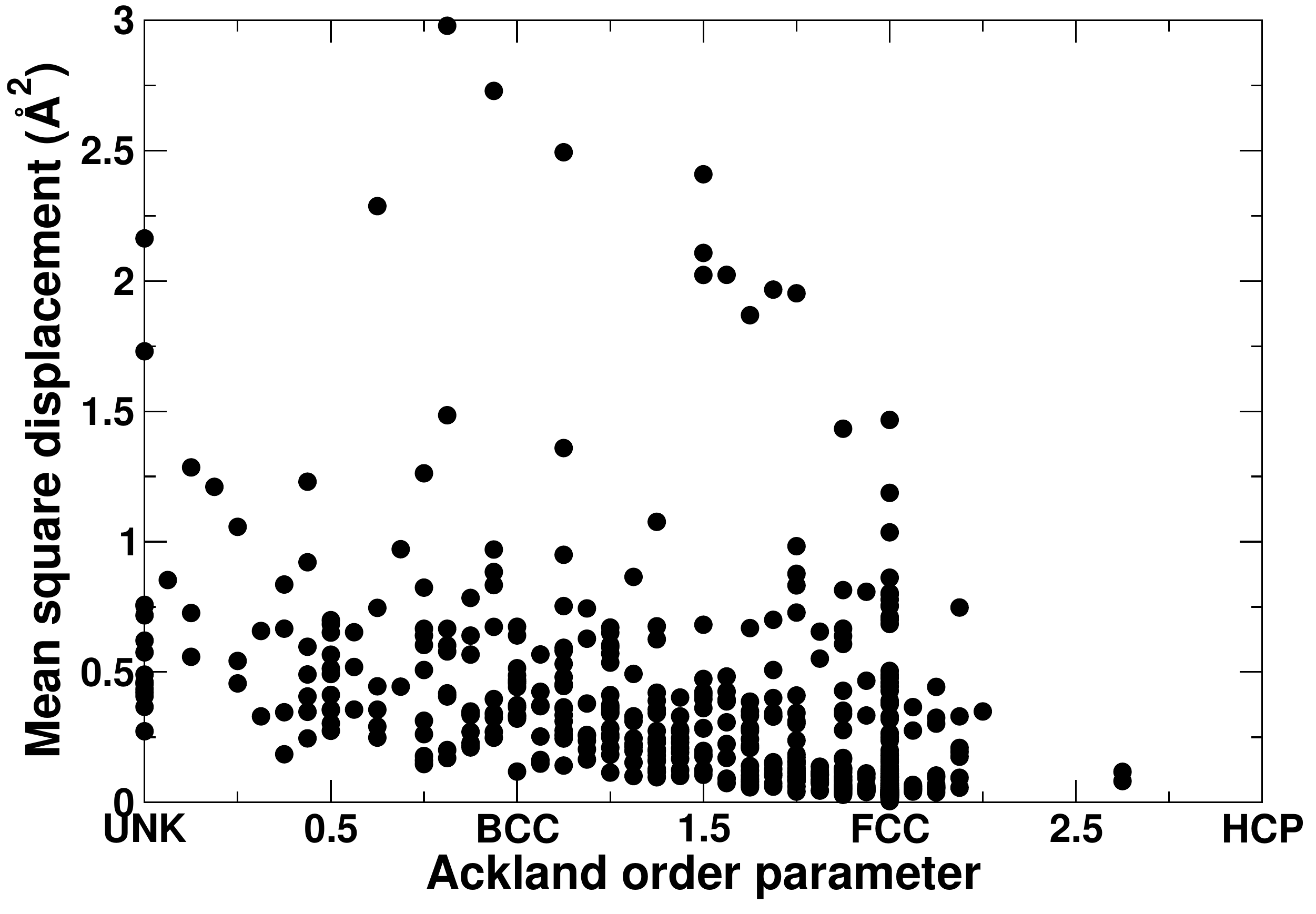}
   \caption{15\% lattice mismatch}
    \label{fig:15percent-lm-SW} 
    \end{subfigure}
  \caption{(color online)  Mapping of structural changes and MSD for SW 
  potential of a FCC lattice. The top panel shows the relaxation for a system 
  with 5\% lattice mismatch, while the bottom graph show relaxation of a 15\% 
  lattice mismatch system. Higher lattice mismatch equals higher mean-squared 
  displacement.}
  \label{fig:Relaxed-SW-ACK-FCC}
\end{figure}

For example, Figure \ref{fig:Relaxed-SW-ACK-FCC} shows the mapping of MSD and 
Ackland order parameter for each structure at 5\% (top plot) and 15\% (bottom 
plot).  Overall, we can see that the MSD increases with higher lattice 
mismatch.  The spread of the Ackland's order parameter is also affected.  
Going from 5\% (system A in table \ref{SW-parameter}) to 15\% (system C in 
table \ref{SW-parameter}), the CE fitting error increases from 41.2\% (110 
clusters) to 63.5\% (156 clusters) for the BCC. When the lattice mismatch 
increases, the mean-squared displacement also increases.  Ackland's order 
parameter and centro-symmetry are useful since they provide information about 
individual atoms. However, Ackland's order parameter and centro-symmetry is 
too specific and it does not provide a useful measure of relaxation.
   
\begin{figure}[]
 \centerline{\includegraphics[width=0.5\textwidth]{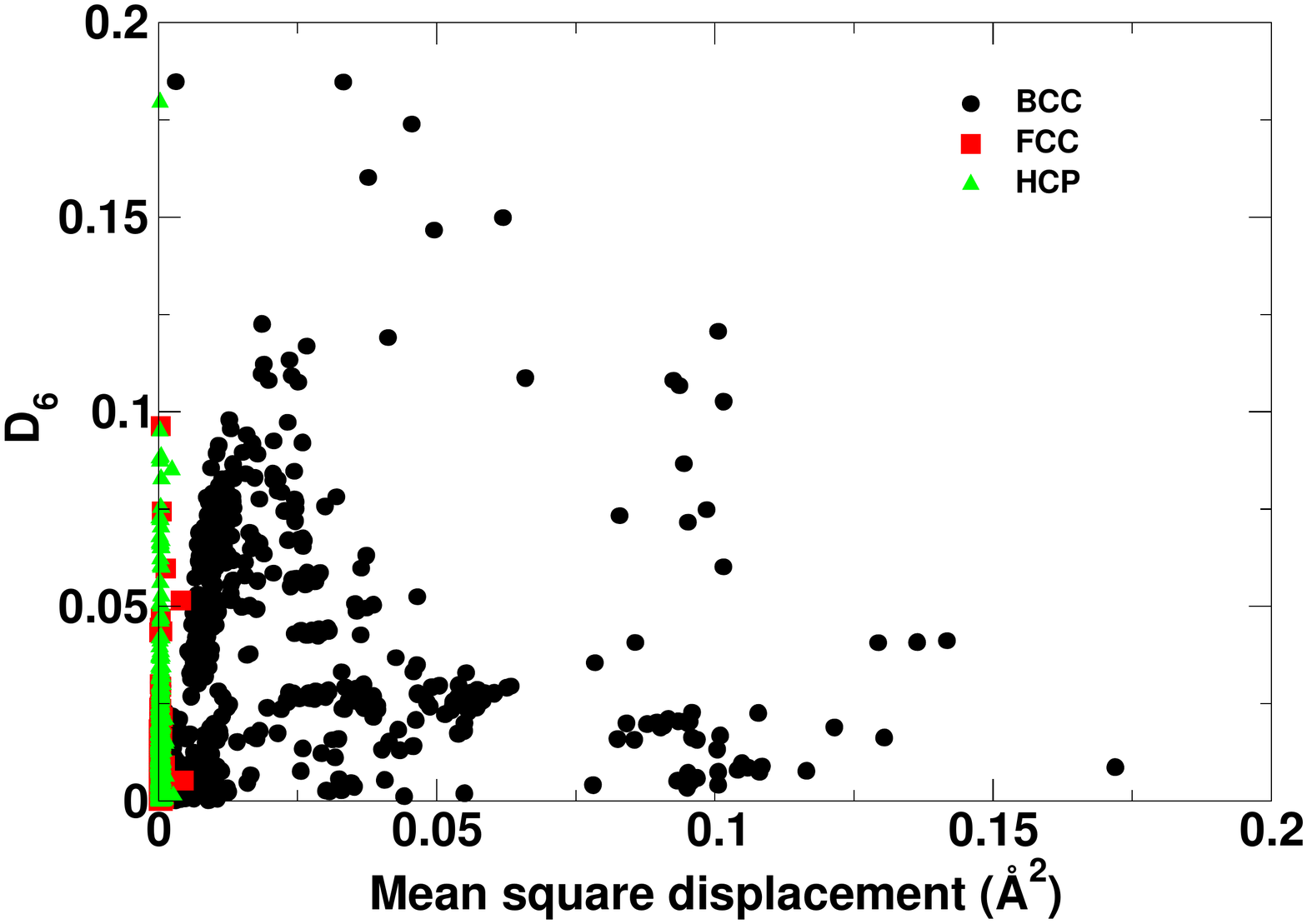}} 
\caption{ (color online) $D_6$ as a measure of the relaxation using a Lennard-
Jones potential. We show that the $D_6$ metric does not correlated with the 
displacement. We show the relaxation of three crystal lattice.  LJ favor FCC/
HCP; thus, we should not observe high relaxation (this is indicated by the 
displacement which is less than 0.001 $\mathrm{\AA^2}$. However, the $D_6$ 
metrics show a very broad range from 0.0 (identical configuration) up to 0.1. 
Although we only show this plot for LJ, the results of SW and EAM potential 
reveal the same conclusion, that is, $D_6$ is not a sufficient metric to 
analyze the various crystal lattices and potentials.}
\label{fig:D6-op}
\end{figure}

\begin{figure}[]
 \centerline{\includegraphics[width=0.5\textwidth]{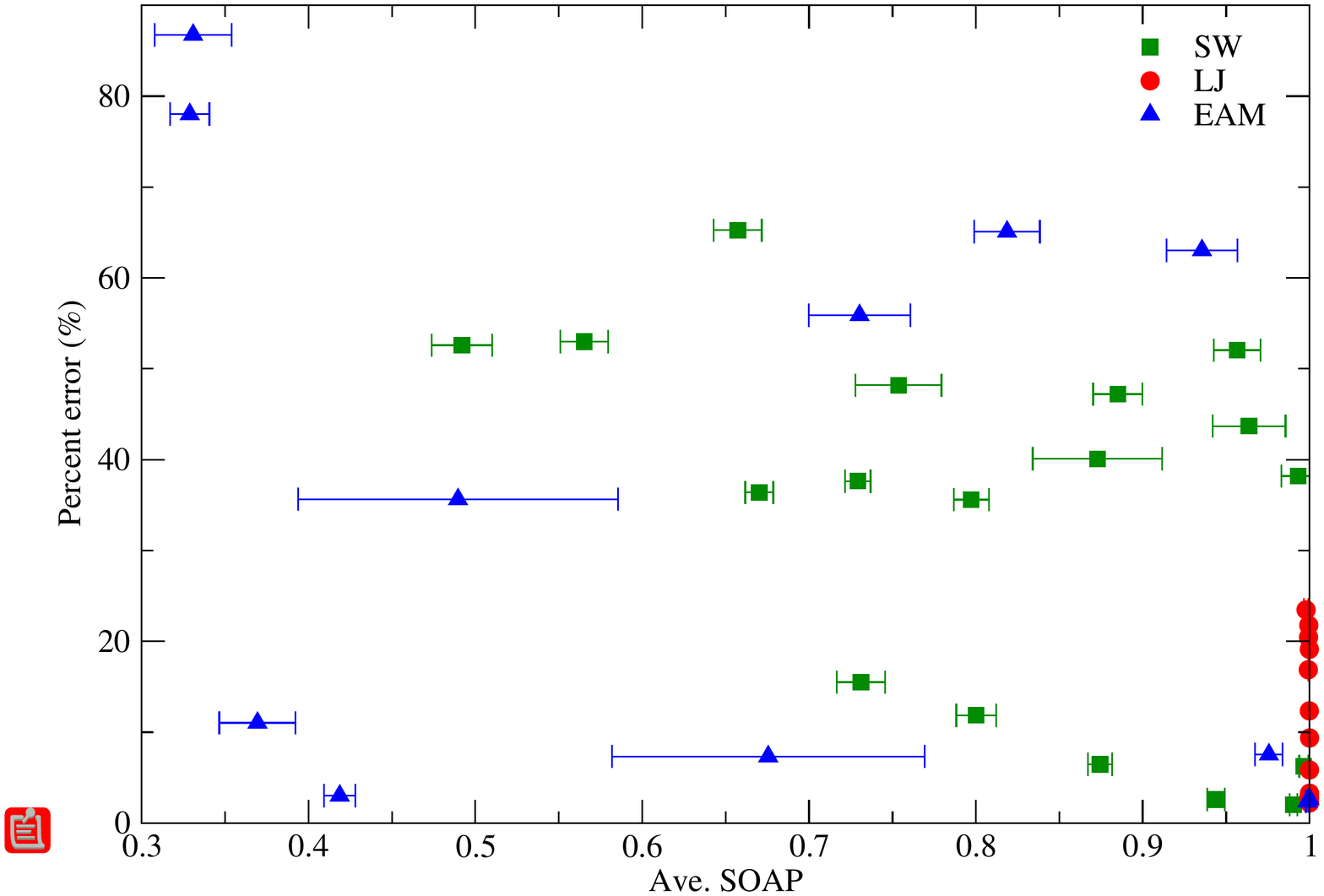}} 
\caption{ (color online) SOAP similarity kernel vs error.  The plot shows the 
error as a function of the SOAP kernel.  SOAP measure the similarity between 
the unrelaxed and relaxed configurations. When SOAP value is close to 1, the 
relaxed and unrelaxed configuration is similar (identical if it is 1). 
However, the plot shows that it is not a robust measure of relaxation as LJ 
has a very high range of prediction error for a very narrow range of SOAP 
value. }
\label{fig:soap-error}
\end{figure}

In addition to using the crystallographic information as a measure of 
relaxation, we used a variant of the Steinhardt's bond order parameter 
\cite{Steinhardt1983} that we called the $D_6$ order parameter or the $D_6$ 
metric, which is a measure of the difference between the local atomic 
environment (relaxed and unrelaxed).  We computed the local atomic environment 
using the $q_6$ (spherical harmonic with $l = 6$) for the unrelaxed and 
relaxed configuration. We averaged the difference of the two configurations, 
$D_6 =\frac{1}{N_{\mathrm{atom}}} \sum_{\mathrm{atom}} (q_{6,\mathrm{rel}}-
q_{6,\mathrm{unrel}})$. Figure \ref{fig:D6-op} shows the $D_6$ metric as a 
measure of relaxation vs the mean-squared displacement (MSD). We observe that 
$D_6$ metric does not correlate withe MSD.  As relaxation increases (higher 
MSD), we expect that the $D_6$ value also increase. However, this metric is not 
robust for all systems, that is, we cannot compare the relaxation across all 
Hamiltonians (potentials and crystal lattices). Similar to the $D_6$ metrics, 
we used another metrics known as the SOAP (smooth overlap of atomic position) 
similarity kernel.  The SOAP similarity kernel measures the difference in 
configuration (1 when it is identical and decreasing as the difference 
increases).  The SOAP kernel is invariant to rotation and translation 
\cite{Bartok2013}; however, this metric is not applicable for multiple species 
cases. Fig. \ref{fig:soap-error} shows the prediction error vs SOAP. Similar 
to $D_6$, the SOAP value does not correlate with the prediction error or 
displacement, that is, these metrics are too broad and vary too much for small 
displacements.  This problem lies in the normalization of SOAP and $D_6$ values.  

\subsection{\label{sectionS3} Numerical Error}

None of the normal quantifying descriptions of distribution shape (e.g., 
width, skewness, kurtosis, standard deviation, etc.) show a correlation with 
the CE prediction error. The error increased proportionally with the level of 
error in each system (2, 5, 10 and 15\% error).

\begin{figure}[H]
\centering
    \begin{subfigure}{0.47\textwidth}
  \includegraphics[width=\textwidth]{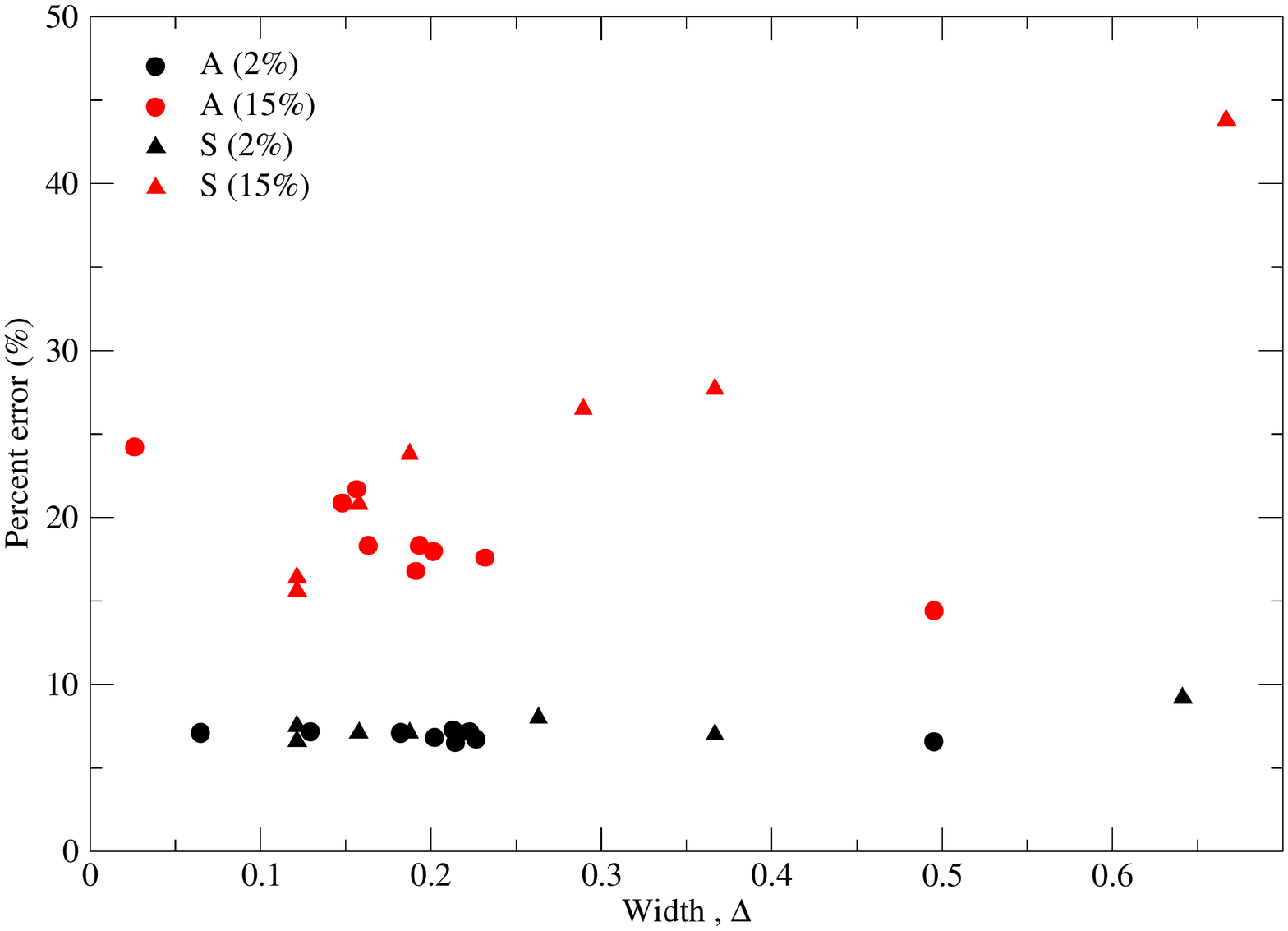} 
   \caption{Error vs Width}
    \label{fig:ErrvsWidth}
    \end{subfigure} 
    \begin{subfigure}{0.47\textwidth}
  \includegraphics[width=\textwidth]{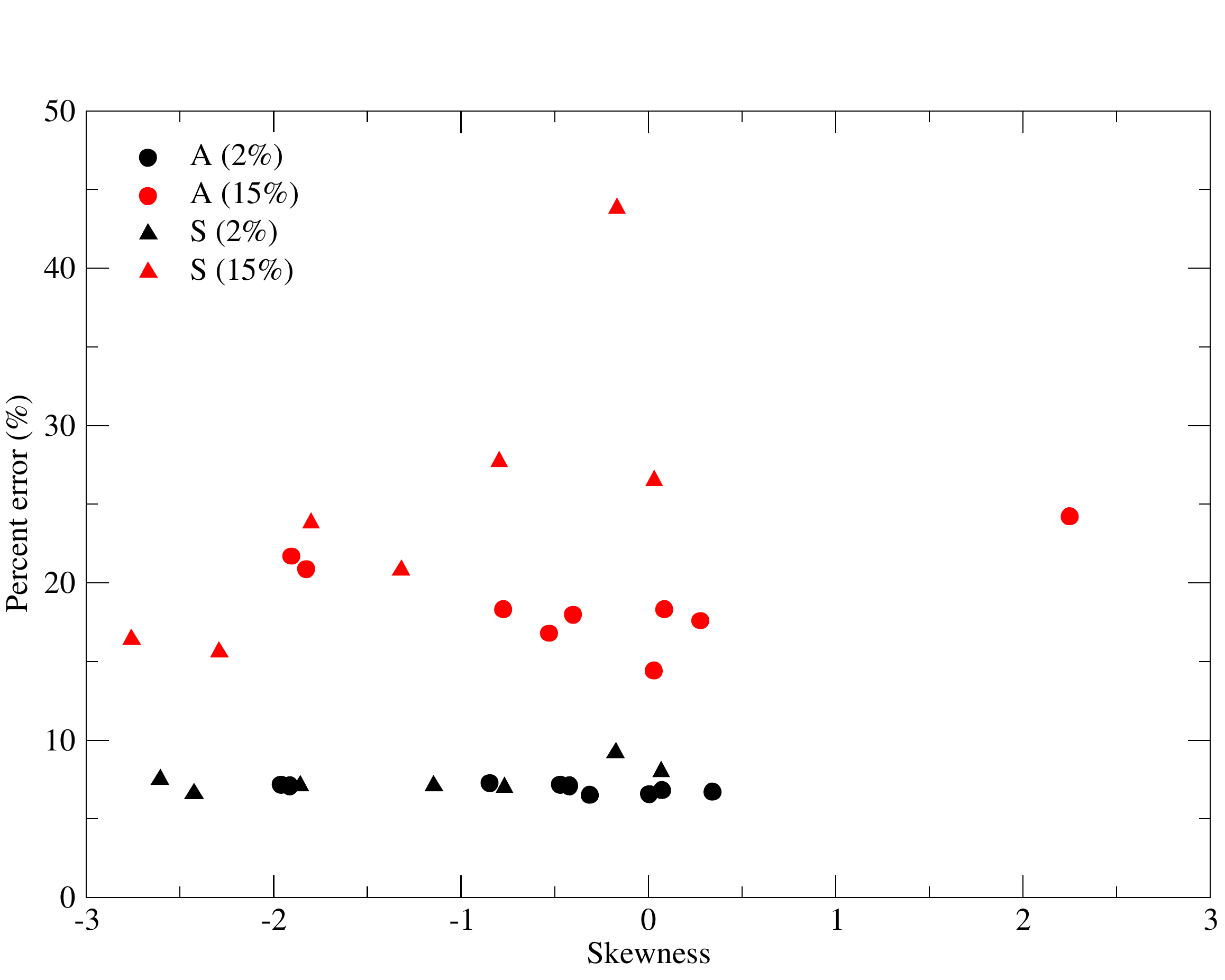} 
   \caption{Error vs Skewness}
    \label{fig:ErrvsSkew}
    \end{subfigure}
    \begin{subfigure}{0.47\textwidth}
  \includegraphics[width=\textwidth]{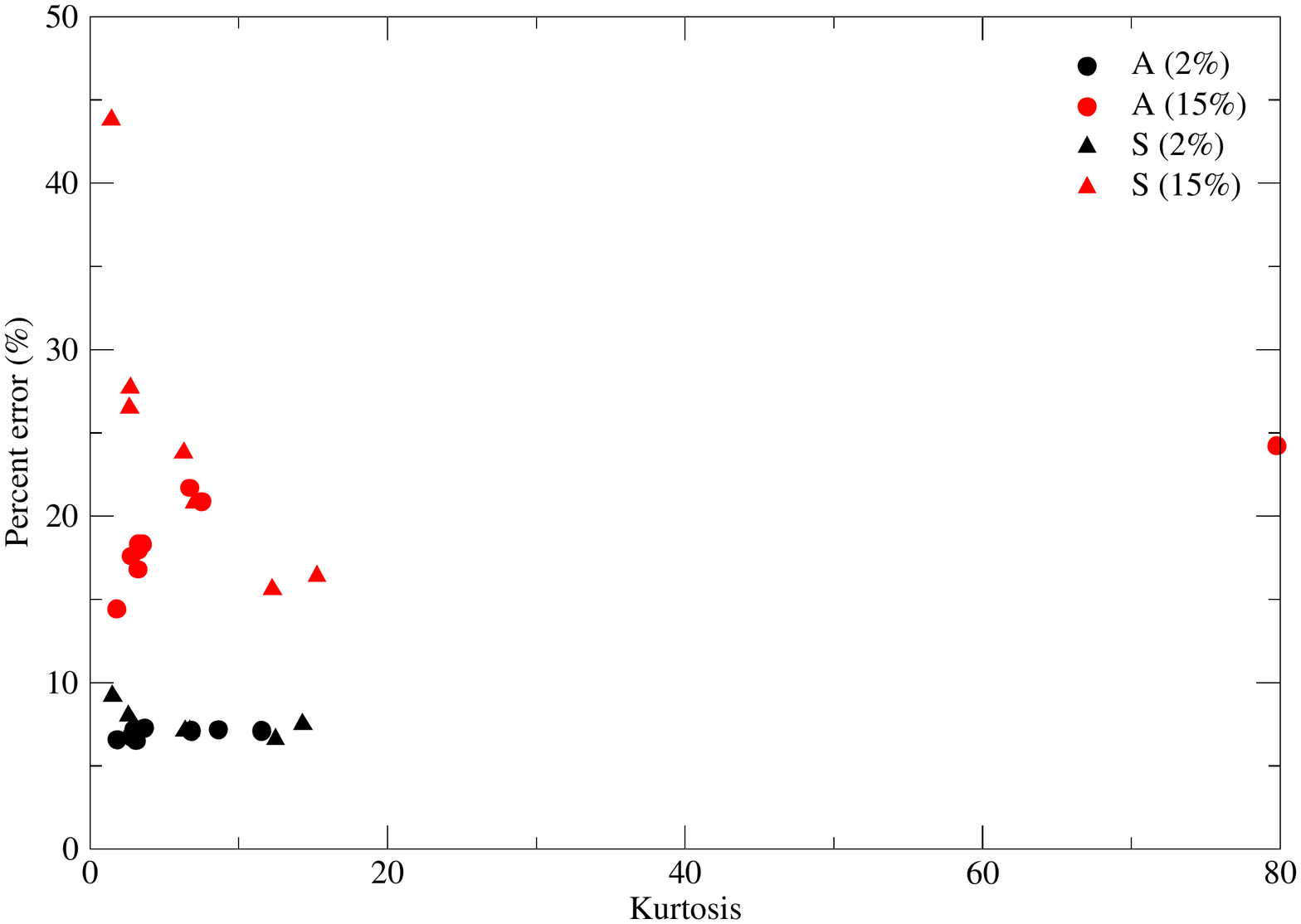} 
   \caption{Error vs Kurtosis}
    \label{fig:ErrvsKurt}
    \end{subfigure}
\caption[]{\label{fig:EqualWidthSummary} (color online) Width, skewness and 
kurtosis using the relaxation energies. The relaxation energies are obtained 
by taking the absolute difference of unrelaxed and relaxed energies. We show 
only the 2\% and 15\% error instead of all four error levels.  This allows us
 to illustrate the effect of error level on the width, skewness and kurtosis 
 of the distribution.}
\end{figure}

\begin{figure}[H]
\centering
    \begin{subfigure}{0.5\textwidth}
  \includegraphics[width=\textwidth]{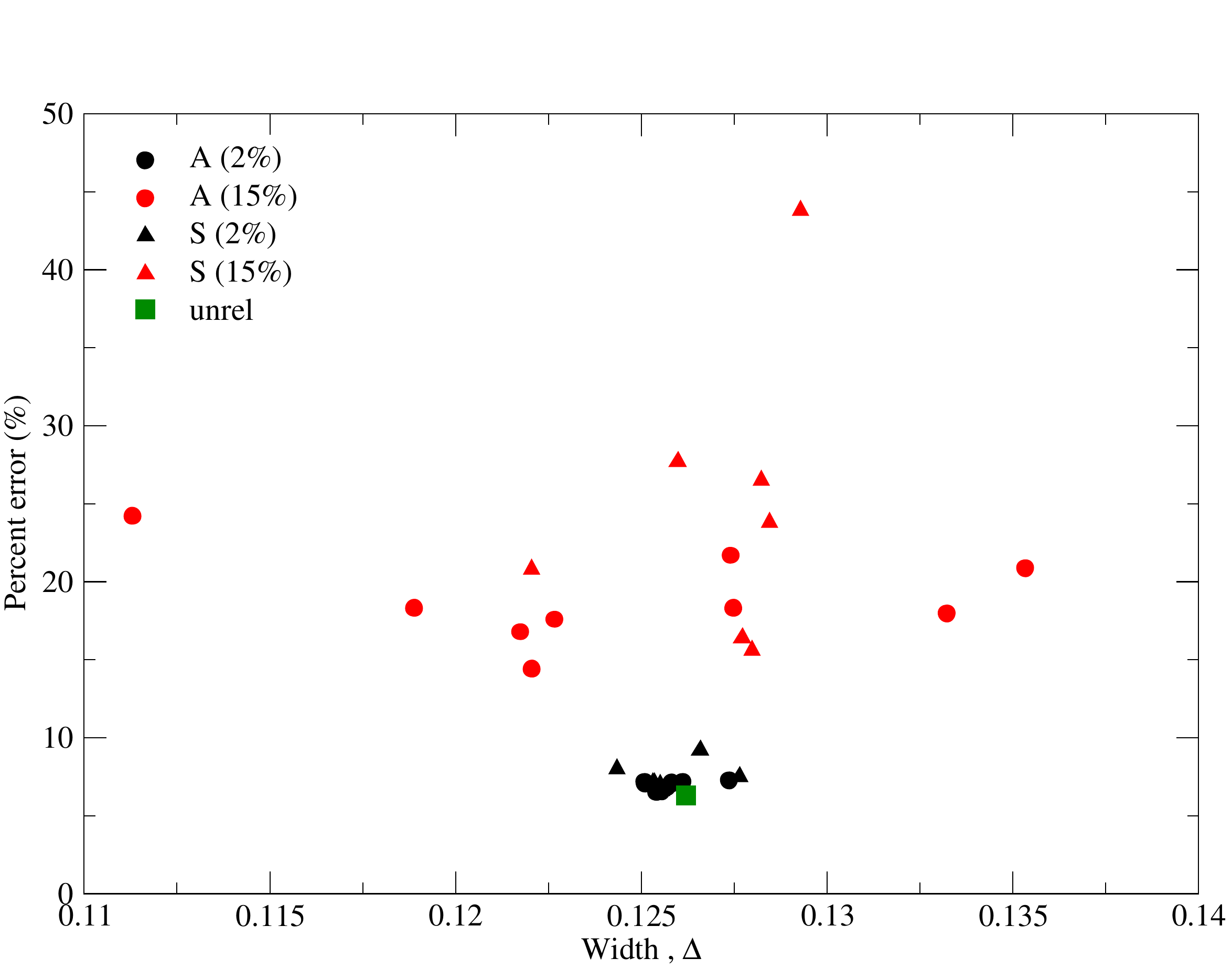} 
   \caption{Error vs Width}
    \label{fig:ErrvsWidth-ur}
    \end{subfigure} 
    \begin{subfigure}{0.5\textwidth}
  \includegraphics[width=\textwidth]{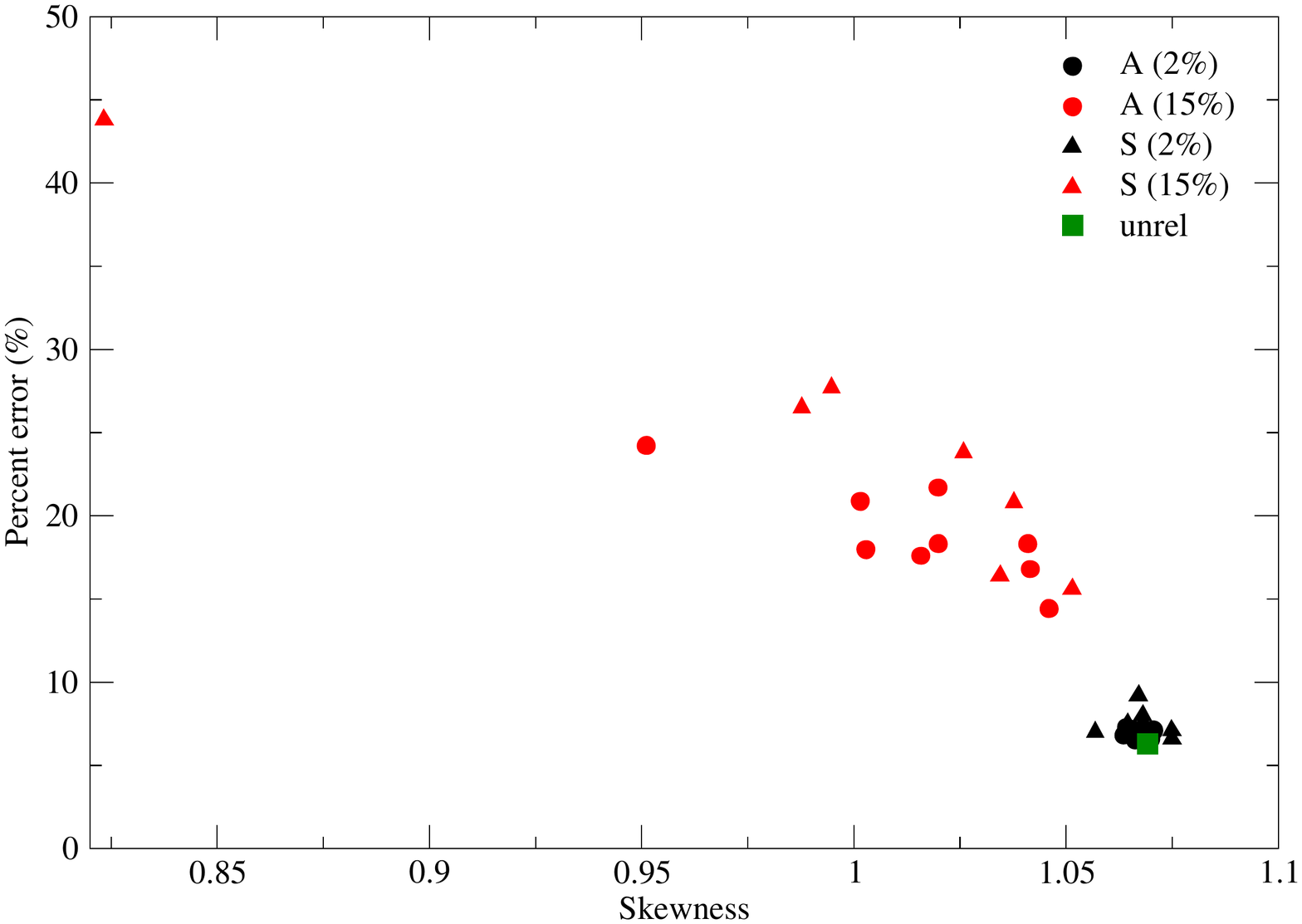} 
   \caption{Error vs Skewness}
    \label{fig:ErrvsSkew-ur}
    \end{subfigure}
    \begin{subfigure}{0.5\textwidth}
  \includegraphics[width=\textwidth]{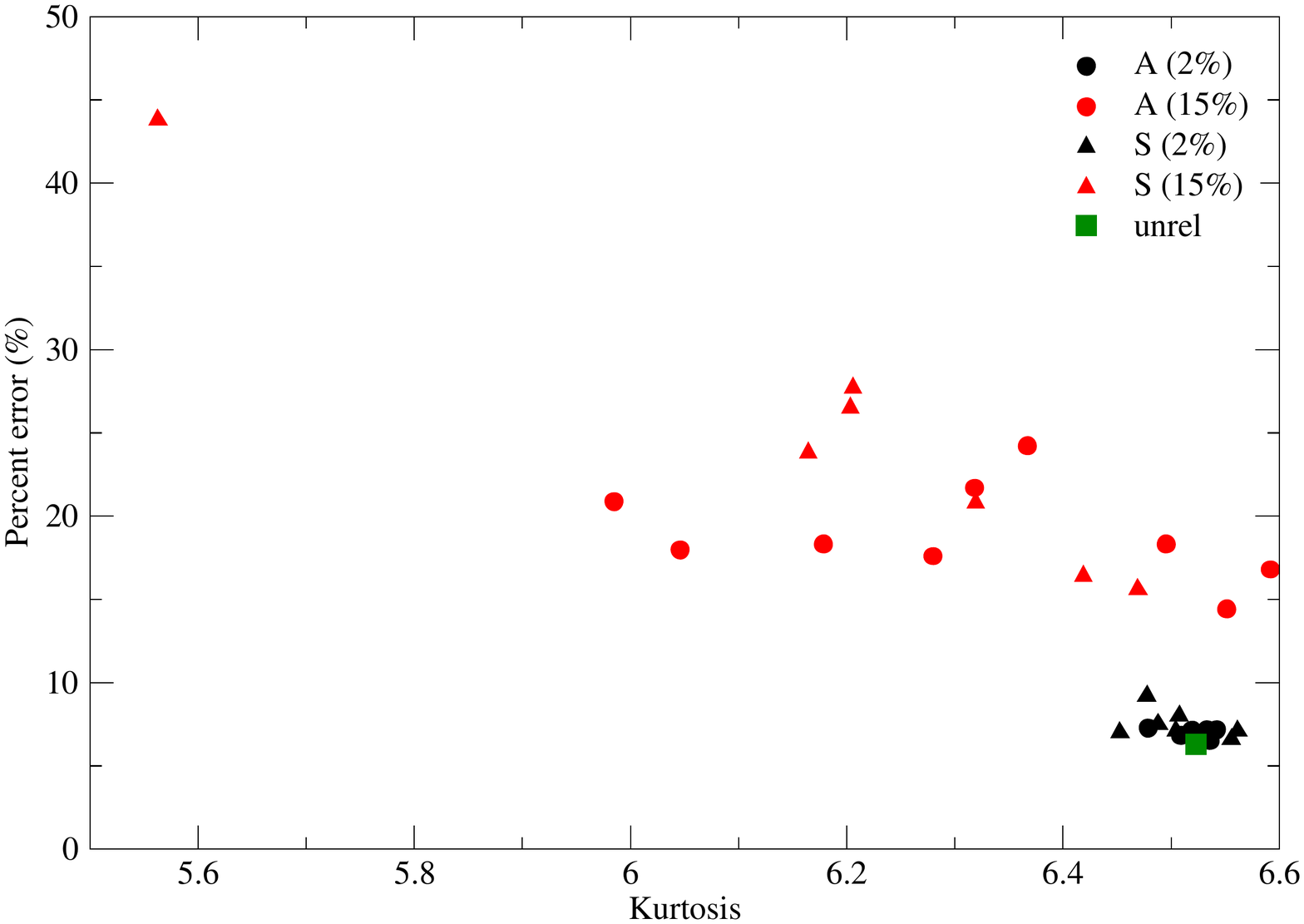} 
   \caption{Error vs Kurtosis}
    \label{fig:ErrvsKurt-ur}
    \end{subfigure}
\caption[]{\label{fig:EqualWidthSummary} (color online) Width, skewness and 
kurtosis using total energies. The energies is the total relaxed energies. The 
green symbol represents the unrelaxed system. }
\end{figure}

\begin{figure}[H]
\centerline{\includegraphics[width=0.5\textwidth]{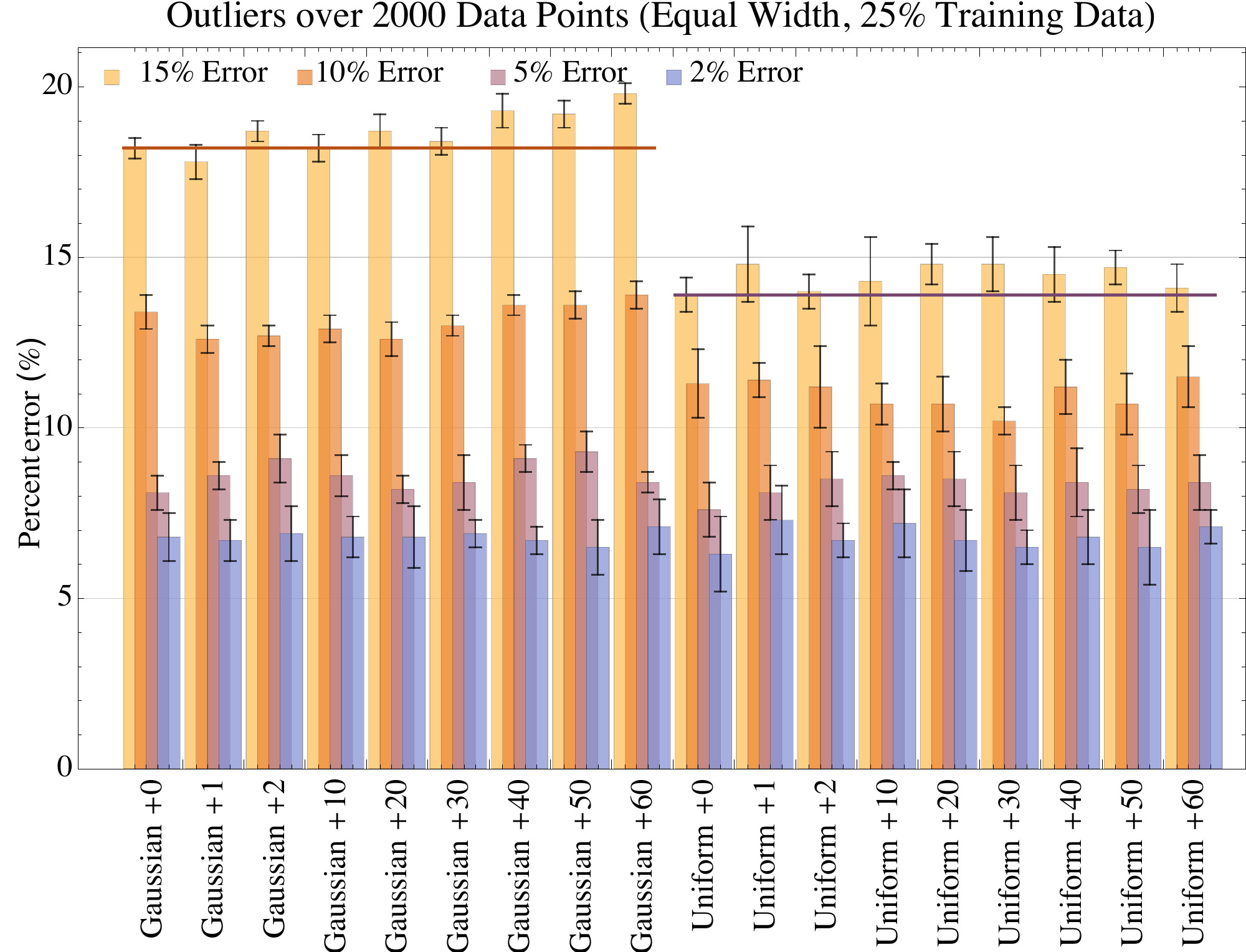}}
\caption[]{\label{fig:OutlierSummary} (color online)  
  Fitting errors as outliers are
  added to the error profile. The difference between fits as the
  number of outliers changes is comparable to the variance in the
  individual fits.}
\end{figure}

One other possibility is that the presence of outliers has a large
impact on the performance of the BCS fit. To rule out that
possibility, we performed fits with 0, 1, 2, 10, 20, 30, 40, 50 and 60
outliers added to the error (representing between 0 and 3\% of the
total data). Outliers were selected randomly from between 2 and 4
standard deviations from the mean and then appended to the regular
list of errors drawn from the distribution (the total number of values
equaling 2000 again to match the number of structures). The summary is
plotted in Figure \ref{fig:OutlierSummary}. The difference between
fits as the number of outliers changes is comparable to the variance
in the individual fits. We conclude then that outliers have no direct
effect on the error profile's performance.\\

\subsection{Evolution of $J$ Coefficients on Relaxation}

To further elucidate the claims relative to the CE framework's failures, we 
investigated whether we could measure the change in configuration upon 
relaxation. Since the $J$ values selected in the model are backed by geometric 
clusters, the presence or absence of certain clusters has some correlation to 
the configuration of the physical system. When the physics is mostly dependent 
on configuration, the function can be sparsely represented by the CE basis. 
Thus, we expect that the sparsity will be a good heuristic in determining when 
the CE breaks down. When the expansion terms do not decay well in the 
representation, it shows a misapplication of the CE basis to a problem that is 
not mostly configurational.

We define three new quantities:
\begin{enumerate}
\item\label{item:XiDefinition} $\Xi$: total number of unique clusters used 
over 100 CE fits of the same dataset. We also call this the model complexity 
as shown in Fig. \ref{fig:ErrorVsModelComplexity}. 
\item\label{item:NotInDefinition} $\not\in$: number of ``exceptional'' 
clusters. These are clusters that show up fewer than 25 times across 100 fits, 
implying that they are not responsible for representing any real physics in 
the signal, but are rather included because the CE basis is no longer a sparse 
representation for the relaxed alloy system (shown in Fig. 
\ref{fig:ErrorVsSingleShow}. It is sensitive to the training subsets. 
\item\label{item:LambdaDefinition} $\Lambda$: number of \emph{significant} 
clusters in the fit; essentially just the total number of unique clusters 
minus the number of ``exceptional'' clusters, $\Lambda = \Xi - \not\in$ (see 
Fig. \ref{fig:ErrorVsSignificantTerms}).
\end{enumerate}

\begin{figure*}[h]
    \centering
    \begin{subfigure}[h]{0.49\textwidth}
        \includegraphics[width=\textwidth]{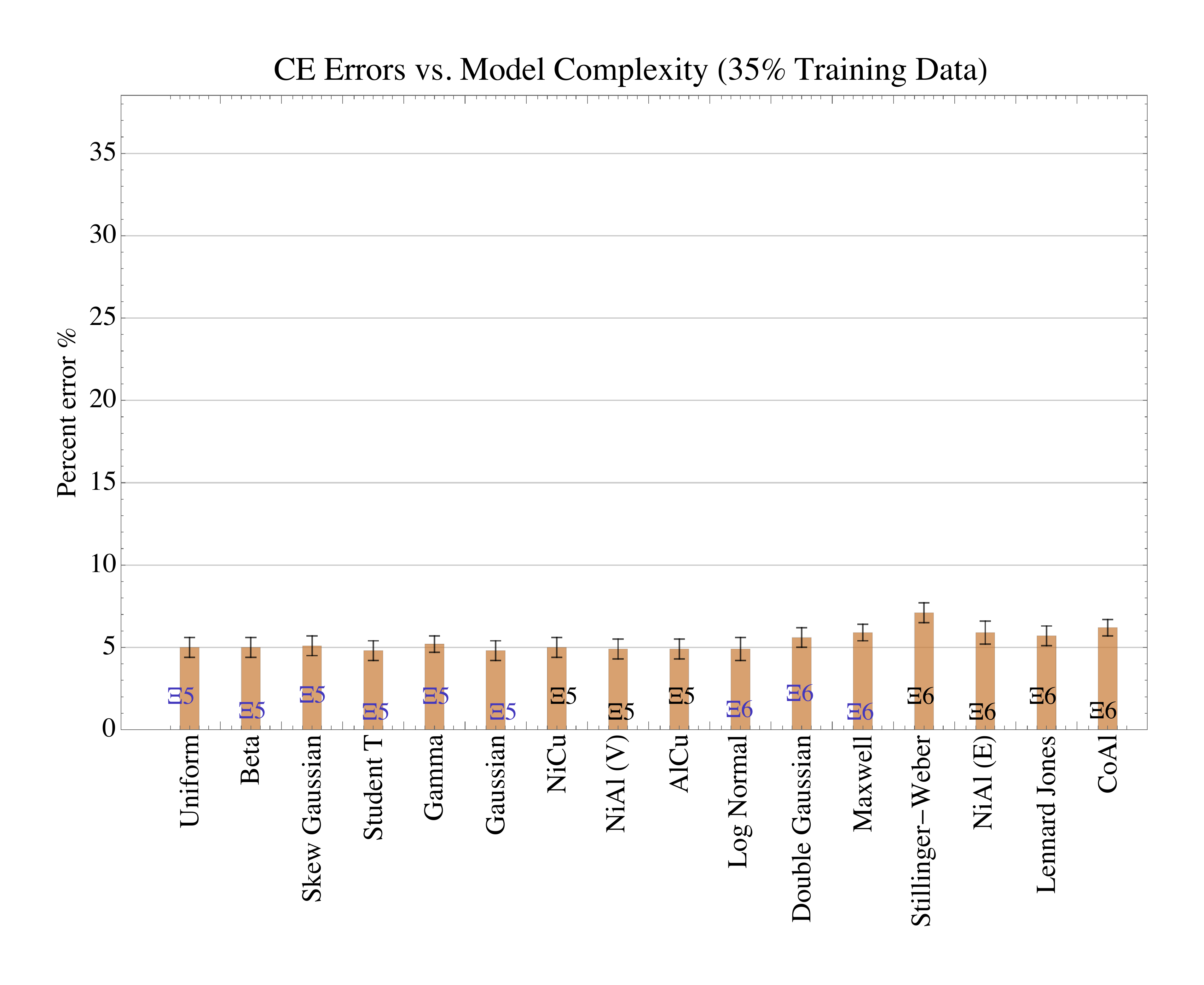}
        \caption{2\% error added}
        \label{fig:errC2}
    \end{subfigure}
    ~
    \begin{subfigure}[h]{0.49\textwidth}
        \includegraphics[width=\textwidth]{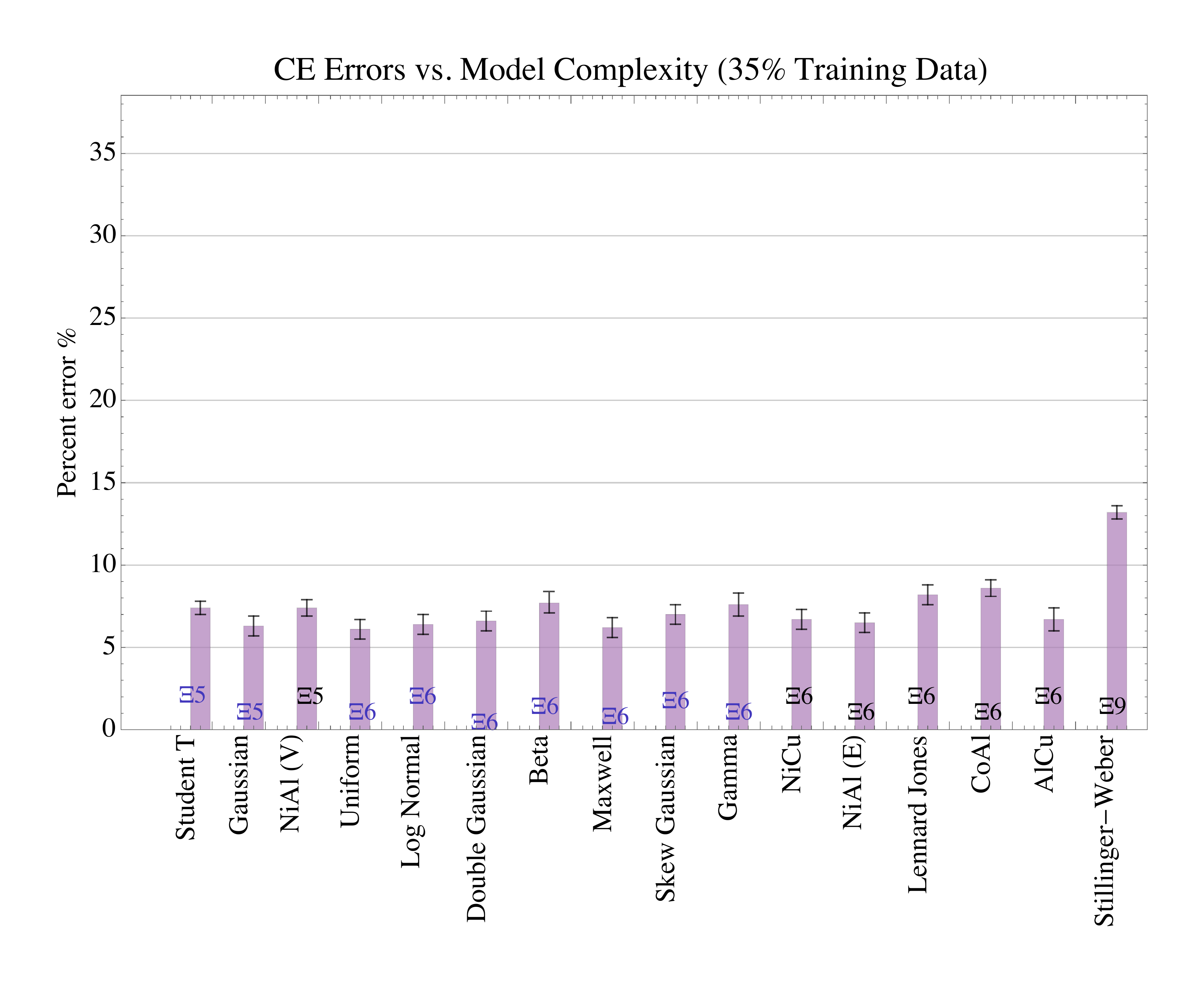}
        \caption{5\% error added}
        \label{fig:errC5}
    \end{subfigure}
    
    \begin{subfigure}[h]{0.49\textwidth}
        \includegraphics[width=\textwidth]{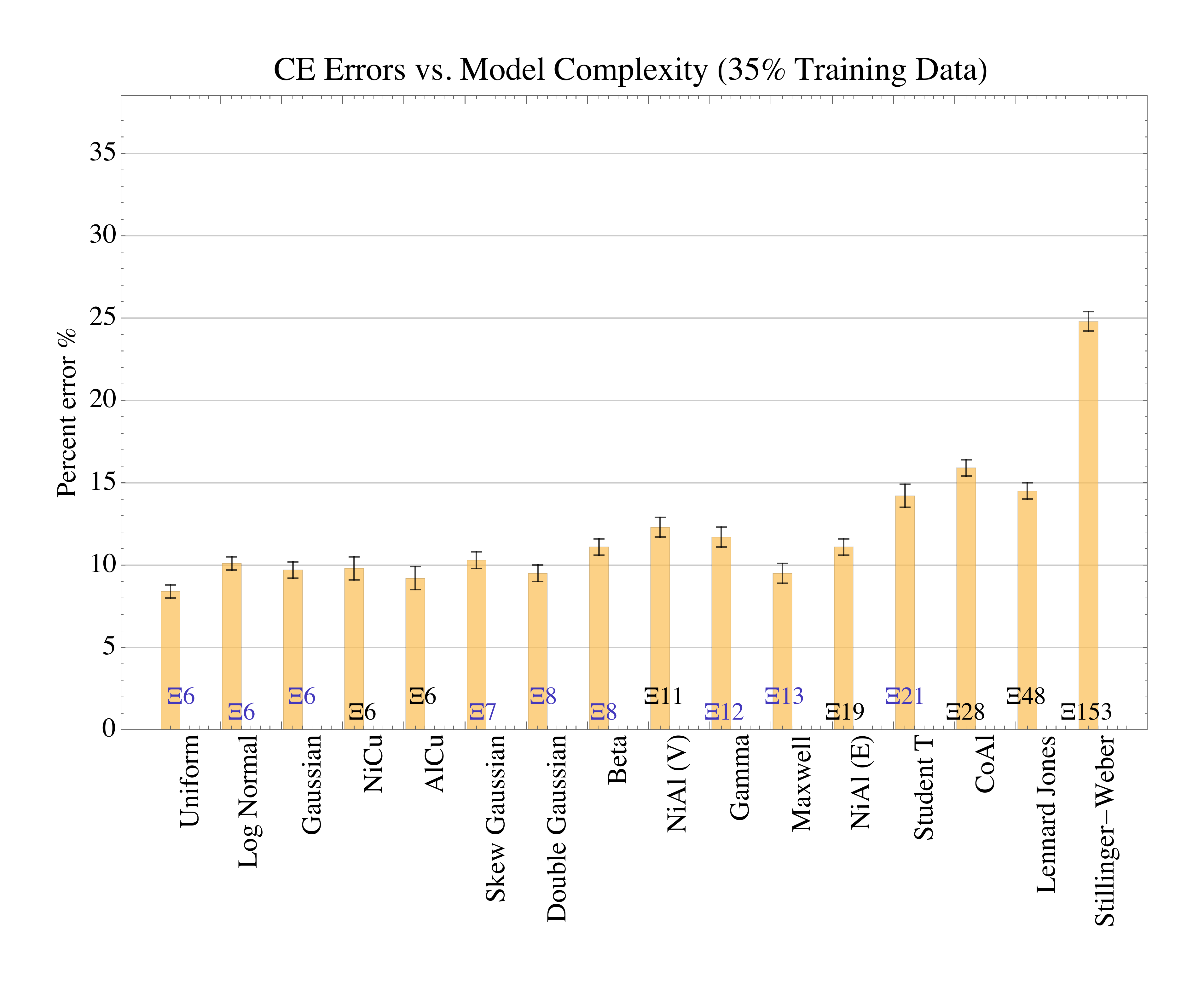}
        \caption{10\% error added}
        \label{fig:errC10}
    \end{subfigure}
~
    \begin{subfigure}[h]{0.49\textwidth}
        \includegraphics[width=\textwidth]{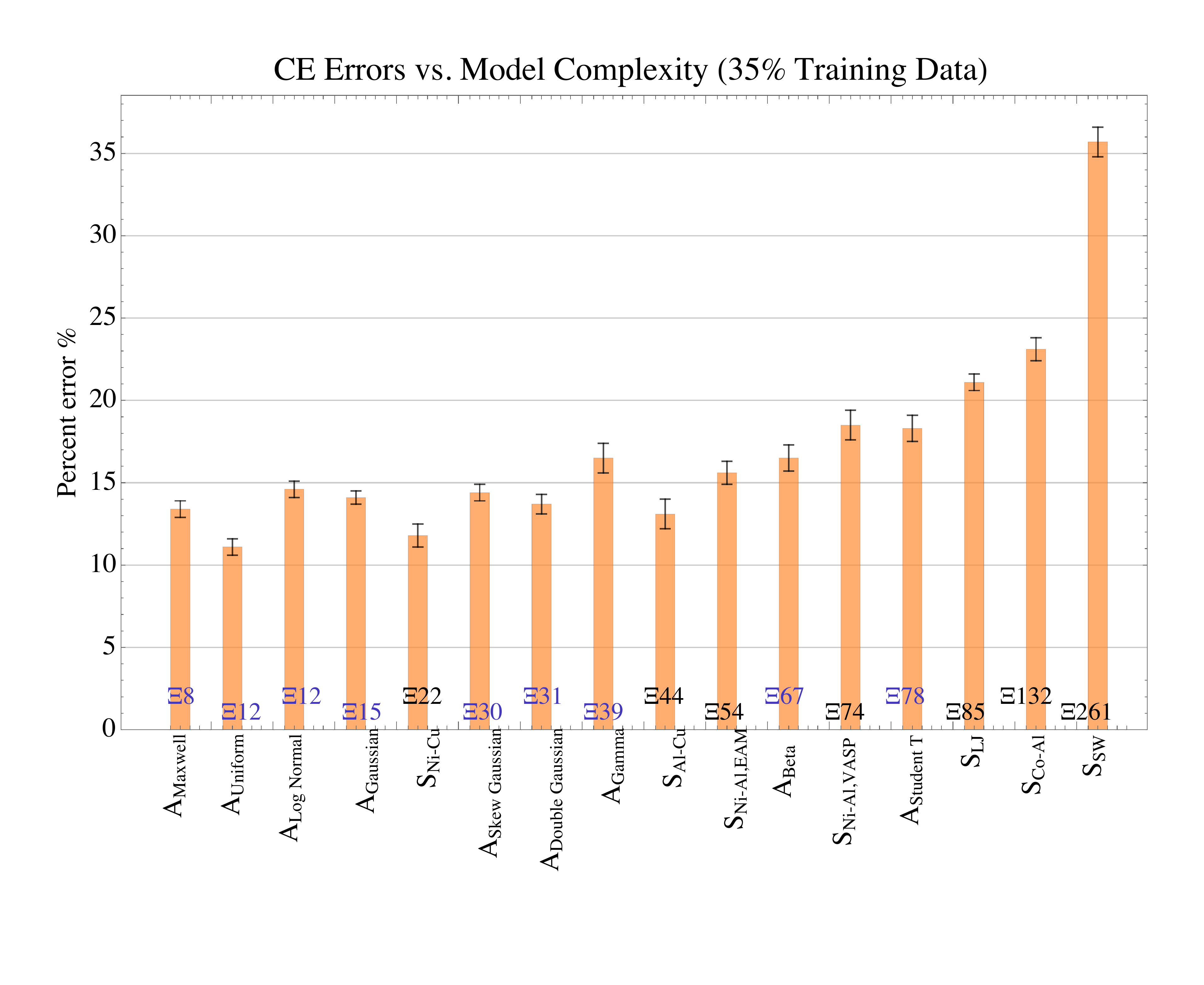}
        \caption{15\% error added}
        \label{fig:errC15}
    \end{subfigure}
    \caption{(color online) Prediction error over 65\% of the structures for 
    the toy CE system. The systems are ordered by $\Xi$, which is the total 
    number of unique clusters used by any of the 100 CE fits for the system. 
    This ordering shows a definite trend with increasing $\Xi$. }
\label{fig:ErrorVsModelComplexity}
\end{figure*} 

\begin{figure*}[h]
    \centering
    \begin{subfigure}[h]{0.49\textwidth}
        \includegraphics[width=\textwidth]{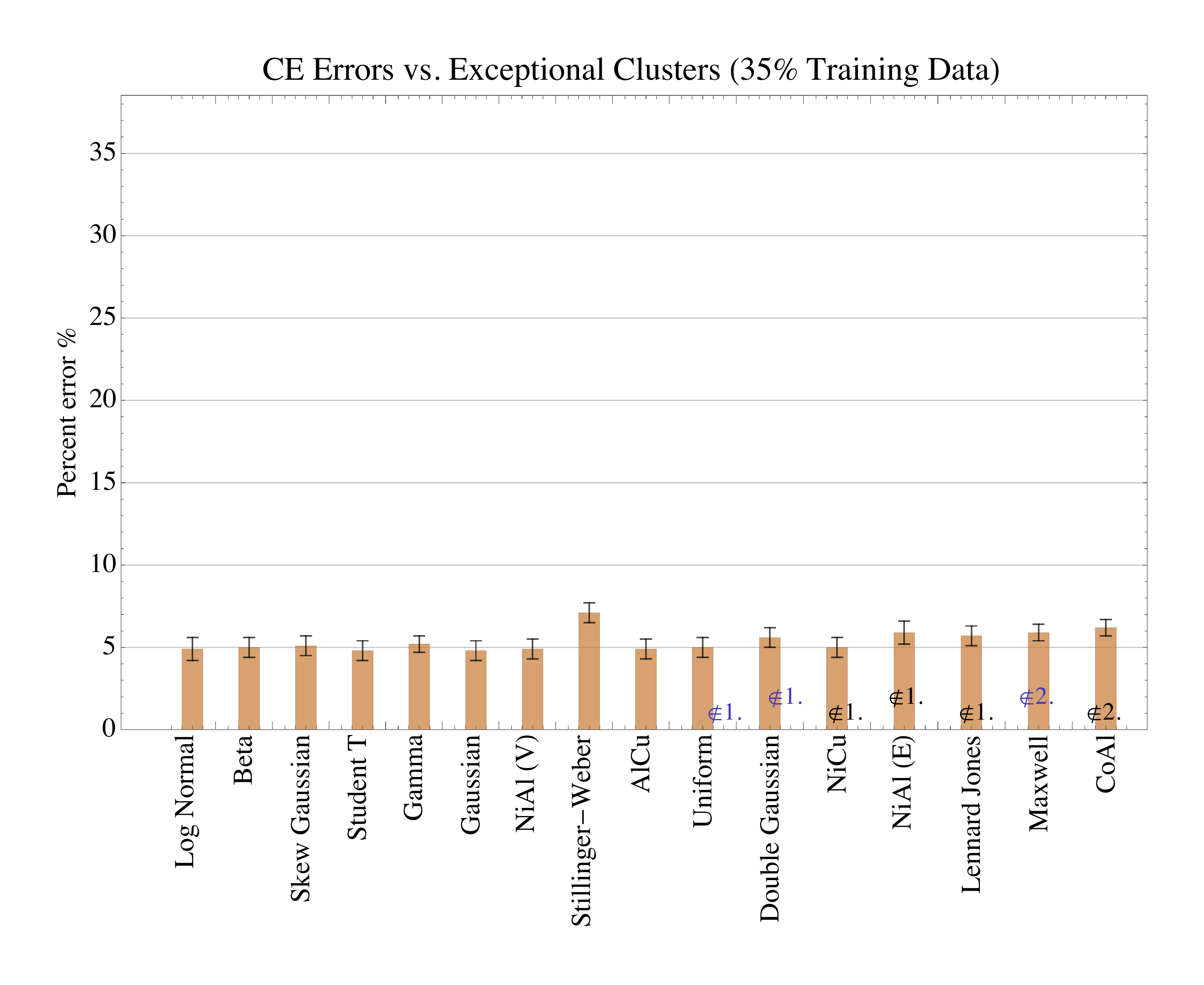}
        \caption{2\% error added}
        \label{fig:errE2}
    \end{subfigure}
    ~
    \begin{subfigure}[h]{0.49\textwidth}
        \includegraphics[width=\textwidth]{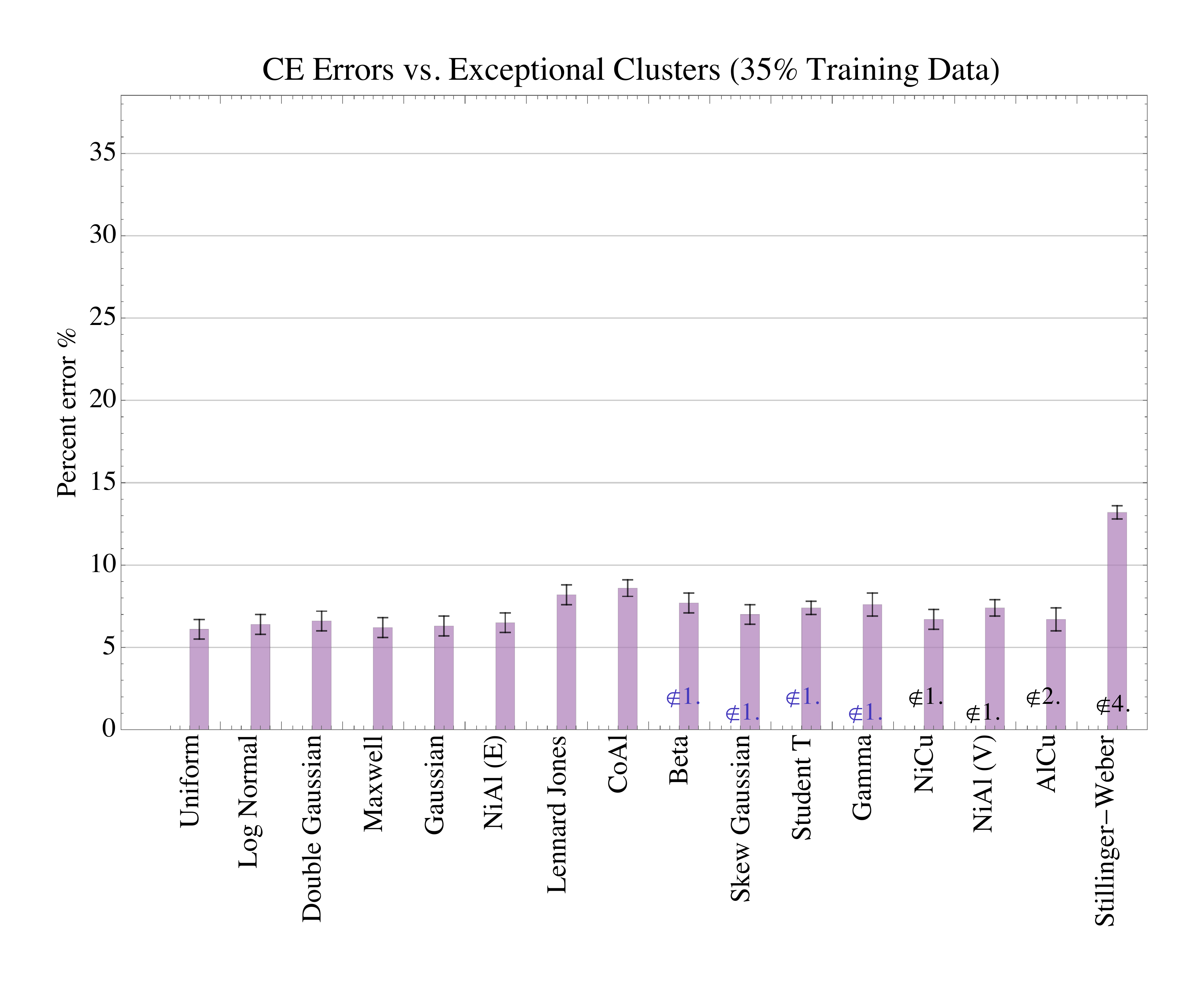}
        \caption{5\% error added}
        \label{fig:errE5}
    \end{subfigure}
    
    \begin{subfigure}[h]{0.49\textwidth}
        \includegraphics[width=\textwidth]{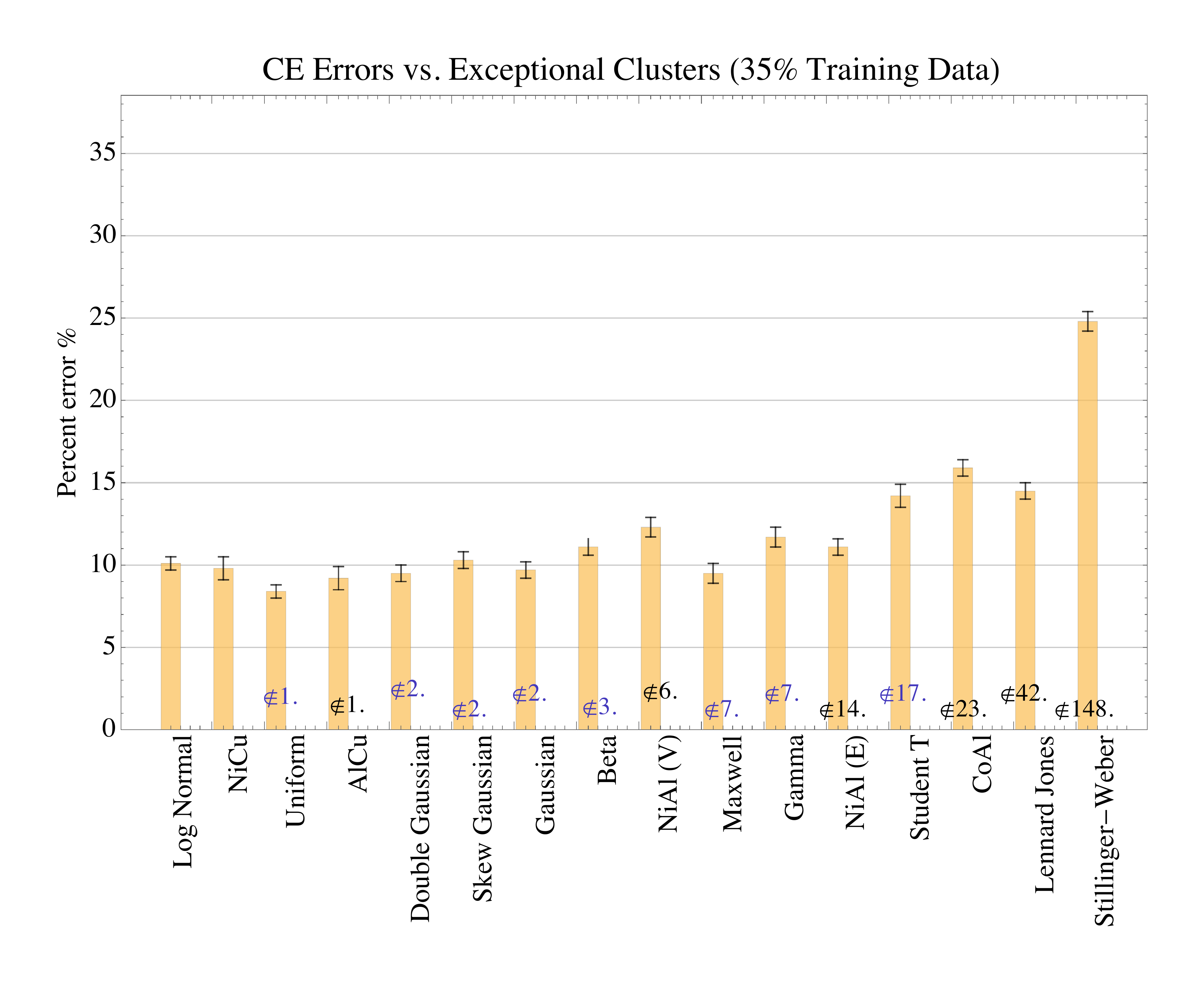}
        \caption{10\% error added}
        \label{fig:errE10}
    \end{subfigure}
~
    \begin{subfigure}[h]{0.49\textwidth}
        \includegraphics[width=\textwidth]{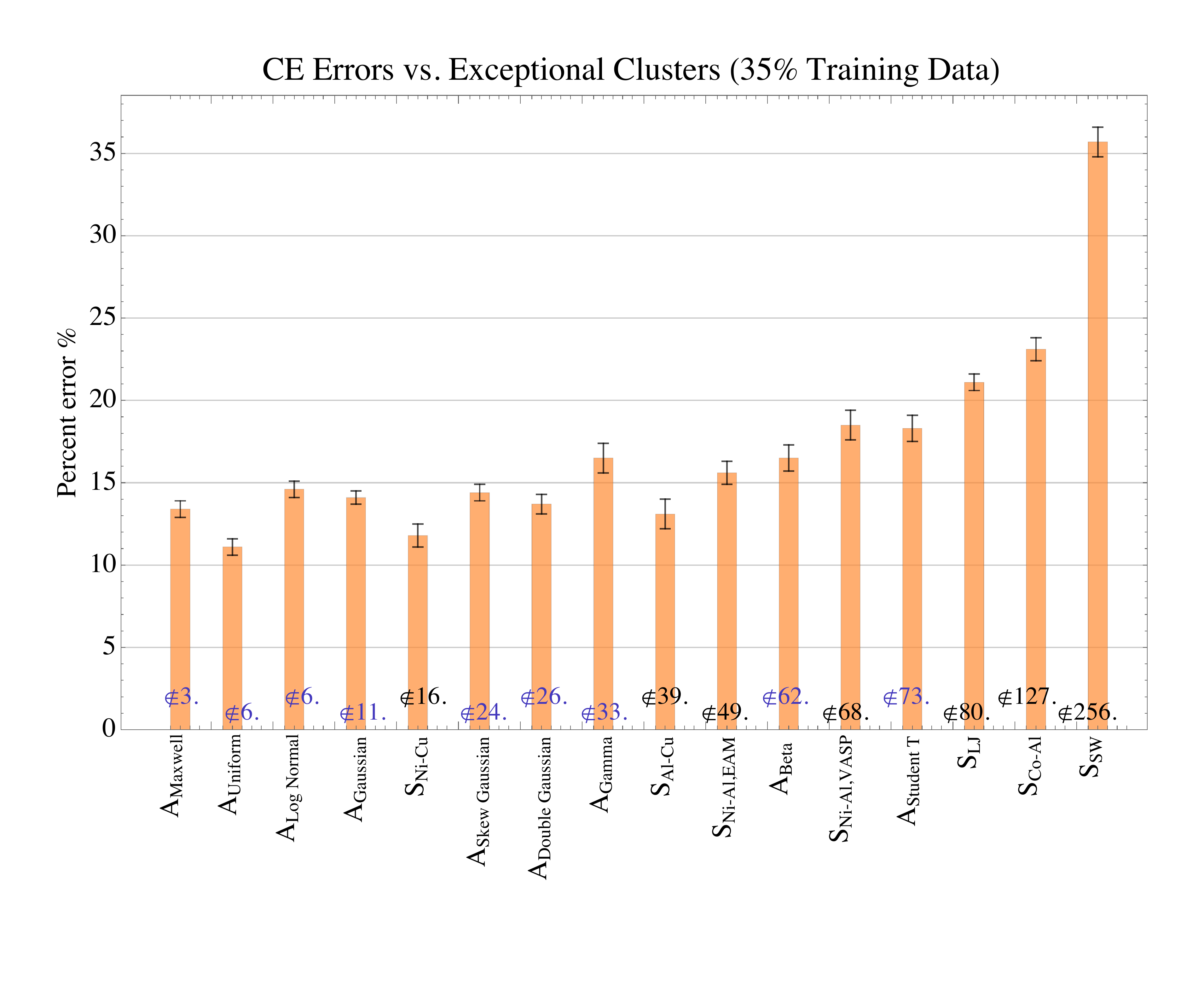}
        \caption{15\% error added}
        \label{fig:errE15}
    \end{subfigure}
    \caption{(color online)  Plot of predictive error over 65\% of the 
    structures for the toy model. The systems are ordered by $\not\in$ the 
    number of clusters that were used less than 25 times across all 100 CE 
    fits. These are considered exceptions to the overall fit for the system. 
    As for Figure \ref{fig:ErrorVsModelComplexity}, there is a definite trend 
    toward higher error for systems with more cluster exceptions. When 
    $\not\in$ is equal to zero, CE only uses the significant terms.} 
 \label{fig:ErrorVsSingleShow}
\end{figure*}

\begin{figure*}[h]
    \centering
    \begin{subfigure}[h]{0.49\textwidth}
        \includegraphics[width=\textwidth]{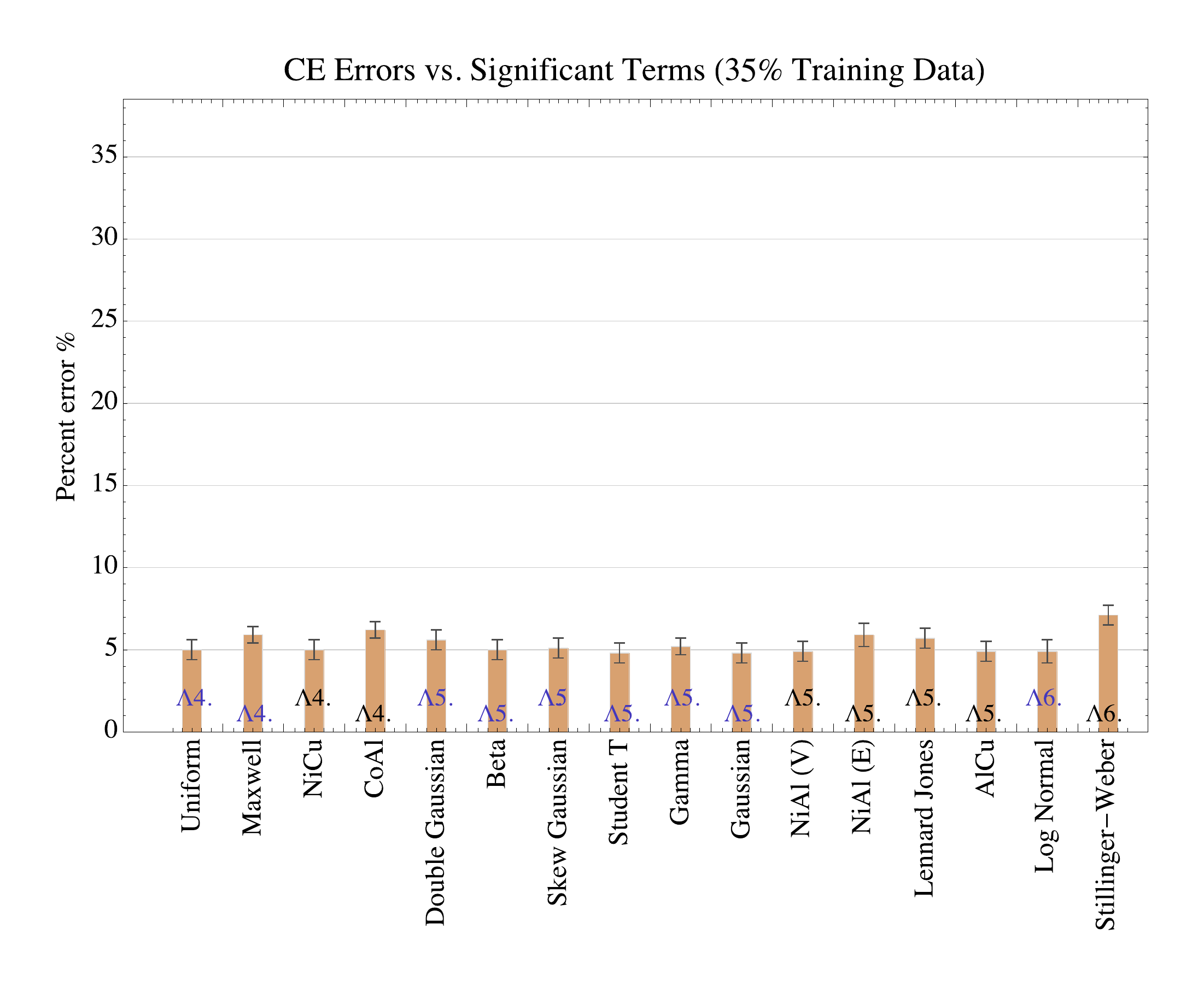}
        \caption{2\% error added}
        \label{fig:errS2}
    \end{subfigure}
    ~
    \begin{subfigure}[h]{0.49\textwidth}
        \includegraphics[width=\textwidth]{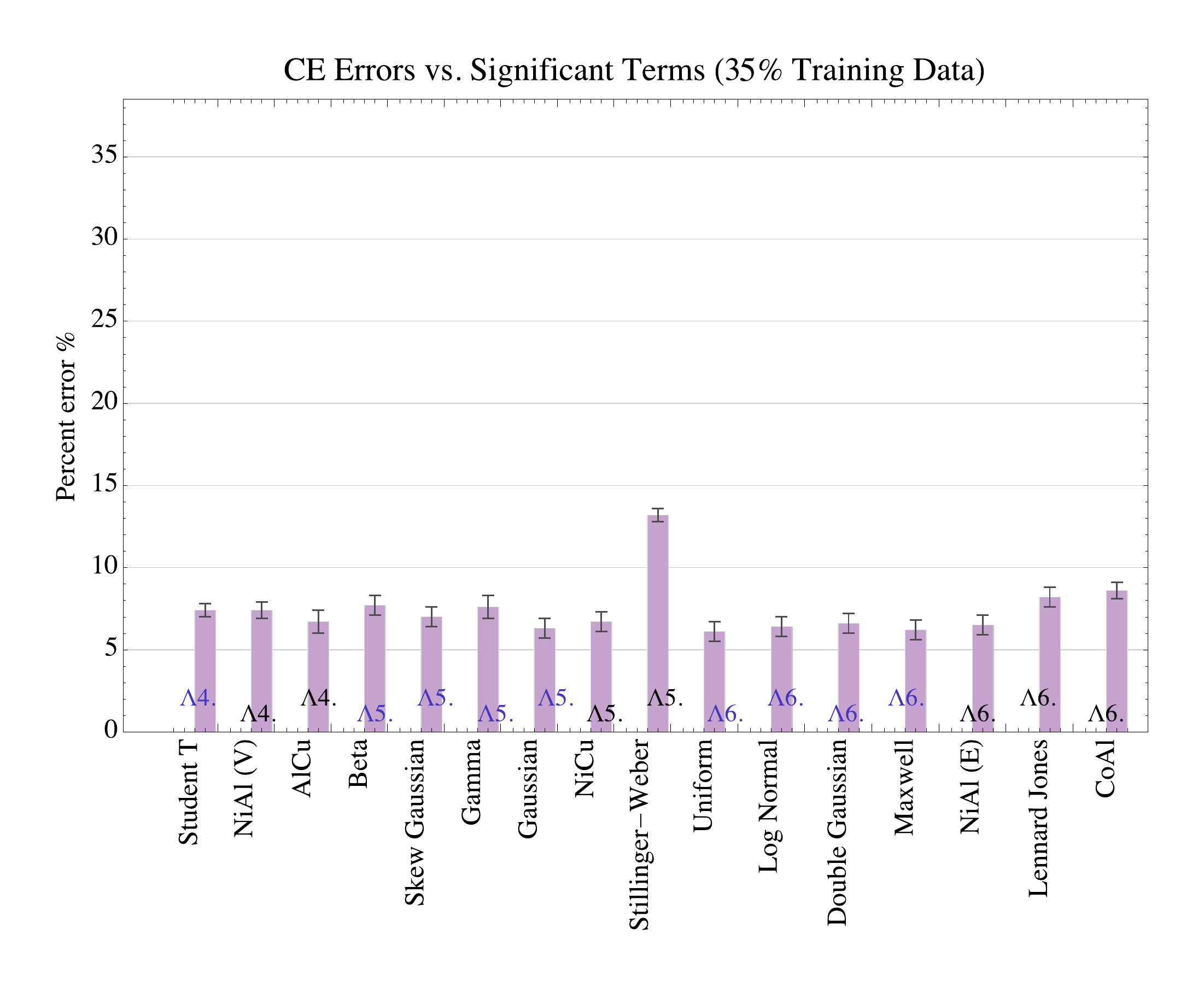}
        \caption{5\% error added}
        \label{fig:errS5}
    \end{subfigure}
    
    \begin{subfigure}[h]{0.49\textwidth}
        \includegraphics[width=\textwidth]{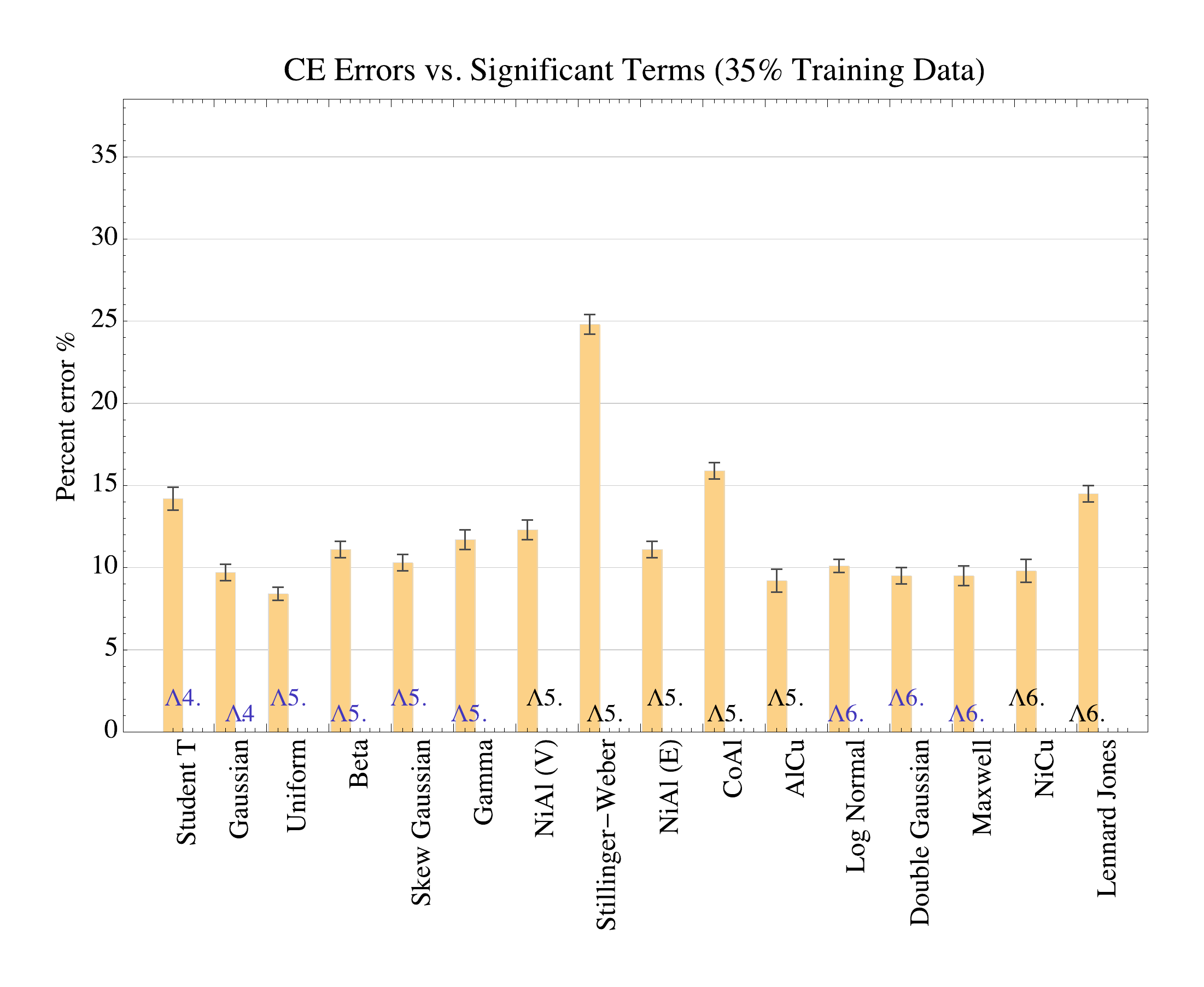}
        \caption{10\% error added}
        \label{fig:errS10}
    \end{subfigure}
~
    \begin{subfigure}[h]{0.49\textwidth}
        \includegraphics[width=\textwidth]{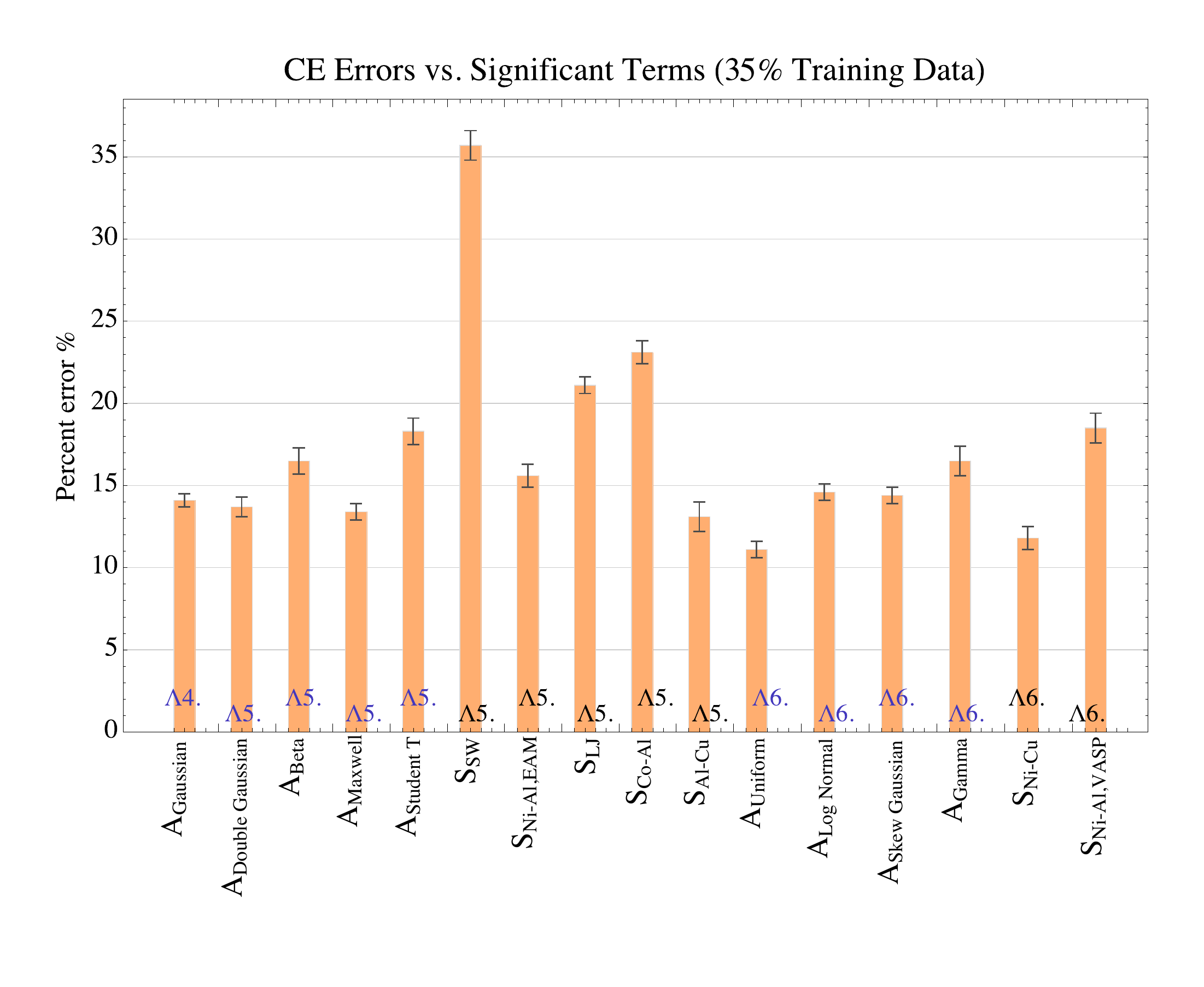}
        \caption{15\% error added}
        \label{fig:errS15}
    \end{subfigure}
    \caption{(color online)  Prediction error over 65\% of the structures with 
    the toy CE model. The errors are ordered by $\Lambda$, the number of 
    significant terms in the expansion. As expected, the values are close to 
    the known model complexity (5 terms) and the ordering once more appears 
    random. 
}\label{fig:ErrorVsSignificantTerms}
\end{figure*}

Fig. \ref{fig:ErrorVsModelComplexity}, \ref{fig:ErrorVsSingleShow},
\ref{fig:ErrorVsSignificantTerms} display the additional plots at different
error level that were not shown in the main text.

\subsection{\label{sec:levelS4}References}

\bibliographystyle{unsrt}

%

\newpage

\end{document}